\documentclass[useAMS,fleqn,usenatbib]{mnras}

\usepackage{newtxtext,newtxmath}

\usepackage[T1]{fontenc}
\usepackage{ae,aecompl}

\usepackage{graphicx}	

\newcommand{\be}{\begin{equation}}
\newcommand{\ee}{\end{equation}}
\newcommand{\bea}{\begin{eqnarray*}}
\newcommand{\eea}{\end{eqnarray*}}

\newcommand{\HH}{$H_2$}
\newcommand{\kms}{km s$^{-1}$}
\newcommand{\Spitzer}{\textit{Spitzer}}

\newcommand{\Herschel}{\textit{Herschel}}
\newcommand{\Planck}{\textit{Planck}}
\newcommand{\ALMA}{\textit{ALMA}}
\newcommand{\SPT}{\textit{SPT}}
\newcommand{\ACT}{\textit{ACT}}

\newcommand{\IRAS}{\textit{IRAS}}
\newcommand{\WISE}{\textit{WISE}}
\newcommand{\HST}{\textit{HST}}

\title[PASSAGES: \textit{Planck}-selected Luminous IR Galaxies]{PASSAGES: The Large Millimeter Telescope and ALMA Observations of Extremely Luminous High Redshift Galaxies Identified by the \textit{Planck} }

\author[D. Berman et al.]{
Derek A. Berman$^{1,2}$,
Min S.~Yun$^1$\thanks{E-mail: myun@umass.edu (MSY)},
K. C. Harrington$^3$, 
P. Kamieneski$^1$, 
J. Lowenthal$^4$,
\newauthor
B. L. Frye$^5$,
Q. D. Wang $^1$,
G. W. Wilson$^1$, 
I.~Aretxaga$^6$,
M. Chavez$^6$, 
R.~Cybulski$^1$,
\newauthor
V.~De~la~Luz$^{7}$, 
N.~Erickson$^1$,
D.~Ferrusca$^6$, 
D. H.~Hughes$^6$, 
A.~Monta{\~n}a$^{6,8}$, 
G.~Narayanan$^1$,
\newauthor
D.~S\'{a}nchez-Arg\"{u}elles$^6$,
F.~P.~Schloerb$^1$,
K.~Souccar$^1$,
E. Terlevich$^6$,
R. Terlevich$^{6,10}$, 
\newauthor
J.~A.~Zavala$^{11}$ 
\\
$^1$Department of Astronomy, University of Massachusetts, Amherst, MA 01003, USA\\
$^2$Department of Earth and Atmospheric Sciences, Cornell University, Ithaca, NY 14853, USA\\
$^3$European Southern Observatory, Alonso de C{\'o}rdova 3107, Vitacura, Casilla 19001, Santiago de Chile, Chile\\
$^4$Smith College, Northampton, MA 01063, USA\\
$^5$Department of Astronomy/Steward Observatory, 933 North Cherry Avenue, University of Arizona, Tucson, AZ 85721, USA\\
$^6$Instituto Nacional de Astrof\'{i}sica, \'{O}ptica y Electr\'{o}nica, Tonantzintla, Puebla, M\'{e}xico\\
$^{7}$Escuela Nacional de Estudios Superiores Unidad Morelia,
Universidad Nacional Aut\'onoma de M\'exico, Morelia, 58190, M\'exico\\
$^8$CONACYT Research Fellow - SCiESMEX, Instituto de Geof\'{i}sica, Unidad Michoac\'{a}n, Universidad Nacional Aut\'{o}noma de M\'{e}xico, \\
Morelia, Michoac\'{a}n, M\'{e}xico CP 58190\\
$^{9}$Consejo Nacional de Ciencia y Tecnolog\'{i}a, Av. Insurgentes Sur 1582, Col. Cr\'{e}dito Constructor, Del. Benito Ju\'{a}rez, C.P.: 03940, M\'{e}xico, D.F.\\
$^{10}$Institute of Astronomy, University of Cambridge, Madingley Road, Cambridge CB3 0HA, UK\\
$^{11}$National Astronomical Observatory of Japan, 2-21-1 Osawa, Mitaka, Tokyo 181-8588, Japan
}

\date{Accepted XXX. Received YYY; in original form ZZZ}

\pubyear{2022}

\begin{document}
\label{firstpage}
\pagerange{\pageref{firstpage}--\pageref{lastpage}}
\maketitle

\begin{abstract}
The Planck All-Sky Survey to Analyze Gravitationally-lensed Extreme Starbursts (PASSAGES) project aims to identify a population of extremely luminous galaxies using the Planck All-Sky Survey and to explore the nature of their gas fuelling, induced starburst, and the resulting feedback that shape their evolution. Here, we report the identification of 22 high redshift luminous dusty star forming galaxies (DSFGs) at $z=1.1-3.3$ drawn from a candidate list constructed using the \Planck\ Catalog of Compact Sources (PCCS) and \WISE\ All-Sky Survey. They are confirmed through follow-up dust continuum imaging and CO spectroscopy using AzTEC and the Redshift Search Receiver (RSR) on the Large Millimeter Telescope Alfonso Serrano (LMT).  
Their {\em apparent} IR luminosities span $(0.1-3.1)\times 10^{14} L_\odot$ (median of $1.2\times10^{14}L_\odot$), making them some of the most luminous galaxies found so far.  They are also some of the rarest objects in the sky with a source density of $\lesssim0.01$ deg$^{-2}$.  Our Atacama Large Millimeter/submillimeter Array (ALMA) 1.1 mm continuum observations with $\theta\approx0.4\arcsec$ resolution show clear ring or arc morphologies characteristic of strong lensing.  
Their lensing-corrected luminosity of $L_{\rm IR}\gtrsim 10^{13}L_\odot$ ($SFR\gtrsim10^3 M_\odot$ yr$^{-1}$) indicates that they are the magnified versions of the most intrinsically luminous DSFGs found at these redshifts. Our spectral energy distribution (SED) analysis finds little detectable AGN activity despite their enormous luminosity, and any AGN activity present must be extremely heavily obscured.
\end{abstract}



\section{Introduction}
\label{sec:intro}
 
An important unanswered question in our understanding of the cosmic star formation history is how a large quantity of cold gas is continuously and rapidly funnelled into the central 1-10 kiloparsec regions of dark matter halos to form $10^{11} M_\odot$ or more of stars over just a few hundred million years, only 2-3 billion years after the Big Bang.   
A related question is how this flow is regulated so that the stellar mass growth is shut off just at the right moment so that theoretical predictions and observational data on mass functions are in agreement at both high and low mass ends.  Observations have shown that the bulk of the stellar mass build-up during the peak of the cosmic star formation density at $z\approx 2-4$ (``Cosmic Noon") has occurred in massive galaxies with star formation rate (SFR) of 10s to 100s of solar masses per year with a characteristic gas depletion time of only $\sim10^8$ yrs  \citep[see][]{madau14,casey14}.  Their apparent local analogs, luminous infrared galaxies (LIRGs) and ultraluminous infrared galaxies (ULIRGs) with a similarly high star formation rate and gas depletion time, are rare in the current epoch, and their large gas inflow rates are associated with strong tidal disruptions and major mergers \citep[see the review by][]{sanders96}.  Strong tidal disruptions and mergers were likely important in early epochs as well, but the average gas accretion rate was also generally much higher in earlier epochs because mean gas mass density was higher.   The extremely luminous starburst galaxies with $L_{\rm IR}\gtrsim10^{12-13}L_\odot$ ($SFR \gtrsim 100-1000\, M_\odot$ yr$^{-1}$) discovered by submillimetre and far-infrared wavelength surveys \citep[``submillimetre galaxies" or SMGs; see reviews by][]{blain02,casey14} have no counterparts in the local universe.  Their extreme luminosity requires an uncertain extrapolation of our current knowledge on gas inflow, and an entirely different process might be at work.

One way to gain new insight into the gas inflow powering this rapid galaxy growth is to identify a population with the most extreme luminosity.  The maximal gas accretion associated with these objects might offer a way to discriminate among different competing processes.   The key feedback processes that regulate the galaxy growth, either radiative or mechanical, are also likely operating maximally at the same time, making them excellent laboratories for studying the nature and efficiency of the feedback processes as well. The main aim of our \Planck\ All-Sky Survey to Analyze Gravitationally-lensed Extreme Starbursts (PASSAGES) project is to identify a population of the most luminous galaxies using the \Planck\ All-Sky Survey and to explore the nature of the gas inflow, its induced starburst, and the resulting feedback and its impact.  As described below, nearly all of the objects we have identified are strongly magnified by gravitational lensing, and they also offer an exceptional opportunity to probe the relevant physical  processes on spatial scales of 100 parsecs or better, which is normally beyond the reach of the existing astronomical facilities ($0.1\arcsec \approx 1$ kpc at $z>1$).

A systematic search for high redshift objects hosting an extremely luminous starburst and/or a supermassive blackhole (SMBH) should be most productive in the infrared wavelengths, as they are expected to be heavily dust-obscured.  Considering the expected trend of decreasing {\em mean} metallicity with look-back time, this is not an obvious choice.  However, \citet{cen99} and others have shown that metallicity evolution is driven by local density, rather than by cosmic time, and the highest density regions quickly approach near-solar metallicity even at high redshifts.  This theoretical prediction is now supported by a growing list of systematic observational studies.  By analysing the 3D-HST treasury program and the CANDELS data, \citet{whitaker17} found a strong dependence of the fraction of obscured star formation on stellar mass, with remarkably little evolution with redshift out to $z=2.5$, with over 50\% and 90\% of star formation being obscured at log($M_*/M_\odot$) = 9.4 and 10.5, respectively.  \citet{deshmukh18} and others have shown that the importance of dust-obscured SF activity extends to $z\approx6$.  Among the submillimetre-selected high redshift galaxies  to $z\sim4$ and beyond, objects with little or no detectable optical/near-IR emission in even the deepest \HST\  exposures are common, presumably because of extinction and the red colour of their stellar hosts \citep{wang09,yun12,wang16,yamaguchi19,smail21}. 

Prior to the analysis of the compact sources identified by the \Planck\ all-sky multi-wavelength surveys, the most luminous high redshift population of objects identified through a {\em systematic} study were the hyperluminous infrared galaxies identified by the all-sky \WISE\ survey \citep{tsai15}.  A sampling of compact sources identified in the \Planck\ all-sky survey using the \Herschel\ Space Observatory has yielded some of the most luminous galaxies identified thus far \citep{canameras15,harrington16}, with most of their luminosity seen in the rest frame IR.\footnote{While \Herschel\ surveys have been extremely successful in finding thousands of high redshift DSFGs, their limited survey coverages (less than 10\% of the sky) have led to only a handful of $L_{\rm IR}\gtrsim10^{14} L_\odot$ sources -- see \citet{harrington16}.} Wide area cosmic microwave background (CMB) surveys at millimetre wavelengths, such as by the South Pole Telescope \citep[SPT;][]{carlstrom11} and the Atacama Cosmology Telescope \citep[ACT;][]{fowler07}, have also yielded a population of strongly lensed high redshift DSFGs that are less luminous but at higher redshifts to $z\gtrsim6$ \citep{greve12,marsden14,spilker16,su17,gralla20}.

In this paper, we report the identification of 22 luminous high-redshift sources at $z=1.1-3.3$ with apparent IR luminosity of $L_{\rm IR}\ge10^{13-14} L_\odot$ selected from the \Planck\ Catalog of Compact Sources \citep[PCCS;][]{planck13} and confirmed using the AzTEC 1.1 mm continuum imaging and CO spectroscopy obtained using the Redshift Search Receiver on the Large Millimeter Telescope.  We also report \ALMA\ 1.1 mm continuum imaging and photometry of a subset of 12 sources obtained at $\theta\sim0.4\arcsec$ resolution to confirm that all of these sources are strongly lensed high redshift dusty star forming galaxies.  We first describe how the high-redshift candidates are identified from the parent PCCS catalog using Wide-field Infrared Survey Explorer (WISE) photometry.  Their confirmation as high-redshift sources using the follow-up LMT and ALMA follow-up observations is described in \S~\ref{sec:obs}, and the characterisation of their spectral energy distribution (SED) and apparent IR luminosity is discussed in \S~\ref{sec:SEDanalysis}.  Finally, the physical origin of their enormous luminosity is explored by examining the evidence for luminous AGN and gravitational lensing, as well as the gas content and gas consumption time, in \S~\ref{sec:discussion}.

\section{Identification of \Planck\ High-Redshift Candidates}
\label{sec:Planck}

One of the data products the \Planck\ Collaboration has published as part of their general data release is the \Planck\ Catalogue of Compact Sources (PCCS), which is a set of single-frequency catalogues of Galactic and extragalactic compact sources covering the entire sky in the frequency range 30-857 GHz.  Following the first release catalogue based on the initial 15 months of mission \citep[PCCS1,][]{planck13}, the second release catalogue including the full survey was released in 2015. The second data release (DR2) consists of two sub-catalogues: the PCCS2 (4891 \& 1694 sources in the 857 GHz \& 545 GHz bands, respectively) with known reliability and the larger PCCS2E  (43290 \& 31068 sources in the 857 GHz \& 545 GHz bands, respectively, mostly within the Galactic mask) with ``unknown reliability"  \citep{planck26a}\footnote{See \citet{planck55} for more detailed discussions of the reliability of the PCCS2 sources.}.  Both \Planck\ compact source catalogs are heavily confusion limited so that the flux density and the reliability of each catalog entry are limited by its location on the sky and data reduction methods, rather than by thermal noise.  
The PCCS2 is deemed more reliable and thus supersedes the PCCS1, but we found it useful to examine both catalogs since we are exploring the faintest sources in each catalog where the catalog completeness is poor.

When we started this project, only the PCCS1 was available, and we developed and built the high redshift source identification technique using the PCCS1 only.  When the PCCS2 and PCCS2E became available, we applied the same technique developed earlier and found similar results but also with some notable differences.  One of the key limitations we encountered with both the PCCS1 \& the PCCS2 is that the photometry for a given high redshift candidate source is frequently incomplete for the full set of frequency bands.  The PCCS catalogs are published by the \Planck\ project for each frequency band separately, and we had to create a band-merged catalog ourselves through position-matching.  Because these candidate high-redshift sources are typically among the faintest sources near the reliability limit, missing a flux density entry in one or more bands in our band-merged catalog is not uncommon.  This is an important limitation because any sources with missing photometry in the critical high frequency bands have to be excluded in our selection method, which is based on a colour-magnitude filter (see below for further discussions).   In the end, we constructed our master band-merged photometry catalogue using primarily the PCCS2 and filling in any missing photometry from the PCCS1 if possible.  This band-merged catalog is used in the subsequent filtering of the data to identify the candidate high redshift \Planck\ sources.

We note that the \Planck\ collaboration has published its own \Planck\ High-redshift Source Catalog \citep[PHZ;][]{planck39} using a different source identification and photometric and colour analysis.  We find a surprisingly small overlap between the PHZ and our own high-redshift candidate catalog, and we discuss this further in \S~\ref{sec:discussion}.

\subsection{Filtering the \Planck\ Catalogues} \label{sec:PlanckFilter}

\begin{figure*}
\includegraphics[width=10cm]{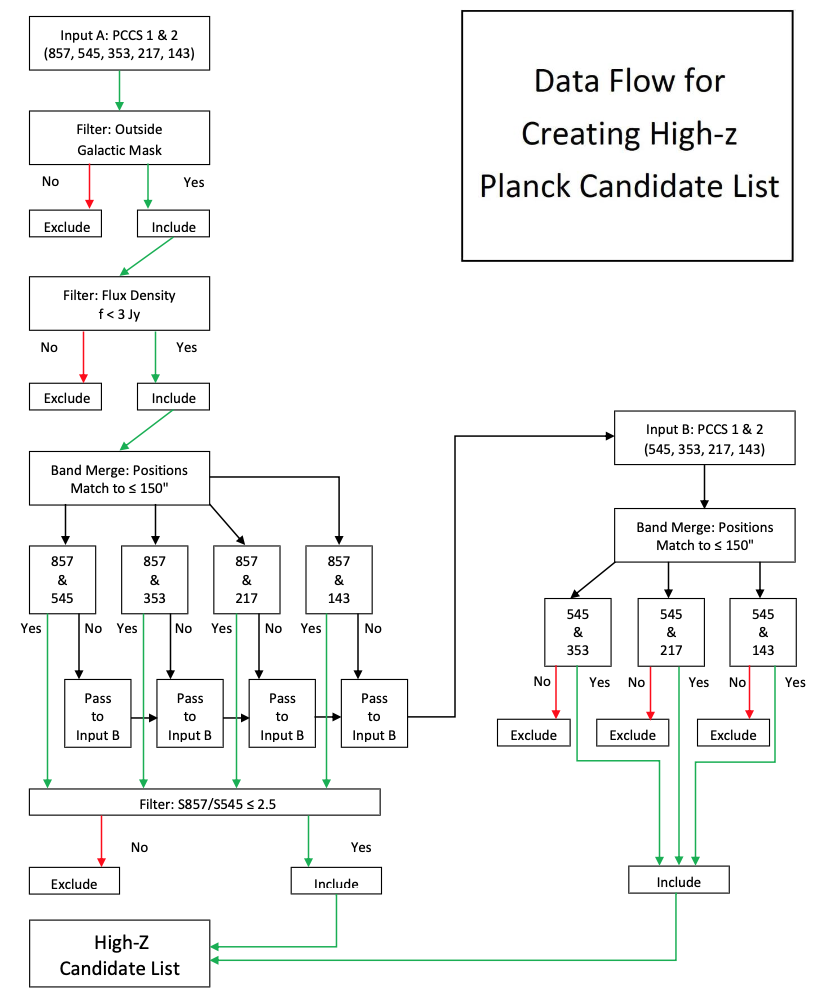}
\caption{A data flow diagram for creating the final \Planck\ high-redshift candidate list from the PCCS1 \& PCCS2 catalogues.  The second pass through the colour-magnitude filter on the right side of the diagram is for the higher redshift ($z\gtrsim3$) sources whose dust peak is shifted out of the 857 GHz band and thus are undetected in that band.}
\label{fig:flowchart}
\end{figure*}

The majority of the sources in the PCCS catalogues are expected to be radio AGNs (including blazars) in the 217 GHz and lower frequency bands and star forming galaxies and Galactic cirrus in the 353 GHz and high frequency bands \citep{planck7,planck13}.  To identify high redshift dusty star forming galaxies, we construct a set of filters that will remove foreground Galactic cirrus sources with $T_{\rm d} \approx 20$ K and radio AGNs.  The former is accomplished by adopting the Galactic Foreground filter established by the Planck Collaboration.  Removing radio AGNs is achieved by constructing a classic ``colour-magnitude" filter that exploits the distinct shape differences between thermal dust emission and power-law non-thermal  AGN SEDs.  The main ideas behind these filters are fairly simple and intuitive, and they are nearly the same as those described by \citet{planck39}.  An important difference in our approach is that we kept the filtering simpler so as to minimise the negative impact of relatively low SNR in the photometry and to prevent introducing too many assumptions and potential biases.  Instead, we exploit the wealth of multi-wavelength data available to refine the selection of high redshift galaxy candidates and ultimately to prioritise the targets for follow-up observations using the Large Millimeter Telescope (see \S~\ref{sec:obs}).  The flow of the data through this filtering process is summarised as a diagram in Figure~\ref{fig:flowchart}, and the key steps are explained in greater detail below.  

\subsubsection{Galactic Foreground Mask}

The removal of foreground Galactic cirrus sources is accomplished primarily by excluding sources in the directions of the Galactic Plane and known filamentary Galactic foreground structures.  Specifically, this is accomplished by (1) applying a Galactic latitude limit of 30 degrees (i.e., $|b|\ge30^\circ$); and (2) excluding all sources falling within the Galactic Zone and Filament masks\footnote{\url{https://irsa.ipac.caltech.edu/data/Planck/release_2/ancillary-data/}} derived by the \Planck\ team using neutral hydrogen column density \citep[see ][for further details]{planck11}.   The Galactic mask we adopted is essentially the same as the one adopted by the Planck team for their high-redshift source candidate identification \citep[see the Figure~1 by][]{planck39}, except our coverage is slightly larger since we are not bound by the IRAS survey coverage.  The resulting total sky coverage for our high redshift galaxy search is 27.9\% or 11,500 square degree in area.

The total number of sources remaining after the Galactic mask is applied to the 857 GHz based band-merged catalogue is 4825, 3965, and 1827 sources for the PCCS1, the PCCS2, and the PCCS2E, respectively (see Table~\ref{tab:sourcestat}).   The fraction of the PCCS1 sources found outside the Galactic mask is only 20\%, suggesting that a large majority of the 857 GHz selected sources are associated with Galactic structure.  This is also obvious in the source distribution shown in Figure~2 by \citet{planck26a}.  The total combined number of 857 GHz sources in the PCCS2 and PCCS2E is nearly twice as large as in the PCCS1, but the number outside the Galactic mask region increases only by 20\%.  Therefore, most of the new sources identified in the DR2 catalogues are within the Galactic mask region.  

The situation is similar in the 545 GHz band in that the total number of sources is doubled in the DR2 compared to DR1, but the increase outside the Galactic mask region is almost negligible (1\%).  The fraction of sources outside the Galactic mask region is much smaller, 1064/16933 = 6.3\% and (403+672)/(1694+31068) = 3.3\% for the DR1 and the DR2 catalogues.  The total number of 545 GHz band-merged catalogue sources that are undetected in the 857 GHz band remains virtually the same between the two data releases.  It is also interesting that the fraction of ``highly reliable" 545 GHz band sources in the PCCS2 catalogue drops to only 24\% of the total outside the Galactic mask region, in contrast to the 857 GHz band (75\%).  This illustrates the complex way the foreground confusion plays an important role in shaping these \Planck\ compact source catalogues.

\begin{figure}
\includegraphics[width=7cm]{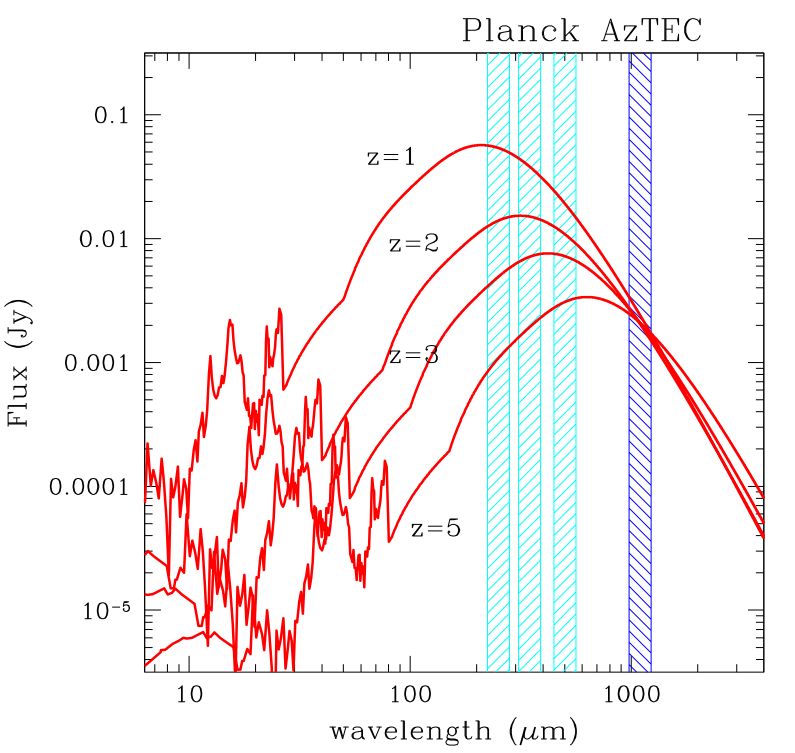}
\caption{A model spectral energy distribution of a dusty star forming galaxy observed at a redshift of $z=1$, 2, 3, \& 5.  The dust peak of a $z=1$ SFG is located near $\lambda \approx 200\ \mu$m, leading to a monotonic decrease in observed flux density in the 857 GHz band, to 545 GHz band, and then to 353 GHz band of the \Planck.  As a galaxy is observed at an increasingly higher redshift, the dust peak shifts to a longer wavelength.  Exploiting this systematic trend is the basic idea behind our high-redshift source selection algorithm. The template SED shown is that of the average $z=2$ SFG observed by \Herschel\ and \Spitzer\ with a derived star formation rate $SFR = 344\, M_\odot$ yr$^{-1}$ \citep{kirkpatrick12}.   
This figure illustrates how the flux density ratio of a DSFG flattens across the \Planck\ and AzTEC bands with increasing redshift.    
The detection limit of \Planck\ PCCS catalogs is about 300 mJy in 857 GHz and 545 GHz (at the top edge of this plot), so an object must have an apparent $SFR \ge 3000\, M_\odot$ yr$^{-1}$ in order to make it into our sample.}
\label{fig:SED}
\end{figure}

\begin{table}
 \centering
 \caption{Summary of Planck Catalogue Filtering}
 \label{tab:sourcestat}
  \begin{tabular}{lccc}
  \hline
 Frequency & Total & Mask only & Mask+Filter \\
  \hline
  857 GHz Band: & & &  \\
  \ \ \ \ \ PCCS1  &  24381 &  4825  &  311 \\
  \ \ \ \ \ PCCS2  &  4896   &  3965  &  218 \\
  \ \ \ \ \ PCCS2E & 43290 &  1827  &   97  \\
  \hline
  545 GHz Band: & & &  \\
  \ \ \ \ \ PCCS1  &  16933 &  1064  &  111 \\
  \ \ \ \ \ PCCS2  &   1694  &    403  &  112 \\
  \ \ \ \ \ PCCS2E & 31068 &   672   &    19 \\
  \hline
\end{tabular}
\end{table}

\subsubsection{Selection by Spectral Energy Distribution and Apparent IR Luminosity \label{sec:Pcolmag}}

The intrinsic spectral energy distribution (SED) of dusty star forming galaxies (DSFGs) is characterised by a prominent dust peak with average dust temperature of $T_{\rm d} \approx 30-40$ K \citep[see the review by][]{casey14}. The observed dust peak shifts to a longer wavelength with increasing redshift with apparent characteristic temperature of $T_{\rm d}/(1+z)$ as shown in Figure~\ref{fig:SED}.  As a consequence, the peak of the dust SED shifts through the three highest frequency \Planck\ bands at $z\approx 1.5$, $z\approx 2.5$, and $z\approx 4$, respectively.  In contrast, the Galactic cirrus emission with $T_{\rm d}\ \approx 15-25$ K peaks at a much shorter wavelength, around $\lambda\approx 100-150\, \mu$m, and this difference in SED can be an important discriminator between high-redshift DSFGs and Galactic cirrus sources.

Another potentially important discriminator is the measured flux density in these \Planck\ bands.  The DSFG SED template with a star formation rate of $SFR = 344 M_\odot$ yr$^{-1}$ by \citet{kirkpatrick12} predicts flux densities of $\sim50$ mJy \&  $\sim30$ mJy in the \Planck\ 857 GHz and 545 GHz bands at $z=1$ and less than 20 mJy in both bands at $z\ge2$ (see Figure~\ref{fig:SED}).  Since the flux density sensitivity for these bands in the PCCS1 \& the PCCS2 is $\sim100$ mJy, only the DSFGs with apparent $SFR\gtrsim1000\, M_\odot$ yr$^{-1}$ can be detected by \Planck\ at $z\lesssim 1.5$.   At $z\ge2$, detection by \Planck\ would require $SFR\ge2000\, M_\odot$ yr$^{-1}$, which is near the upper bound of the most active SFGs discovered previously \citep[e.g.,][]{yun12,yun15}.  The predicted flux density falls rapidly in the longer wavelength bands as $S \propto \lambda^{-3.5}$, making a detection much less likely in the 353 GHz and lower frequency bands.

\begin{figure*}
\includegraphics[height=7cm]{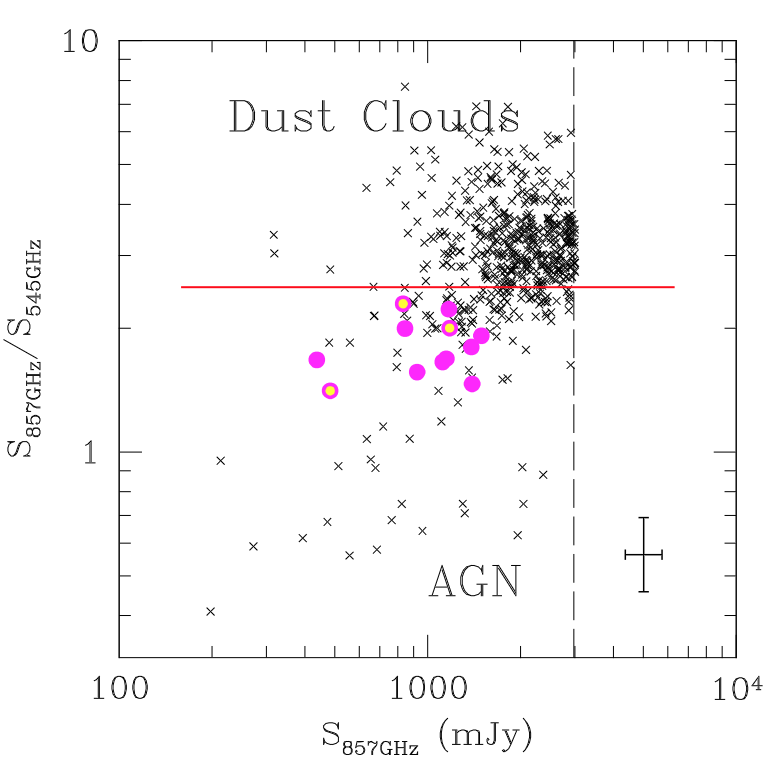}
\includegraphics[height=7cm]{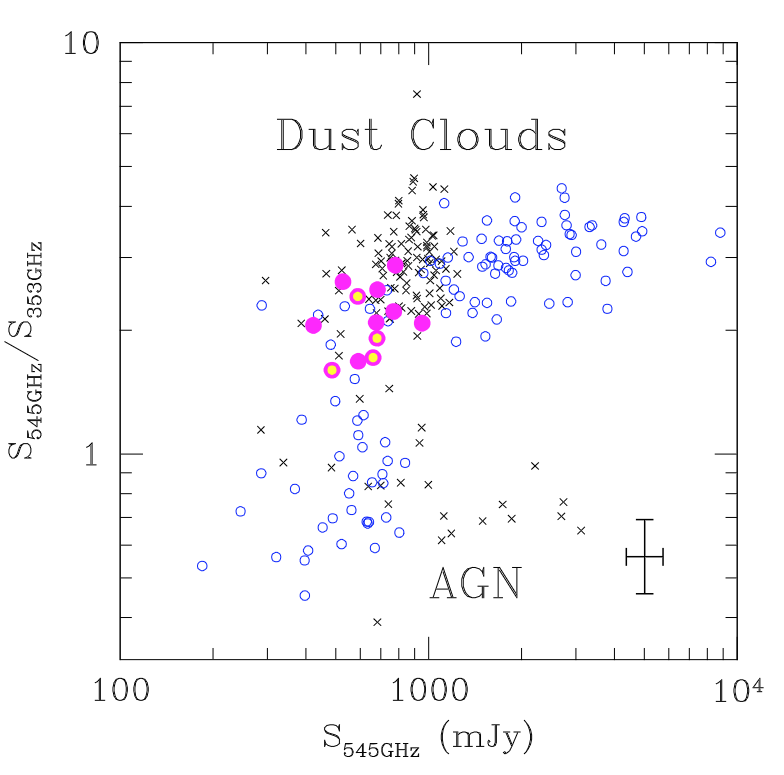}
\caption{Plots of flux density ratios as a function of flux density for \Planck\ PCCS sources.  The flux ratios of $S_{\rm 857GHz}/S_{\rm 545GHz}$ for the PCCS sources identified in the 857 GHz band are shown as small crosses on the left panel.  The flux density cutoff of $S_{\rm 857GHz}=3$ Jy is shown with a vertical dashed line.  The red horizontal line at $S_{\rm 857GHz}/S_{\rm 545GHz}\ge 3$ is used to exclude PCCS sources associated Galactic dust clouds (also low-z galaxies) from the flat spectrum radio AGNs and high redshift dusty galaxies \citep[filled magenta points,][also see \S~\ref{sec:literature}]{herranz13,canameras15,harrington16}. A subsample from the \Planck-\Herschel\ study by \citet{canameras15} is shown as magenta circles. 
The $S_{\rm 545GHz}/S_{\rm 353GHz}$ ratios for the PCCS sources identified in the 545 GHz band are shown as small crosses on the right panel, and all sources undetected in the 857 GHz band are shown as blue circles.  Confirmed high redshift sources (magenta dots) are on the bottom range of the $S_{\rm 545GHz}/S_{\rm 353GHz}$ ratio for the dust sources.  Almost all sources with $S_{\rm 545GHz}/S_{\rm 353GHz}\lesssim 2$ are known flat spectrum radio AGNs and blazars such as 3C~84, 3C~279, and 3C~454.3, with a characteristic flat spectrum of $S_{\rm 857GHz}/S_{\rm 545GHz}\lesssim 1$.  Typical uncertainties for the photometry and flux ratio are shown on the bottom right corner of each figure.
}
\label{fig:Pcolmag}
\end{figure*}

Our high redshift DSFG selection algorithm is illustrated on the left panel of Figure~\ref{fig:Pcolmag}.   Since the great majority of sources in the PCCS catalogs are foreground Galactic cirrus sources, our primary aim here is to exclude as many of the Galactic dust clumps as possible by (1) excluding sources with a high flux density ratio between the 857 GHz band to the 545 GHz band (i.e., $S_{\rm 857GHz}/S_{\rm 545GHz}\ge 2.5$) and (2) also excluding all sources brighter than 3 Jy in the 857 GHz band.  The first criterion is a simple and effective way to capture the SED differences illustrated in Figure~\ref{fig:SED}, while the second criterion excludes sources that are too bright to be extragalactic even in the presence of lensing (see below).  The precise details of the selection criteria are somewhat arbitrary. The specific values adopted do change between independent studies, in each case resulting in a successful set of \Planck\ and \Herschel\ selected high redshift sources (filled magenta points in  Figure~\ref{fig:Pcolmag})\footnote{The magenta points in Figure~\ref{fig:Pcolmag} also include 5 additional \Planck\ sources already known in the literature, as discussed in \S~\ref{sec:literature}.  They are not used in the definition of the colour-magnitude cuts since they are identified from the high-redshift candidate list produced by this study.}.  Adopting the template SED for the average $z=2$ dusty star forming galaxy by \citet{kirkpatrick12} shown in Figure~\ref{fig:SED}, the upper flux density limit of $S_{\rm 857GHz}=3$ Jy translates to a star formation rate of $SFR\approx 20,000\, M_\odot$ yr$^{-1}$ ($L_{\rm IR}\approx 10^{14.3}\, L_\odot$) at $z=1$ and $SFR\gtrsim 70,000\, M_\odot$ yr$^{-1}$ ($L_{\rm IR}\approx 10^{14.9}\, L_\odot$) at $z\ge2$.  These $SFRs$ are unusually large, explainable only with the aide of the largest lensing magnification ever seen, as discussed by \citet{harrington16}.  There is a limit to a plausible magnification factor for an extended object like a galaxy,  which is rarely much larger than 20-30 \citep[e.g.,][]{bussmann13,bussmann15,spilker16}.  Therefore, an apparent $SFR$ of 30,000 to 60,000 $M_\odot$ yr$^{-1}$ is the upper bound of the most luminous far-IR sources \Planck\ might detect.  The flux density limit we chose is twice as large as the brightest \Planck\ high-redshift source discovered thus far, and it is 7 times higher than the measured flux density\footnote{$S_{350\mu} = 437\pm14$ mJy measured by the \Herschel\ SPIRE instrument, from NASA/IPAC Infrared Science Archive (IRSA).} of the most luminous IR QSO known, APM 08279+5255 \citep{irwin98,weiss07} -- see further discussions on possible dusty quasars in \S~\ref{sec:AGN}. 

As the dust peak shifts out of the 857 GHz band and into the 545 GHz band at $z\approx2$, the observed flux density in the 857 GHz band begins to fall rapidly with increasing redshift, as shown in Figure~\ref{fig:SED}, and $z\gtrsim3$ DSFGs might drop out of the \Planck\ 857 GHz band source catalog entirely.  Instead, these higher redshift sources would appear in the band-merged catalog based on the \Planck\ 545 GHz catalog.  Indeed four (out of 15) high-redshift \Planck\ sources previously identified using the \Herschel\ data are  857 GHz dropouts.  As summarised in Table~\ref{tab:sourcestat}, the number of 545 GHz band selected sources are far fewer than the 857 GHz detected sources, and the confirmed high redshift \Planck\ sources are much harder to identify by their flux density or by their $S_{\rm 545GHz}/S_{\rm 353GHz}$ flux density ratio (see the right panel of Fig.~\ref{fig:Pcolmag}).  Therefore, the entire 545 GHz selected sources are passed along to the next step without any further filtering.  The fact that they are undetected in the 857 GHz band immediately rules out their possible Galactic cirrus origin.

The net result of applying both the Galactic foreground mask and colour-magnitude filter to the \Planck\ 857 GHz selected sources is that a total of 311 and 315 high-redshift candidate sources are identified from the PCCS1 and PCCS2 catalogues, respectively (see the far right column of Table~\ref{tab:sourcestat}).  For the 545 GHz selected sources without an 857 GHz band detection, a total of 111 and 131 high-redshift candidate are identified without any further colour or magnitude cut.  The PCCS1 \& PCCS2 high-redshift candidate lists largely overlap as expected, but there are also notable differences.  Therefore, both lists are carried through for the remainder of the sample selection process in parallel.  

The total number of high redshift \Planck\ source candidates we have identified here is much smaller than the total of 2151 high redshift \Planck\ source candidates identified in essentially the same area by the \Planck\ project team using their own SED modelling \citep{planck39}.  This discrepancy arises from differences in the parent catalogs and data filtering, which we discuss in detail in \S~\ref{sec:PHZ}.

\subsubsection{Removal of Nearby Galaxies and Radio AGNs \label{sec:foreground}}

Although we are able to remove more than 98\% of likely foreground Galactic sources through the filtering described above, two important classes of contaminating sources remain: (1) nearby galaxies that are bright enough to be detected by \Planck, and (2) radio AGNs that are strongly beamed.   A check on the coordinates revealed that a large number of the remaining \Planck\ sources are in the direction of nearby galaxies such as the Large and Small Magellanic Clouds, M31, and M33.  Some of the 545 GHz selected sources are also well known 3C radio sources and bright radio quasars and blazars.  

We remove potentially contaminating nearby galaxies and radio AGNs by performing a cone search for any known extragalactic sources within the 2.5 arcmin radius \Planck\ beam in the NASA Extragalactic Database (NED)\footnote{\url{https://ned.ipac.caltech.edu}}.  Specifically, we removed any \Planck\ sources from our candidate list if one or more optically bright galaxies (typically a galaxy at $z\lesssim 0.1$ in the NGC or an IC catalogue) or a bright radio source (e.g., brighter than 20 mJy in the 1.4 GHz NVSS or FIRST Survey) is found in the NED cone search.

\subsection{Identifying and Prioritising the Optical and Near-IR Counterpart Candidates \label{sec:OIR}}

\begin{figure}
\centering
\includegraphics[height=7cm]{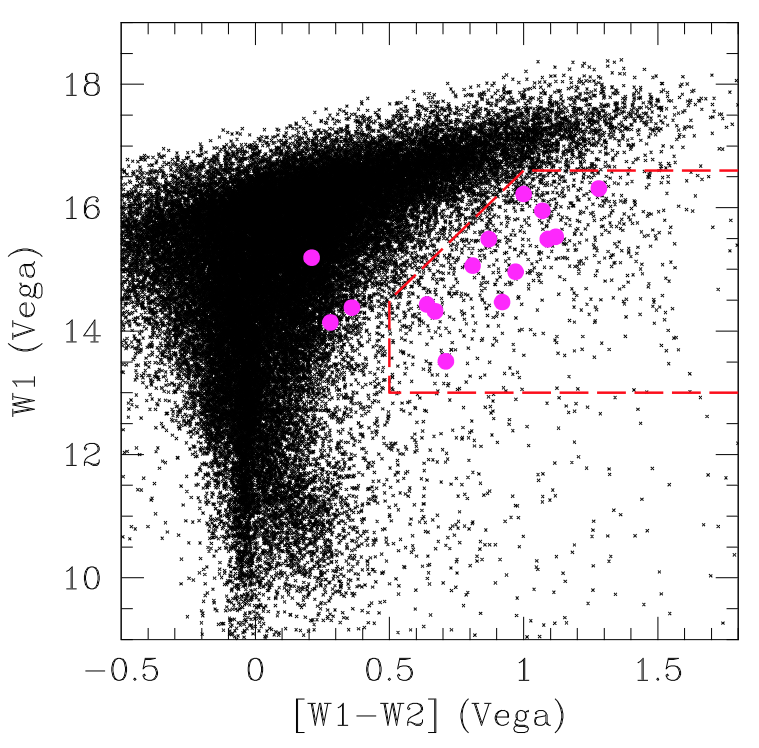}
\caption{A colour-magnitude plot of \WISE\ $W1$ (3.4 $\mu$m) \&  $W2$ (4.6 $\mu$m) photometry for the sources found in the \Planck\ high-redshift candidate source fields.  Large filled (magenta) circles are the 15 \Planck\--\Herschel\ sources reported by \citet{herranz13}, \citet{canameras15}, and \citet{harrington16}, while the small dots show more than 50,000 \WISE\ sources found within the 2.5\arcmin\ search radius from the \Planck\ PCCS catalogue positions of the high-redshift candidate sources.  A long dashed line marks the selection boundary for the area of candidate high-redshift sources -- see the text for the details.
}
\label{fig:Wcolmag}
\end{figure}

Identifying specific individual high redshift dusty star forming galaxy or galaxies responsible for the submillimetre continuum emission is another highly challenging task.  Each 2.5 arcmin radius \Planck\ beam area includes $\approx$20,000 galaxies at the depth of the \HST\ survey of the Hubble Ultra-Deep Field \citep{beckwith06}, and additional information is required to accomplish this task.  Here, we utilise the \WISE\ All-Sky Survey \citep{wright10} to identify the likely optical and near-IR counterparts to our high-redshift DSFGs by exploiting their characteristic red SED in these wavelength bands.

High redshift dusty star forming galaxies at $z\ge1$ are exceedingly faint and red in their rest frame UV bands, and their intense star forming regions can be heavily obscured by dust \citep[see][and references therein]{yun08,yun12,whitaker17,smail21}.  \citet{yun08} have shown that these galaxies have an extremely red UV to near-IR SED, characteristic of heavily obscured young stellar populations, and yet they are bright and are easily detected in their rest frame photospheric emission traced by the \Spitzer\ 3.6 $\mu$m and 4.5 $\mu$m bands.  The combined effects of their intrinsic red colours and a favourable $k$-correction make these high redshift dusty star forming galaxies easily distinguishable from the foreground population -- see Figures~12 \& 13 of \citet{yun12}.  Unfortunately, the necessary \Spitzer\ photometry is not available in most parts of the sky, and the \WISE\ All-Sky Survey \citep{wright10} offers the next best photometric data in the near-IR bands.  Although significantly shallower and 3 times worse in angular resolution ($\sim$6$\arcsec$ FWHM) than the \Spitzer\ IRAC data, the \WISE\ All-Sky survey is sensitive enough to detect the high redshift \Planck\ sources (see below and Section~\ref{sec:SED}). 

The near-IR colour-magnitude filter we adopted for separating dusty star forming galaxies at $z\ge1$ from the foreground galaxy population is shown in Figure~\ref{fig:Wcolmag}. The basic principle behind this filter is to identify and exclude \WISE\ sources that are too blue and too bright to be $z\gtrsim1$ DSFGs, with a magnitude cutoff to exclude all low SNR sources in the \WISE\ catalog.  As was first shown by \citet{wright10}, nearly all of the nearby galaxies have the characteristic \WISE\ colours of [$W1-W2$] $\approx$ 0.0 in Vega magnitudes because their light is dominated by stellar photospheric emission.  Our colour selection of [$W1-W2$] $\ge 0.5$ effectively removes all but the most dust obscured galaxies and AGNs \citep{wright10}.  The lower magnitude cutoff of $W1 \le 16.6$ is the $5\sigma$ completeness limit of the \WISE\ All-Sky Source Catalog\footnote{\url{http://wise2.ipac.caltech.edu/docs/release/allsky/}}, and a diagonal wedge between ($W1$,$W2$) = (+0.5, +14.5) and ($W1$,$W2$) = (+1.0, +16.6) was added to mitigate the impact of the noisier $W2$ band photometry ($5\sigma$ completeness limit of $W2 \le 15.5$).  Lastly, a bright end magnitude cutoff of $W1 \ge 13.0$ is added as an empirical filter to exclude galaxies that are too bright to be at $z\ge1$ (see below).  This bright end cutoff was designed to remove any confusing bright foreground galaxies, but it is still not restrictive enough to remove any of the \WISE-selected HyLIRGs reported by \citet{tsai15}, which is important in our IR AGN discussion (see \S~\ref{sec:AGN}).  

A direct outcome from this \WISE\ colour-magnitude filtering is that the individual galaxies likely responsible for the \Planck\ detection are identified with  one arcsecond positional accuracy.  With this localisation, these \WISE-selected \Planck\ sources (``\Planck-\WISE\ sources" hereafter) can now be examined further through targeted follow-up observations and SED analysis.

The effectiveness of this colour-magnitude filter and the impact of a possible foreground confusion is tested by examining those of the \Herschel-selected \Planck\ sources, shown as large filled circles in Figure~\ref{fig:Wcolmag}.  A total of 12 (out of 15) \Herschel-selected sources fall within the selection filter area while three sources fall within the foreground galaxy cloud.  Essentially {\em all} of these sources are high redshift dusty SF galaxies magnified by a foreground lensing galaxy or a galaxy group, but the lensed DSFGs dominate the combined photometry and yield the characteristic extreme red colour in most cases.  For the three sources outside the selection region, the foreground lensing galaxies are brighter than the lensed DSFGs, shifting them outside the selection region.  This represents the primary mode by which our \WISE\ Colour-magnitude filter might exclude otherwise good high redshift DSFG candidates.  This \Herschel-selected \Planck\ sample is entirely independent of \WISE\ data, and therefore a loss of $\sim$20\% of high redshift galaxy candidates from this analysis is a reasonable estimate for our colour-magnitude filter.  We note that \citet{iglesiasgroth17} developed a similar near-IR colour selection method for high-redshift DSFGs, and \citet{diazsanchez17} used this method to discover independently one of our sources PJ132934.2 as a strongly lensed high-redshift DSFG without any \Planck\ prior (see the discussion in Appendix~B).

\subsection{Sample Statistics and Completeness}

After removing all high-redshift \Planck\ source candidates associated with nearby galaxies and radio AGNs (\S~\ref{sec:foreground}) and applying our \WISE\ band colour-magnitude filter, the final list of likely \Planck\ sources associated with a high-redshift dusty star forming galaxy includes 118 \WISE\ sources.  This corresponds to a mean source density of $10^{-2}$ per square degree, which is  smaller than that of the strongly lensed high redshift DSFGs found by \Herschel\ \citep[0.02-0.1 per square degree;][]{vieira13,weiss13,wardlow13} and is among the rarest population of extragalactic sources known.   Here, we report the AzTEC 1100 \micron\ photometry and CO spectroscopy of the confirmed high-redshift sources so far.  The entire candidate source catalog and follow-up observations will be published elsewhere when the LMT follow-up observations are completed.

The main shortcoming of using the \Planck\ data is the severe confusion noise associated with its large survey beam size.  While all PCCS sources are highly statistically robust ($\ge7\sigma$), the PCCS photometry carries a large systematic uncertainty (see Table~\ref{tab:smm}).  This means the completeness of the sample depends on the foreground and background confusion, and this is the leading limitation for the sample.  Follow-up studies using \Herschel\ and other telescopes have shown that extragalactic PCCS sources are mixes of bright individual sources as well as groups of multiple fainter sources \citep{planck26,canameras15,harrington16}, and only a detailed follow-up study can verify their nature.  Three of our \Planck\ candidates are associated with two distinct \WISE\ counterparts that are high redshift DSFGs (see Appendix~B), and the number of ``confirmed" sources depends on whether one is counting the PCCS sources or the \WISE\ sources.  

By keeping the catalog filtering as general as possible, our sample definition is relatively free of any systematic bias, but the large systematic uncertainty in the \Planck\ photometry is a major limitation in defining the source completeness.  As a result, any source statistics quoted here are strictly lower limits.  The work presented here focusses mainly on identifying high redshift DSFGs with apparent $L_{\rm IR}\ge10^{13-14}L_\odot$ and understanding their nature.

\subsection{Comparison with Independent \Planck-\Herschel\ Samples}

Our initial exploratory study of high-redshift \Planck\ sources using the \Herschel\ selection and the LMT follow-up observations have been reported by \citet{harrington16}.   Eight PCCS1 \Planck\ sources associated with individual \Herschel\ sources brighter than 100 mJy in archival \Herschel\ SPIRE 857 GHz images were confirmed to be high redshift DSFGs using the AzTEC and RSR instruments on the LMT.  Their apparent IR luminosity of $(0.1-2.9)\times 10^{14} L_\odot$ and inferred $SFR=(1.5-30)\times 10^3 \, M_\odot$ yr$^{-1}$ are so extreme that strong magnification by gravitational lensing is naturally suspected and is later confirmed with the \HST\ (Lowenthal et al., in prep.) and \ALMA\ observations (this study).

All eight high-redshift \Planck\ sources identified using the \Herschel\ data are detected by \WISE\ with $SNR\ge5$ in both $W1$ and $W2$ bands, and our \WISE\ colour-magnitude filter discussed above would have recovered seven of the sources (see Figure~\ref{fig:Wcolmag}).  The only exception is PJ132302.9, whose SED shown in Figure~5 by \citet{harrington16} indicates that its \WISE\ $W1$ and $W2$ band photometry is dominated by the foreground $z\approx0.5$ lensing galaxy.  This outcome suggests that our \WISE\ colour filtering may exclude a small fraction of candidates with a bright foreground lensing galaxy.  

Among the 11 \Planck\ sources analysed by \citet{canameras15}, six sources are drawn from the PCCS1 catalog, and five of those overlap with the Harrington et al. sample.  The lone exception, PJ105322.5, did not make the Harrington et al. sample because its dust peak shifted out of the \Planck\ 857 GHz band due to its high redshift ($z=3.55$).  This source is recovered by this study as one of the 545 GHz selected (857 GHz drop-out) high-redshift \Planck\ source that also satisfies our \WISE\ colour-magnitude selection (see below).  Two additional sources studied by Ca{\~n}ameras et al. are found in the PCCS2 catalog, and they are also recovered by this study (see below) while the remaining three are missed by both PCCS1 and PCCS2 catalogs.

\subsection{Other \Planck-selected DSFGs in the Literature \label{sec:literature}}

Aside from the \Planck-\Herschel\ sources studied by \citet{canameras15} and \citet{harrington16}, and one \Planck-\Herschel\ source found in the \Planck\ Early Release Compact Source Catalog studied by \citet{herranz13}, cross-checking our final list of \Planck-\WISE\ selected high-redshift source candidates using the NASA/IPAC Extragalactic Database (NED) has yielded several matches with known high redshift galaxies.  One well studied example is \Planck-\WISE\ source PJ213511.6$-$010252 which is the $z=2.326$ lensed SMG also known as ``Cosmic Eyelash" \citep{ivison10,swinbank10}.  The \Planck-\WISE\ source PJ213512.7$-$010144 also falls within the same PCCS error ellipse, and this $z=3.074$ lensed SMG found in a Lyman break galaxy search \citep[``Cosmic Eye",][]{smail07,coppin07} likely contributes to the \Planck\ detection as well.  
The \Planck-\WISE\ source PJ090403.9+005619 is a candidate lensed high redshift source found by the \Herschel-ATLAS survey with a photometric redshift of $z_{\rm ph}=1.7$ \citep{gonzaleznuevo12}.  Archival \Herschel/SPIRE images show an extended, clumpy source in several \Herschel\ bands, and it is likely an over-density of sources, rather than a single bright source (see Table~\ref{tab:smm} and \S~\ref{sec:lensing}). The \Planck-\WISE\ source PJ132630.3+334407, which we report here with a detailed study, was first studied as a $z=2.951$ lensed \Herschel-ATLAS source by \citet[][see Appendix~B]{bussmann13}.
Given the extremely small number density of $\sim10^{-2}$ per square degree for the \Planck-selected DSFGs, finding only a handful of counterparts in the {\em entire} extragalactic fields studied by \Herschel\ covering $\approx$1000 sq. degree in total area is about as expected.  It also follows that the vast majority of the known high redshift \Herschel\ sources are too faint to be detected individually by \Planck\ (see further discussions in \S~\ref{sec:discussion}).  

Other large area submillimetre cosmology experiments such as the South Pole Telescope (SPT) survey and the Atacama Cosmology Telescope (ACT) survey should include a few \Planck-\WISE\ sources as well.  The $z=2.515$ \Planck-\WISE\ source PJ012507.1$-$472356 (SPT0125$-$47), the $z=2.783$ source PJ053816.8$-$503052 (SPT0538$-$50), and the $z=2.726$ source PJ233226.5$-$535839 (SPT2332$-$53) are three of the 81 dust-dominated sources identified in the 2500 square-degree survey fields \citep{vieira10,greve12,weiss13,reuter20}.  The list of 30 brightest DSFGs found in the \ACT\ survey \citep[480 sq. degree in area,][]{gralla20} includes three \Planck-\WISE\ sources that are already known: Cosmic Eyelash (PJ213511.6$-$010252, discussed above), PJ020941.3+001559 \citep{geach15,harrington16}, and PJ231356.6+010918 (this study).

The \Planck\ and \Herschel\ photometry of the four additional high redshift \Planck-selected DSFGs identified in the literature are included in Table~\ref{tab:smm}, and they are also shown in Figure~\ref{fig:Pcolmag} along with the \Herschel-selected sources to show the parameter space they occupy in these colour-magnitude plots.  They are included in the entire \Planck-\WISE\ sample used to characterise the properties of the \Planck-selected DSFG population as a whole later in \S~\ref{sec:discussion}.  
While this paper is undergoing a referee review, \citet{trombetti21} posted on the arXiv.org preprint server a paper reporting a search for candidate strongly lensed dusty galaxies in the \Planck\ catalogues using a similar SED analysis method utilising more recent \Planck\ photometry products, trained using a sample of known \Planck\ selected DSFGs including most of the sources reported here. Their follow-up confirmation work has started, and a future comparison of the two methods should prove interesting. 

\begin{table*}
 \caption{Summary of LMT Observations}
 \label{tab:observations}
\begin{tabular}{@{}lccccccc}
  \hline
  ID & RA & DEC & \multicolumn{2}{c}{RSR} & \multicolumn{2}{c}{AzTEC} \\
  & (J2000) & (J2000) & Dates & Int. Time (mins) & Dates & Int. Time (mins) \\
  \hline
  PJ011646.8 & 01h16m46.8s  & $-$24d37m02s & 2015-12-18 & 30 &  &  \\
  PJ014341.2 & 01h43m41.2s  & $-$01d47m26s & 2015-12-08, 2015-12-18 & 75 &  &  \\
  PJ022634.0 & 02h26m34.0s  & +23d45m28s & 2016-02-03 & 15 &  2016-01-24 & 10  \\
  PJ030510.6 & 03h05m10.6s  & $-$30d36m30s & 2016-02-02, 2016-02-03 & 30 &  2016-02-05 & 10 \\
  PJ074851.7 & 07h48m51.7s  & +59d41m54s & 2016-02-01 & 30 &  & \\
  PJ074852.6 & 07h48m52.6s  & +59d42m09s & 2016-01-30, 2016-02-01 & 60 &  2016-01-24 & 10 \\
  PJ084648.6 & 08h46m48.6s  & +15d05m57s & 2016-01-16, 2016-02-02 & 60 &  2016-01-22 & 10 \\
  PJ084650.1 & 08h46m50.1s  & +15d05m47s & 2015-12-15 & 45 &  2014-11-10 & 10 \\
  PJ105322.6 & 10h53m22.6s  & +60d51m47s & 2015-12-16, 2016-02-01 & 60 & 2016-01-23 & 10 \\
  PJ112713.4 & 11h27m13.4s  & +46d09m24s & 2015-12-15 & 30 &  & \\
  PJ113805.5 & 11h38m05.5s  & +32d57m57s & 2015-06-16, 2015-06-18 & 30 &  & \\
  PJ113921.7 & 11h39m21.7s  & +20d24m51s & 2016-02-03 & 30 & 2016-02-05 & 10 \\
  PJ114038.5 & 11h40m38.5s  & +53d21m57s & 2015-12-18 & 30 &  & \\
  PJ114329.5 & 11h43m29.5s  & +68d01m07s & 2015-12-18, 2016-01-27, 2016-02-07 & 60 &  & \\
  PJ132217.5 & 13h22m17.5s  & +09d23m26s & 2016-02-20, 2015-06-10 & 60 &  & \\
  PJ132630.3 & 13h26m30.3s  & +33d44m07s & 2014-05-05, 2014-05-23 & 60 &  & \\
  PJ132934.2 & 13h29m34.2s  & +22d43m27s & 2016-02-17, 2016-02-19 & 50 &  2016-01-22 & 10 \\
  PJ132935.3 & 13h29m35.3s  & +22d43m24s & 2016-02-18 & 30 &  & \\
  PJ133634.9 & 13h36m34.9s  & +49d13m14s & 2016-02-07 & 30 &  2016-01-22 & 10 \\
  PJ141230.5 & 14h12m30.5s  & +50d34m55s & 2015-05-12 & 45 &  & \\
  PJ144653.2 & 14h46m53.2s  & +17d52m33s & 2016-02-19 & 30 &  2016-01-22 & 10 \\
  PJ144958.6 & 14h49m58.6s  & +22d38m37s & 2015-05-12 & 45 &  2016-01-23 & 10 \\
  PJ154432.4 & 15h44m32.4s  & +50d23m44s & 2015-01-29 & 30 & 2016-01-23 & 10 \\
  PJ231356.6 & 23h13m56.6s  & +01d09m18s & 2015-05-20, 2015-06-13, 2015-06-16 & 75 & & \\    
    \hline
\end{tabular}
\break
 \end{table*}

\section{LMT and ALMA Observations}
\label{sec:obs}

An observational program to confirm the 118 \Planck-selected DSFGs candidates identified with the \WISE\ colour-magnitude filter (described in \S~\ref{sec:OIR}) is carried out through AzTEC 1.1mm continuum camera imaging and CO spectroscopy using the Redshift Search Receiver on the Large Millimeter Telescope. This is a long-term observational program, and here we report the initial confirmations of 24 DSFGs.

The design of the LMT confirmation program was to start with AzTEC snapshot observations and follow-up spectroscopy using the RSR, but the limited availability of the AzTEC instrument due to weather conditions and instrument readiness forced us to proceed with RSR spectroscopy without any AzTEC prior.  A total of 24 CO line sources are confirmed so far, including 22 $z\gtrsim1$ DSFGs and 2 ULIRGs at $z\sim0.2$. 
Essentially an independent AzTEC photometry campaign conducted in parallel detected 11 high-redshift DSFGs confirmed with RSR CO detection.

Here, we also present the 1.1 mm continuum imaging results from our Cycle 5 ALMA program that was designed to map the distribution of dust continuum and CO line emission in a selected subset of 12 \Planck-selected DSFGs reported here.  These ALMA observations provide independent 1.1 mm photometry of 12 DSFGs, including six new 1.1 mm continuum measurements, and they are included in the broadband SED analysis.

\begin{figure*}
\includegraphics[width=12cm]{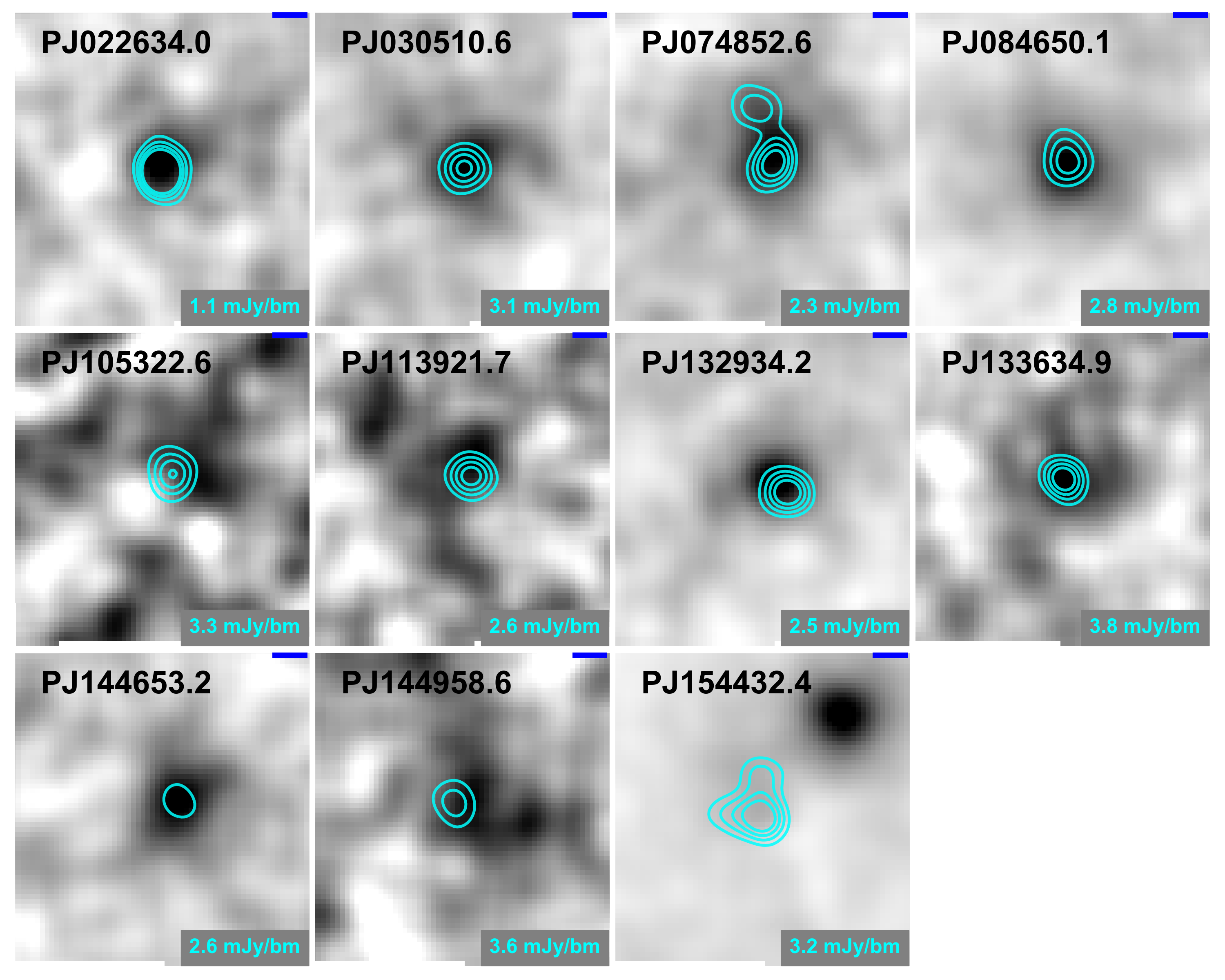}
\caption{AzTEC 1.1 mm images at 8.5 arcsecond resolution shown in contours over the \WISE\ 22 \micron\ images in greyscale.  Each square box is 90 arcsecond on each side. Contour levels correspond to 5$\sigma$, 10$\sigma$, 15$\sigma$, and 20$\sigma$ ($\sigma$ values are given on the bottom right corner of each frame).  The orientation of the images is such that north is up while east is to the left. PJ084648.6 is in the same frame as PJ084650.1, but the peak is less than $5\sigma$ and is not shown.}
\label{fig:AzTEC}
\end{figure*}

\subsection{AzTEC 1.1 mm Continuum Observations}
\label{sec:AzTEC}

AzTEC 1.1 mm continuum observations were made with the AzTEC camera \citep{wilson08}, which is a facility instrument on the Large Millimeter Telescope Alfonso Serrano \citep{hughes20}.  During the observing period between March 2014 and May 2016, only the inner 32-m diameter area of the primary mirror was illuminated, leading to an effective beam of 8.5 arcsecond (FWHM).  A modified Lissajous pattern is used to image a uniform sensitivity area of 1.5 arcminute diameter centred on the \WISE\ source position.  Each source was observed with an on-source integration time of 10 minutes each.  Each target observation was bracketed by observations of a nearby bright quasar ($\le20^\circ$ away) to ensure the pointing accuracy of better than 1.5 arcsecond.  The data for each source were reduced using the AzTEC Standard Pipeline described by \citet{wilson08} and \citet{scott08}, using the Wiener filtering optimised for point source detection \citep{perera08}.  The resulting noise in the AzTEC images is between $\sigma=1.0-3.1$ mJy depending on the total integration time and weather conditions.

A total of 30 sources identified by the \Planck-\WISE\ selection were observed, and 13 sources (43\%) are clearly detected.  The details of the AzTEC observations and the derived 1.1 mm photometry are summarised in Tables~\ref{tab:observations} \& \ref{tab:smm} while the details of the 17 sources without a bright compact counterpart are given in Appendix A (see Table~\ref{tab:NoAzTEC}).  The 1.1 mm flux density measured for the 13 detected sources ranges between 15 and 202 mJy, comparable to those of the 8 \Planck\ sources we previously identified using \Herschel\ data  \citep[][$S_{\rm 1.1  mm}= 8-147$ mJy]{harrington16}.  The 17 sources undetected by AzTEC have $3\sigma$ upper limits between 3.2 and 9.3 mJy, and these measurements rule out the detected \Planck\ emission originating from a single bright IR source.  

The AzTEC 1.1 mm continuum images are shown in contours over the greyscale \WISE\ 22 \micron\ ($W4$) band images in Figure~\ref{fig:AzTEC}.  All but two of the sources have a compact morphology, consistent with being unresolved by the 8.5 arcsecond AzTEC beam.  The two extended AzTEC sources, PJ074852.6 and PJ154432.4, are associated with a crowded foreground galaxy concentration, and their dust continuum emission is extended over ten arcsecond scales (see Appendix~B for further discussions).

This 43\% AzTEC confirmation rate indicates that our candidate list includes a significant number of foreground confusing sources even after the extensive vetting described in \S~\ref{sec:Planck}.  Since our candidate selection requires an inclusion in at least two PCCS bands, each with $\ge7\sigma$ statistical significance, they are not likely spurious sources.  Instead, \Planck\ sources that are not a single bright AzTEC source are likely extended or distributed dust sources -- extended emission is obvious in some cases. Some \Planck\ candidates are likely foreground cirrus clouds with an arcminute scale structure that survived our filtering.  Previous studies of \Planck\ selected sources have also shown some to be over-density of fainter high-redshift DSFGs \citep{clements14,clements16,florescacho16}, rather than a single bright object.

\begin{table*}
\caption{Summary of \Planck, \Herschel, AzTEC, and ALMA photometry}
\scriptsize
\label{tab:smm}
\begin{tabular}{@{}lcccccccccc}
\hline
\hline
Source ID & $Planck_{350}$ & $Planck_{500}$ & $Planck_{850}$ & $SPIRE_{250}$ &   $SPIRE_{350}$ &  $SPIRE_{500}$ & $S_{850}$ & $AzTEC_{1100}$ & $ALMA_{1100}$ & References \\
                   & (mJy) & (mJy) & (mJy) & (mJy) &  (mJy)  & (mJy) & (mJy) & (mJy) & (mJy) & \\
\hline
PJ011646.8 & $513\pm462$ & $555\pm336$ & -- & -- & -- & -- & -- & -- & $66\pm10$ & \\
PJ014341.2 & $441\pm336$ & -- & -- & $210\pm21$ & $149\pm25$ & $82\pm22$ & -- & -- & $4.0\pm0.6$ & \\
PJ022634.0 & $1804\pm1241$ & $1113\pm381$ & $433\pm185$ & -- &	-- & -- & -- & $103\pm15$ & -- &  \\
PJ030510.6 & $706\pm194$ & $375\pm177$ & -- & -- & -- & -- & -- & $50\pm10$ & $48\pm7$ & \\
PJ074851.7 & $997\pm593$ & $737\pm229$ & $351\pm123$ & -- & -- & -- & -- & $66\pm10$ & -- & \\
PJ084650.1 & $1748\pm322$ & $1167\pm240$ & $500\pm91$ & -- & -- & -- & --  & $118\pm15$ & $76\pm11$  & \\
PJ105322.6 & -- & $680\pm127$ & $356\pm52$ & $460\pm 46$ & $739\pm74$ & $771\pm77$ & $360\pm36$  & $202\pm30$  & -- & 1 \\
PJ112713.4 & $195\pm127$ & $<330$ & -- & -- & -- & -- & -- & -- & -- & \\
PJ113805.5 & $232\pm145$ & $93\pm90$ & $57\pm50$ & -- & -- & -- & -- & -- & $9.5\pm1.4$ & \\
PJ113921.7 & $406\pm312$ & $343\pm312$ & -- & $300\pm30$ &	$397\pm40$ & $341\pm34$ & $111\pm12$& $58\pm12$& $49\pm7$  & 1 \\
PJ114038.5 & $413\pm258$ & $185\pm152$ & -- & -- & -- & -- & -- & -- & -- &  \\
PJ114329.5 & $439\pm326$ & $349\pm226$ & $111\pm75$ & -- &	-- & -- & -- & -- & -- &  \\
PJ132217.5 & $717\pm191$ & $322\pm152$ & -- & -- & -- & -- & -- & -- & $13\pm2$ &  \\
PJ132630.3  & $282\pm200$ & -- & $148\pm89$ & $197\pm20$ & $288\pm29$ & $288\pm29$ & $65\pm7$ & -- & $28\pm4$ & 3 \\
PJ132934.1  & $1298\pm219$ & $713\pm139$ & $263\pm88$ & -- & -- & -- & $128\pm20$ & $60\pm10$ & $51\pm8$ & 2 \\
PJ133634.9  & $627\pm140$ & $313\pm154$ & $166\pm82$ & -- & -- & -- & -- & $65\pm12$ & -- & \\
PJ141230.5  & $944\pm195$ & $338\pm156$ & $354\pm75$ & -- & -- & -- & -- & -- & -- & \\
PJ144653.2  & $1149\pm588$ & $615\pm284$ & $113\pm86$ & -- & -- & -- & -- & $15\pm3$ & $17\pm3$ &  \\
PJ144958.6  & $668\pm166$ & $457\pm160$ & $230\pm107$ & -- & -- & -- & -- & $37\pm8$ & $37\pm6$ & \\
PJ154432.4 & -- & $485\pm265$ &$303\pm93$ & $225\pm23$ & $353\pm35$& $353\pm35$ & $87\pm11$ & $35\pm10$ & -- & 1 \\
PJ231356.6  & $393\pm266$ & $234\pm224$ & -- & -- & -- & -- & -- & -- & $26\pm4$ & 6 \\
\hline
 \multicolumn{2}{l}{\Planck\ Sources from Literature} & & & & & & & & & \\
Cosmic Eyelash & $1393\pm416$ & $919\pm201$ &$475\pm81$ & $372\pm22$ & $453\pm26$& $365\pm18$ & $106\pm7$ & -- & -- & 4,5,6 \\
PJ090403.9        & $1493\pm393$ & $708\pm232$ & -- & $104\pm7$ & $87\pm8$& $50\pm9$ & -- & -- & -- & 7 \\
SPT0125$-$47   & $903\pm113$ & $600\pm256$ &$367\pm69$ & $848\pm11$ & $817\pm17$& $546\pm17$ & -- & -- & -- & 8 \\
SPT0538$-$47   & $844\pm215$ & $426\pm126$ & -- & $391\pm10$ & $484\pm14$& $377\pm13$ & -- & -- & -- & 9 \\
\hline
\end{tabular}
\\
Names of the photometry bands are listed in micron rather than GHz in order to be consistent with the data table in our earlier paper by \citet{harrington16}.  The listed uncertainty on the AzTEC and ALMA photometry includes 15\% systematic uncertainty in the absolute flux calibration.  The large uncertainties in the \Planck\ photometry reflects the systematic uncertainty associated with the source confusion (see \S~\ref{sec:SEDdata} for details).
References: (1) \citet{canameras15}; (2) \citet{diazsanchez17}; (3) \citet{bussmann13}; (4) \citet{ivison10}; (5) \citet{swinbank10}; (6) \citet{gralla20}; (7) \citet{gonzaleznuevo12}; (8) \citet{weiss13}; (9) \citet{greve12}
 \end{table*}

\begin{figure*}
\includegraphics[height=10.5cm]{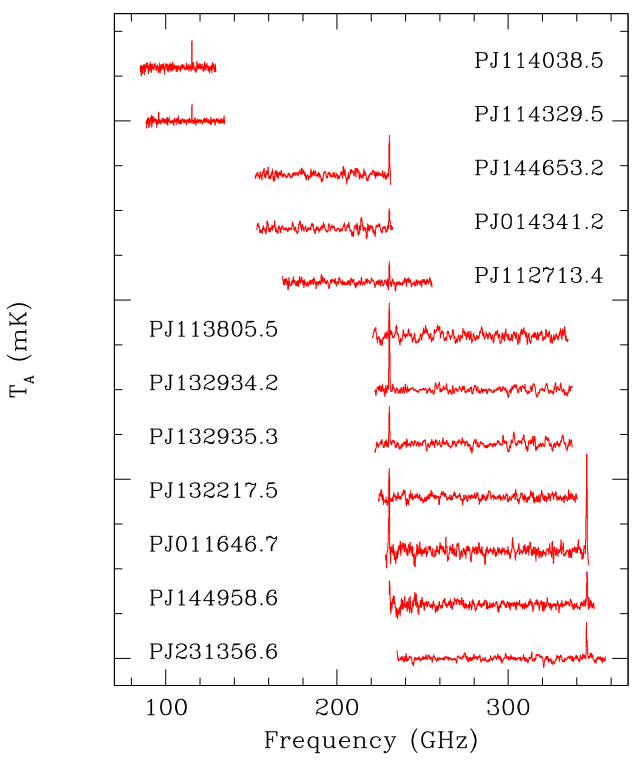}
\includegraphics[height=10.5cm]{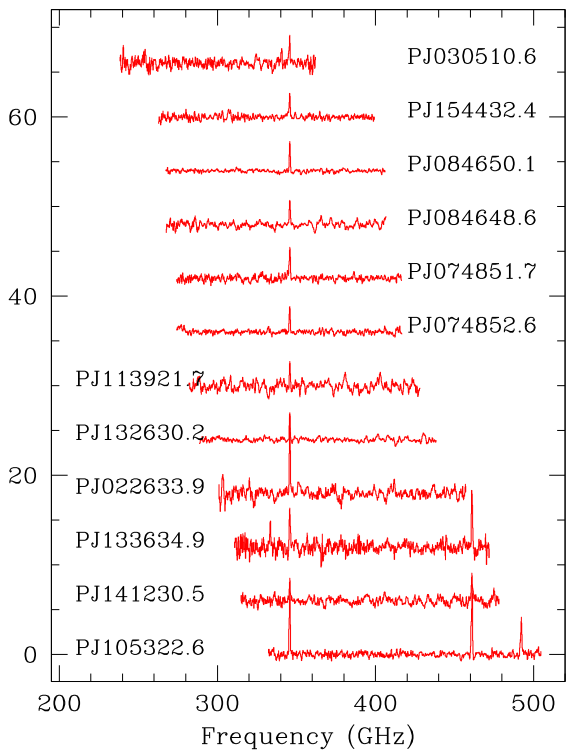}
\caption{The RSR spectra of 24 \Planck\ selected sources in their rest frame.  They are sorted by their redshift, and two CO lines are detected for 3 of the highest redshift sources, yielding a secure redshift.  The highest redshift source also shows the redshifted 492 GHz [C I] line as well. }
\label{fig:RSR}
\end{figure*}

\begin{table*}
 \caption{Summary of RSR Measurements}
 \label{tab:RSR}
 \scriptsize
\begin{tabular}{@{}lccccccccc}
  \hline
   ID & $\nu_{\rm CO}$  & Line & $z_{\rm CO}$ & $\Delta V$  & $S\Delta V$ & $\mu L'_{\rm CO}$ & $\mu M_{\rm H2}$ & $\mu M_{\rm gas}^\dagger$ & notes \\
   & (GHz) & & & (km/s) & (Jy km/s) & ($10^{10}$ K km/s pc$^2$) & ($10^{10} M_\odot$) & ($10^{10} M_\odot$) & \\
 \hline
  PJ011646.8 & 110.660  & CO (3--2) & $2.1249\pm0.0002$ & $352\pm5$ & $33.2\pm2.6$ & $81\pm 4$ & $524 \pm 50$ & $473 \pm 547$ & (1) \\
  PJ014341.2 & 110.002  & CO (2--1) & $1.0956\pm0.0003$ & $622\pm22$ & $10.6\pm0.7$ & $17\pm 2$ & $85\pm 15$ & & (2) \\
  PJ022634.0 & 83.924  & CO (3--2) & $3.1203\pm0.0002$ & $502\pm7$ & $29.1\pm0.8$ & $136\pm 8$ & $886 \pm 133$ & $227 \pm 280$ & (1) \\
  PJ030510.6 &  105.981 & CO (3--2) & $2.2628\pm0.0003$ & $283\pm 7$ & $9.7\pm1.2$ & $26\pm 3$ & $169 \pm 21$  & $145 \pm 35$  &  (1) \\
  PJ074851.7 &  92.079 & CO (3--2) & $2.7554\pm0.0003$ & $578\pm 8$ & $15.0\pm0.8$ & $57\pm 5$ & $371 \pm 40$  & $294 \pm 400$  & (1) \\
  PJ074852.6 &  92.079 & CO (3--2) & $2.7554\pm0.0003$ & $607\pm 12$ & $12.1\pm1.1$ & $46\pm 4$ & $300 \pm 53$  & & (1)  \\
  PJ084648.6 &  94.374 & CO (3--2) & $2.6641\pm0.0004$ & $593\pm14$ & $12.5\pm0.6$ & $45\pm 4$ & $293 \pm 49$  & & (1) \\
  PJ084650.1 &  94.464 & CO (3--2) & $2.6606\pm0.0002$ & $502\pm10$ & $12.8\pm0.4$ & $46\pm 3$ & $300 \pm 53$  & $212 \pm 171$  & (1) \\
  PJ105322.6 & 76.016  & CO (3--2) & $3.5490\pm0.0003$ & $568\pm7$ & $24.4\pm0.7$ & $142\pm 9$ & $928 \pm 139$  & $1550 \pm 360$  & (3) \\       
                     & 101.363  & CO (4--3) & $3.5488\pm0.0003$ & $495\pm7$ & $25.8\pm0.6$ & $85\pm 5$ & $795 \pm 119$  & & \\       
  PJ112713.4 & 100.100  & CO (2--1) & $1.3029\pm0.0002$ & $496\pm12$ & $8.7\pm0.6$ & $19\pm 3$ & $70 \pm 13$& $110 \pm 12$ &  (1) \\
  PJ113805.5 & 76.364  & CO (2--1) & $2.0190\pm0.0002$ & $154\pm9$ & $4.9\pm0.6$ & $15\pm 2$ & $77 \pm 11$ & $36 \pm 38$ &  (1) \\
  PJ113921.7 & 89.624  & CO (3--2) & $2.8583\pm0.0005$ & $496\pm14$ & $11.5\pm1.2$ & $40\pm 6$ & $265 \pm 39$ & $311 \pm 54$ &  (1), (3) \\
  PJ114038.5 & 98.933  & CO (1--0) & $0.1652\pm0.0001$ & $589\pm14$ & $13.0\pm0.6$ & $1.6\pm 0.1$ & $7.1 \pm 1.6$ & &  (4) \\
  PJ114329.5 & 95.125  & CO (1--0) & $0.2118\pm0.0001$ & $279\pm8$ & $4.7\pm0.5$ & $1.3\pm 0.1$ & $5.3 \pm 0.8$ &  & (4) \\
  PJ132217.5 & 75.151  & CO (2--1) & $2.0676\pm0.0001$ & $158\pm4$ & $7.2\pm0.6$ & $38\pm 3$ & $192 \pm 30$ & $50 \pm 96$ &  (1) \\
  PJ132630.3 & 87.519  & CO (3--2) & $2.9511\pm0.0002$ & $292\pm5$ & $9.8\pm0.5$ & $43\pm 3$ & $280 \pm 45$ & $140 \pm 20$ &  (1) \\
  PJ132934.2 & 75.839  & CO (2--1) & $2.0398\pm0.0001$ & $476\pm4$ & $36.1\pm1.6$ & $180\pm 8$ & $910 \pm 137$ & $162 \pm 48$ & (1), (5)  \\
  PJ132935.3 & 75.829  & CO (2--1) & $2.0402\pm0.0002$ & $474\pm9$ & $15.5\pm0.9$ & $77\pm 5$ & $390 \pm 59$  & & (1)  \\
  PJ133634.9 & 81.280  & CO (3--2) & $3.2541\pm0.0004$ & $638\pm12$ & $16.9\pm1.1$ & $86\pm 8$ & $563 \pm85$ & $179 \pm 54$ & (1) \\       
                     & 108.370  & CO (4--3) & $3.2543\pm0.0003$ & $532\pm6$ & $21.9\pm1.1$ & $60\pm 5$ & $557 \pm 84$ & & \\       
  PJ141230.5 & 80.224  & CO (3--2) & $3.3103\pm0.0004$ & $451\pm14$ & $7.1\pm0.6$ & $38\pm 5$ & $247 \pm36$ & & \\       
                     & 106.948  & CO (4--3) & $3.3105\pm0.0005$ & $540\pm22$ & $11.6\pm0.9$ & $33\pm 3$ & $311 \pm 47$ & & \\       
  PJ144653.2 & 110.598  & CO (2--1) & $1.0844\pm0.0001$ & $458\pm8$ & $16.5\pm0.6$ & $42\pm 3$ & $137 \pm 20$ & &  (2) \\
  PJ144958.6 & 109.666  & CO (3--2) & $2.1533\pm0.0002$ & $387\pm8$ & $10.8\pm0.7$ & $26\pm 2$ & $167 \pm 29$  & $152 \pm 114$  & (1) \\
  PJ154432.4 & 96.092  & CO (3--2) & $2.5989\pm0.0003$ & $407\pm12$ & $9.7\pm0.5$ & $32\pm 3$ & $208 \pm 32$  & $326 \pm 33$  & (3) \\
  PJ231356.6 & 107.550  & CO (3--2) & $2.2167\pm0.0002$ & $247\pm14$ & $10.7\pm0.7$ & $28\pm 2$ & $184 \pm 28$  & $299 \pm 524$  & (1) \\
 \hline
 \end{tabular}
 \break
    $^\dagger$ $M_{\rm gas}$ is total cold gas mass derived using a full radiative transfer modelling of the multiple CO, [C I], and dust continuum measurements by \citet{harrington20}.
     Table notes: (1) IRAM 30-m and GBT confirmations  by \citet{harrington20}; (2) for the single line detections of PJ014341.2 and PJ144653.2, alternative redshifts are $z=2.144$ and $z=2.127$ if the detected line is the CO (3--2) transition; (3) \citet{canameras15}; (4) optical redshifts from the Sloan Digital Sky Survey; (5) \citet{diazsanchez17}. \\
 \end{table*}
 
\subsection{Redshift Search Receiver CO Spectroscopy \label{sec:RSR}}

	The Redshift Search Receiver \citep[RSR,][]{erickson07} is an LMT facility broadband spectrometer system that operates in the frequency window of 73--111 GHz, with 4 detector pixels organised in a dual--beam, dual--polarisation configuration. The RSR beam-switches at 1 kHz frequency between the two beams separated by 78 arcsecond in Azimuth direction, and this rapid differencing leads to a stable and flat baseline across its ultra-wide bandwidth.  Its backend spectrometer covers the entire 38 GHz bandwidth with a spectral resolution of 31.25 MHz (102 km s$^{-1}$ at 92 GHz). We expect to detect at least one of the low-$J$ ($J_{\rm up} = 2-4$) CO transitions to yield the redshifts of our targets. Table~\ref{tab:observations} provides information on the integration time for each source. For the majority of the sources a CO line was evident within 15 minutes of integration, and a longer integration (30-75 minutes) was used to yield a secure CO line detection with a $S/N$ ratio $\ge7$ in all cases. 
	
The measured RSR spectra of all 24 observed sources are shown in  Figure~\ref{fig:RSR}, and the measured and derived CO line quantities are summarised in Table~\ref{tab:RSR}.  The design of the experiment was to target the sources confirmed to be a bright AzTEC 1.1 mm source, but high priority candidates without any AzTEC observations were also targeted to facilitate the survey scheduling.  As a result, all 13 sources detected by AzTEC have a CO redshift and luminosity, while 11 additional sources were observed based on our \WISE\ colour selection alone.

As discussed in detail by \citet{yun15}, two or more CO transitions fall within the RSR spectral band at $z\ge3.15$, and an unambiguous redshift for a CO source can be determined from the RSR spectrum alone.  For PJ105322.6, PJ133634.9, and PJ141230.5, both CO (3--2) and CO (4--3) lines are detected, and their redshifts are unambiguously determined.  For PJ011646.7 and PJ144958.6, CO (2--1) and CO (3--2) lines are detected in their RSR spectra, also yielding an unambiguous redshift, although their CO (2--1) is too close to the spectrum edge to yield usable line information.  For the remaining 19 targets, only a single CO line is detected, and their redshifts are assigned based on the photometric redshift support, as discussed by \citet{harrington16}.  The single CO lines detected for PJ114038.5 and PJ114329.5 correspond to the CO (1--0) transition at the optical spectroscopic redshifts of the \WISE\ sources reported by the Sloan Digital Sky Survey (SDSS), and we conclude that they are foreground ULIRGs at $z\approx 0.2$, rather than high redshift sources.  For the remaining sources, their single line redshifts are confirmed for all but two cases by detecting other CO transitions using the IRAM 30-m telescope and the Green Bank Telescope (GBT), and those results are presented elsewhere \citep[][also see  Table~\ref{tab:RSR} and Appendix B]{harrington20}.  The single line redshifts of  PJ014341.2 and PJ144653.2 remain to be confirmed.

Despite the coarse spectral resolution of the RSR, all of the CO lines are clearly resolved.  The CO line-widths $\Delta V$ reported in Table~\ref{tab:RSR} are derived by fitting a Gaussian to the measured spectra and applying a correction for the instrumental resolution (see \citet{harrington16} for details).  The uncertainty in the reported line-widths reflects both the SNR of the spectra and the formal uncertainty in the spectral resolution correction.  The measured CO line-widths range between 150 and 637 km s$^{-1}$, comparable to the values reported for similar objects \citep[e.g.,][]{canameras15,harrington16}.

The CO line luminosity, $L'_{\rm CO}$, is computed using Eq.~(3) by \citet{solomon97}\footnote{The RSR spectra measured in $T_{\rm A}^*$ unit are converted to flux density unit (Jy) using a frequency dependent conversion relation, $Jy/K = 0.027*\nu_{\rm GHz} + 3.36$ for the 32-m diameter area illuminated \citep{yun20}.}.
Molecular gas mass $M_{\rm H2}$ is derived first by converting the line luminosity to $L'_{\rm CO(1-0)}$ using the average ``SMG" ratios of $L'_{\rm CO(2-1)}/L'_{\rm CO(1-0)}=0.85$, $L'_{\rm CO(3-2)}/L'_{\rm CO(1-0)}=0.66$, and $L'_{\rm CO(4-3)}/L'_{\rm CO(1-0)}=0.46$ \citep{carilli13} and then applying the standard CO-to-\HH\ conversion factor of $\alpha_{\rm CO}=4.3\,M_\odot$ [K \kms\ pc$^2$]$^{-1}$ \citep{bolatto13}.  A recent analysis of the literature data by \citet{kirkpatrick19} suggests the mean CO excitation relations can vary by up to 10\%, depending on the sample definition.
While some authors have advocated a ``ULIRG" (e.g., $\alpha_{\rm CO}\approx1$) or other CO-to-H$_2$ conversion relations for various reasons, there are no compelling reasons to believe they are any more accurate \citep[see the detailed discussions by][]{scoville16}. By obtaining new measurements of additional CO rotational transitions and analysing the gas excitation directly, \citet{harrington20} have obtained new gas mass estimates for 16  \Planck\ sources in our sample, largely independent of these standard assumptions, and they are discussed further in \S~\ref{sec:gasmass}.

The derived molecular gas masses are in the range of $M_{\rm H2}=(0.5-93)\times 10^{11}\mu^{-1} M_\odot$ (uncorrected for an unknown magnification factor $\mu$) with a  median value of $M_{\rm H2}=28\times 10^{11}\mu^{-1} M_\odot$.  If they are all strongly magnified with $\mu\sim 10$ (see the discussion in \S~\ref{sec:lensing}), then their intrinsic gas masses are comparable to those of other SMGs reported previously using similar assumptions \citep[see review by][]{carilli13}.  

\begin{figure*}
\includegraphics[width=15cm]{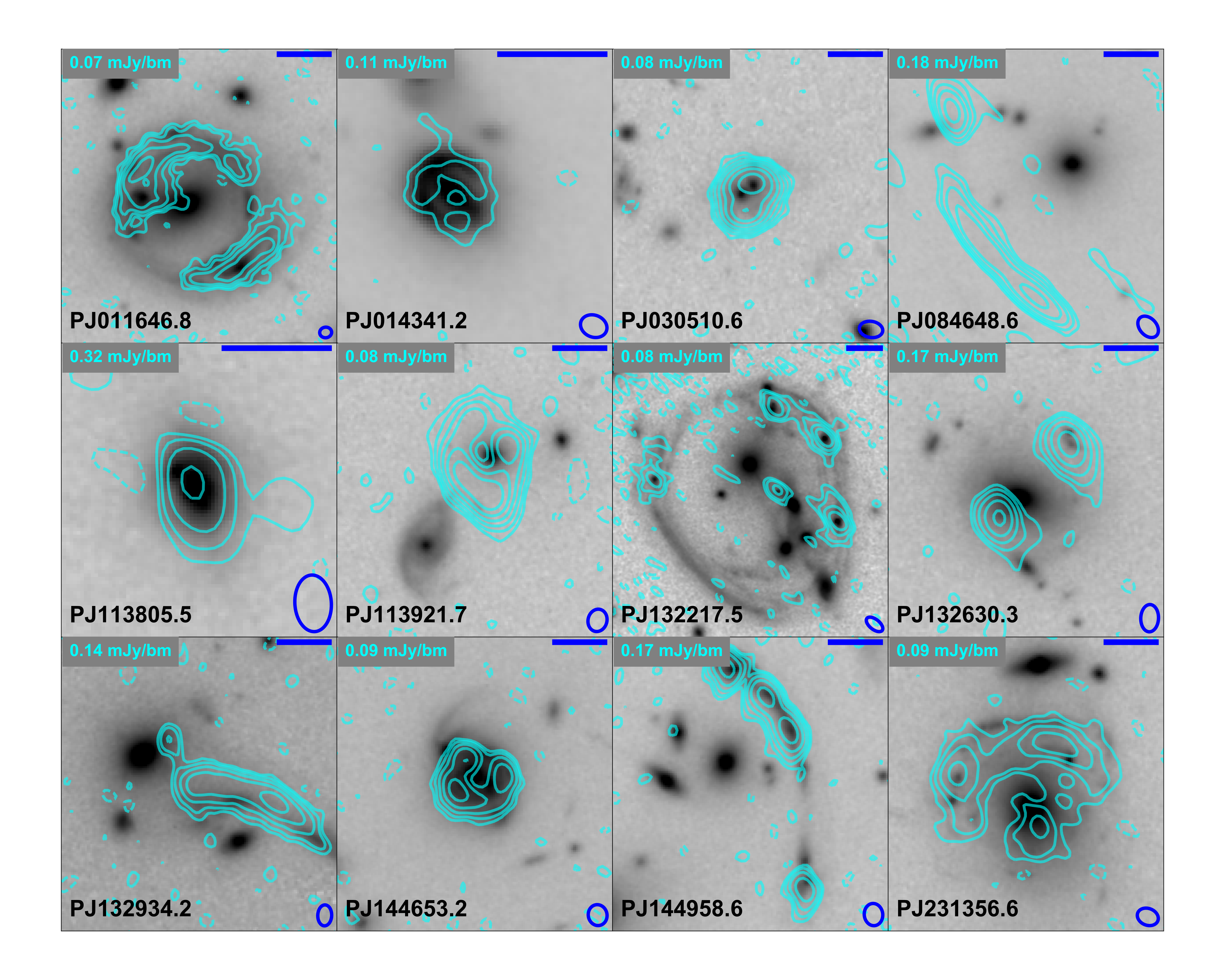}
\caption{ALMA 260 GHz continuum images are shown in contours over the HST/WFC3 F160W greyscale images (Lowenthal et al., in prep.) for the 12 \Planck\ sources targeted by both surveys.  Each frame is 10 arcsecond in size in most cases, and a blue bar on the top right hand corner is 2 arcsecond in size.  The ALMA beams are shown as an ellipse on the bottom right corner.  The contours shown are $-$2, 2, 4, 8, 16, 32, 64, and 128 times $1\sigma$, which is indicated on the top left corner in each frame.}
\label{fig:ALMAstamps}
\end{figure*}

\subsection{ALMA 1.1 mm Continuum Observations}
\label{sec:ALMA}

ALMA 260 GHz (Band 6) observations of the 12 Planck selected DSFGs have been obtained as part of the 2017.1.01214.S program (PI: M. Yun) between March 27 and August 31, 2018.  The default continuum dual polarisation set up with a total bandwidth of 8 GHz (250-254 GHz \& 266-270 GHz) is used, with a target synthesis beam of $\theta \approx0.4\arcsec$.  

The ALMA 260 GHz continuum images are shown in contours over the HST/WFC3 F160W greyscale images (Lowenthal et al., in prep.) in Figure~\ref{fig:ALMAstamps}.  All of the sources are clearly resolved by ALMA, in many cases showing the classic morphology of Einstein rings and lensed arcs. Sources such as PJ011646.7 and PJ014342.2 are lensed primarily by a single massive galaxy, while others such as PJ132217.5 are likely lensed by a small galaxy group.  Some of the large lensed arcs such as PJ084648.5, PJ132934.2, and PJ144958.5 are likely lensed by a massive galaxy group or a galaxy cluster.  Two objects, PJ030510.6 and P113805.5, are compact with a source size of only $\sim$1 arcsecond, but the CO line images obtained in the same ALMA program (not shown here) display the characteristic Einstein ring morphology.  Dust continuum features often have an associated stellar feature in the HST images, but they are frequently displaced from each other.  There are also lensed stellar features without a dust continuum counterpart (e.g., P011646.7 and PJ132217.5), indicating that either the stellar light is more extended than the gas and dust, or the lensed galaxy has multiple components or stellar companions.  The stellar light along the ALMA dust continuum peaks is strongly suppressed or is entirely missing in many cases (e.g., PJ084648.6 and PJ113921.7). 

\begin{figure}
\centering
\includegraphics[width=0.8\columnwidth]{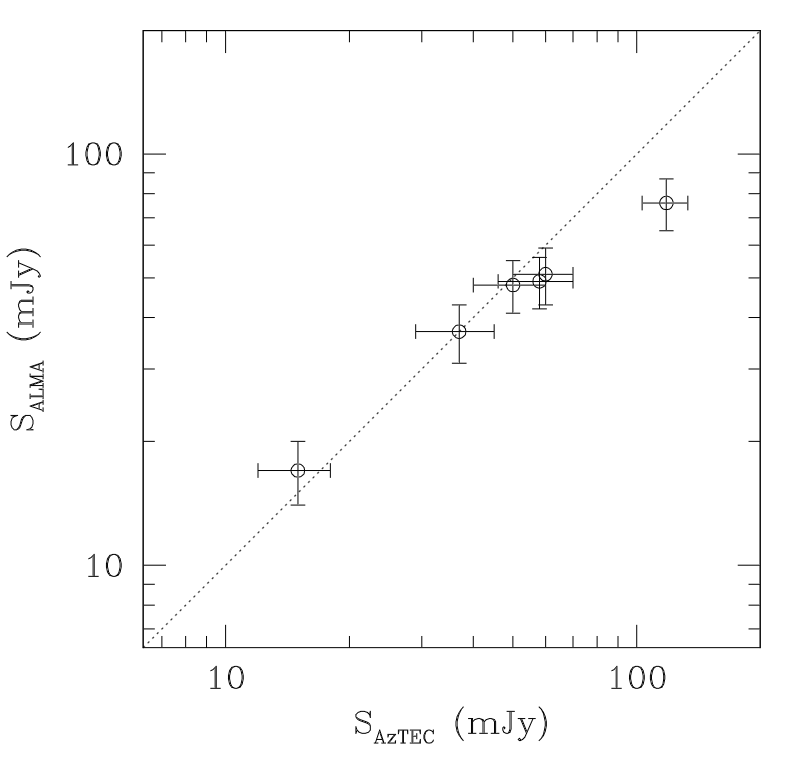}
\caption{A comparison of the total ALMA 260 GHz flux density measured at $\theta\sim0.4\arcsec$ with those measured by AzTEC with the $8.5\arcsec$ beam at the LMT.  The dotted diagonal line is a one-to-one correspondence line shown only to guide the eye.
} 
\label{fig:compare1mmflux}
\end{figure}

The measured ALMA 260 GHz (1100\,\micron) flux densities are listed in the last column of Table~\ref{tab:smm}, adjacent to the AzTEC 1100\,\micron\ flux density.  Six sources are observed by both ALMA and AzTEC (see Figure~\ref{fig:compare1mmflux}), and the total 260 GHz flux density measured by ALMA agrees well with AzTEC measurements in most cases.  This indicates that our ALMA observations are sensitive to and are capable of recovering the entire 1 mm continuum flux detected by AzTEC, despite its nearly 30 times higher angular resolution.
One possible exception is the brightest source PJ084650, which is a large lensed arc system with multiple components including a 6\arcsec\ long arc (see top right panel of Figure~\ref{fig:ALMAstamps}).  Its total measured ALMA 260 GHz flux of $76\pm11$ mJy is 36\% smaller than the total flux measured by AzTEC, and this ``missing flux" is an expected result since the largest angular scale (LAS) recoverable by the ALMA configuration used is $\sim4\arcsec$.  

\section{Spectral Energy Distribution Analysis of the \Planck\ sources}
\label{sec:SEDanalysis}

\begin{table*}
\caption{Summary of WISE Near-IR and VLA 20 cm photometry}
\label{tab:OIR}
 \scriptsize
\begin{tabular}{lccccc}
\hline
\hline
Source ID & $W_{3.4\mu}$  & $W_{4.6\mu} $  & $W_{11\mu}$  & $W_{22\mu}$ & $S_{\rm 20cm}$ \\
          & (mJy) & (mJy) & (mJy) & (mJy) & (mJy) \\
\hline
PJ011646.8 & $0.93\pm0.09$ & $1.17\pm0.16$ & $0.67\pm0.13$ & $13.4\pm2.2$ & $1.4\pm0.5$ \\ 	
PJ014341.2 & $0.63\pm0.02$ & $0.53\pm0.02$ & $0.98\pm0.07$ & $<3.0$ & $<1.5$  \\ 		
PJ022634.0 & $0.64\pm0.09$ & $0.45\pm0.07$ & $0.71\pm0.12$ & $4.9\pm0.9$ & $17\pm3.0$ \\ 		
PJ030510.6 & $0.14\pm0.02$ & $0.20\pm0.03$ & $0.59\pm0.10$ & $5.0\pm0.9$ & $<2.0$   \\ 		
PJ074852.6 & $0.22\pm0.03$ & $0.29\pm0.05$ & $1.00\pm0.15$ & $4.0\pm0.9$ & $5.6\pm1.0$  \\ 		
PJ084650.1 & $0.24\pm0.03$ & $0.31\pm0.05$ & $2.63\pm0.40$ & $17.2\pm3.0$ & $3.5\pm0.5$  \\ 		
PJ105322.6 & $0.19\pm0.03$ & $0.45\pm0.05$ & $0.43\pm0.06$ & $<5.6$ & $<1.5$ \\ 		
PJ112713.4 & $0.40\pm0.06$ & $0.56\pm0.07$ & $1.16\pm0.15$ & $2.86\pm1.00$ & $2.4\pm0.4$\\ 	
PJ113805.5 & $0.14\pm0.02$ & $0.19\pm0.02$ & $<0.83$ & $3.77\pm1.00$ & $1.5\pm0.3$\\ 	
PJ113921.7 & $0.13\pm0.02$ & $0.19\pm0.02$ & $<1.25$ & $<5.5$ & $<0.5$ \\ 	
PJ114038.5 & $1.00\pm0.15$ & $1.11\pm0.20$ & $12.5\pm1.8$ & $96.4\pm15.0$ & $17.6\pm2.5$ \\ 	
PJ114329.5 & $1.36\pm0.19$ & $1.25\pm0.17$ & $11.1\pm1.5$ & $16.6\pm3.0$ & $3.6\pm1.0$ \\ 	
PJ132217.5 & $0.34\pm0.04$ & $0.33\pm0.05$ & $<0.48$ & $4.50\pm1.50$ & $1.5\pm0.5$  \\ 	
PJ132630.3 & $0.26\pm0.03$ & $0.25\pm0.03$ & $<1.3$ & $<9.7$ & $<0.5$  \\ 	
PJ132934.2 & $0.37\pm0.04$ & $0.45\pm0.05$ & $0.71\pm0.10$ & $10.6\pm2.0$ & $3.6\pm0.5$ \\ 	
PJ133634.9 & $0.37\pm0.06$ & $0.33\pm0.05$ & $0.29\pm0.10$ & $3.63\pm0.9$ & $0.60\pm0.15$ \\ 	
PJ141230.5 & $0.22\pm0.03$ & $0.15\pm0.03$ & $0.55\pm0.08$ & $3.34\pm0.7$ & $1.5\pm0.3$  \\ 	
PJ144653.2 & $0.81\pm0.08$ & $0.79\pm0.08$ & $1.70\pm0.20$ & $7.4\pm1.0$ & $1.6\pm0.3$  \\ 	
PJ144958.6 & $0.35\pm0.04$ & $0.37\pm0.04$ & $0.28\pm0.10$ & $2.07\pm0.63$ & $1.3\pm0.3$  \\ 	
PJ154432.4 & $0.25\pm0.03$ & $0.17\pm0.02$ & $<0.69$ & $2.3\pm0.3$ & $0.50\pm0.15$  \\ 	
PJ231356.6 & $0.53\pm0.05$ & $0.56\pm0.06$ & $0.49\pm0.15$ & $6.3\pm1.3$ & $4.48\pm0.45$\\ 	
\hline
\end{tabular}
\break
\end{table*}

\subsection{Summary of Multi-wavelength Photometry \label{sec:SEDdata}}

\begin{figure*}
\includegraphics[width=1.8\columnwidth]{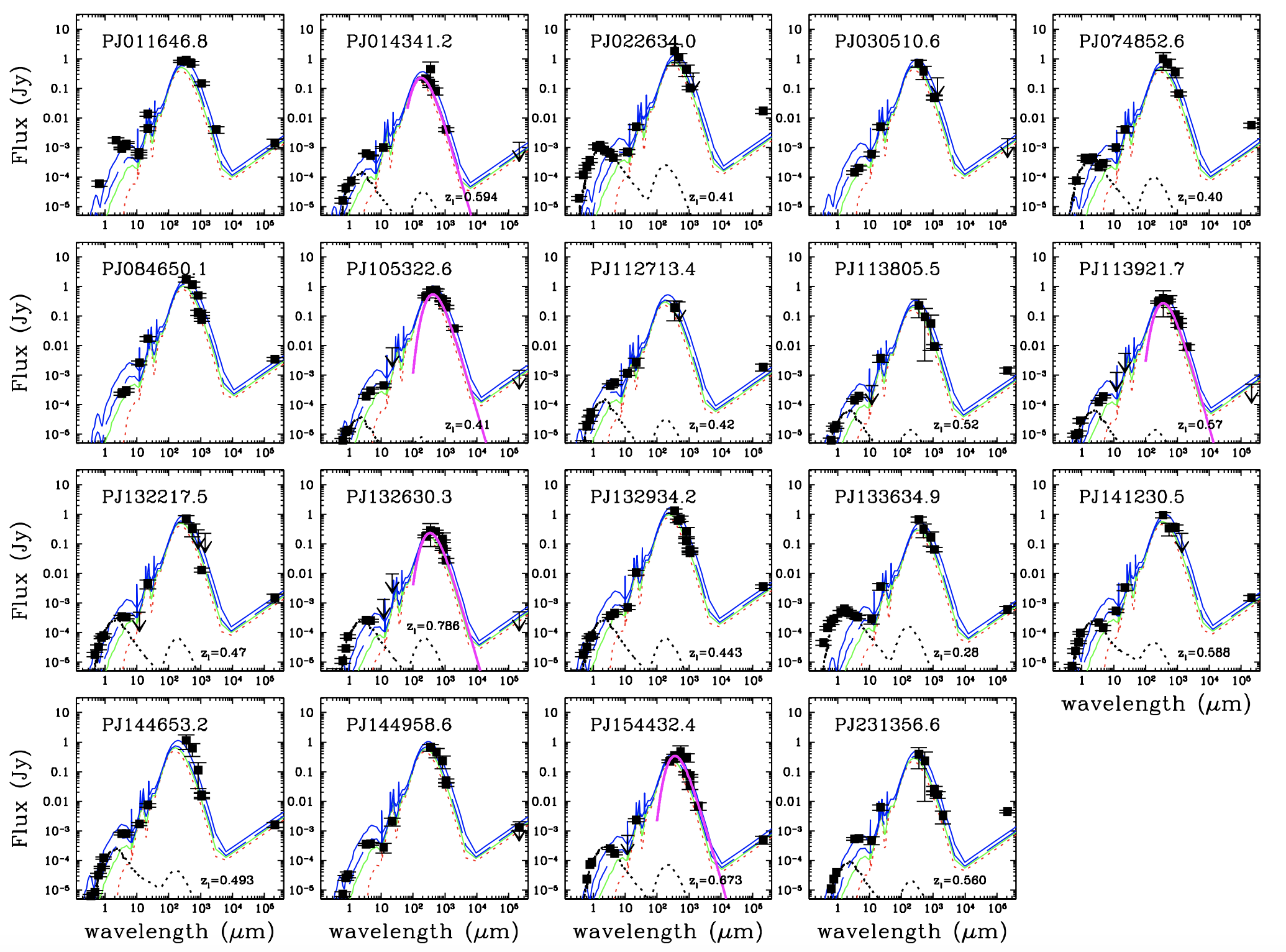}
\caption{Spectral Energy Distributions (SEDs) of the 19 new $z\ge1$  \Planck\ sources with photometry listed in Table~\ref{tab:smm}. Two sources, PJ112713.4 and PJ114038.5, have only the RSR spectral line data aside from the \Planck\ photometry, and their SEDs are now shown.  The models shown are the best fit dusty starburst SEDs by \citet{efstathiou00} with an extension to the radio wavelengths \citep{yun02} at their CO redshifts (see Table~\ref{tab:RSR}).  Modified blackbody models fitting only the \Herschel\ and AzTEC photometry are shown in thick magenta lines.   
An elliptical galaxy SED template by \citet{polletta07} covering the optical wavelength bands is also shown using a dashed line for the sources with an obvious foreground galaxy in the SDSS plates. 
}
\label{fig:SEDs}
\end{figure*}

The \Planck, \Herschel, AzTEC, and ALMA photometry of the \Planck-\WISE\ DSFGs is summarised in Table~\ref{tab:smm}. Their spectral energy distributions (SEDs) mapped by these bands contain the bulk of their bolometric luminosity, as shown in Figure~\ref{fig:SEDs}.  The AzTEC and ALMA 1.1 mm photometry (``$AzTEC_{1100}$" and ``$ALMA_{1100}$") is already discussed in some detail above.  Here, we discuss briefly the \Planck, \Herschel, \WISE, and 20cm radio continuum photometry obtained from archives and literature (see Table~\ref{tab:OIR}).  In most cases, the uncertainty on photometry reported is dominated by systematic uncertainties, such as the source confusion for the \Planck\ photometry and absolute flux calibration uncertainty for all other photometry.

\smallskip
\noindent{\bf Planck Photometry}.  The \Planck\ 857 GHz (350 \micron, ``$Planck_{350}$"), 545 GHz (500 \micron, ``$Planck_{500}$"), and 353 GHz (850 \micron, ``$Planck_{850}$") flux densities reported in Table~\ref{tab:smm} are the \Planck\ aperture photometry (APERFLUX) values in the PCCS2\footnote{The nomenclature for the \Planck\ photometry in Table~\ref{tab:smm} uses the wavelength (i.e., 350 \micron, 500 \micron, 850 \micron), rather than the frequency designation, to be consistent with the convention used by \citet{harrington16}.}.  The PCCS2 includes four different estimates of source flux: detection pipeline photometry (DETFLUX), aperture photometry (APERFLUX), PSF fit photometry (PSFFLUX), and Gaussian fit photometry (GAUFLUX).  DETFLUX is suggested as the flux estimation method of choice for unresolved sources in regions of low background, given its greater sensitivity. The internal consistency check has shown that DETFLUX is subject to a greater scatter and a significant flux bias.  In addition, an external consistency check by comparing \Planck\ and \Herschel\ data indicates a greater reliability for APERFLUX over DETFLUX \citep{planck26a}.  Therefore, we adopt the APERFLUX for \Planck\ band photometry in our analysis.  For the five \Planck-\WISE\ sources with published \Herschel\ photometry, \Planck\ photometry is consistent with the higher resolution \Herschel\ data (see Table~\ref{tab:smm}).  The agreement between the \Planck\ 353 GHz photometry and the published SCUBA-2 photometry \citep{canameras15,diazsanchez17} is not as good for the five sources in common, and there might be some systematic problems with either set of photometry data.  As discussed in \S~\ref{sec:literature}, two \Planck\ sources found in the literature (Cosmic Eyelash, PJ090403.9) included in Table~\ref{tab:smm} are in crowded fields with multiple \Herschel\ sources, and the discrepancy between \Planck\ and \Herschel\ can be explained by this source blending.

\smallskip
\noindent{\bf Herschel Photometry}.  The \Herschel\ Space Observatory and its dedicated ``large area" surveys such as the \Herschel\ ATLAS \citep[H-ATLAS,][]{eales10} and the Herschel Multi-tiered Extragalactic Survey \citep[HerMES,][]{oliver12} covered less than 10\% of the extragalactic sky in total.  Therefore, only a small fraction of \Planck-selected high-redshift candidates are expected to have \Herschel\ photometry.  We found \Herschel\ photometry of a total of 5 \Planck-selected sources in the NASA/IPAC Infrared Science Archive (IRSA), and they are included in Table~\ref{tab:smm}.  PJ014341.2 and PJ132630.3 are located in the survey areas covered by the \Herschel\ Stripe 82 Survey \citep{viero14} and the H-ATLAS \citep{eales10}, respectively, while the three remaining sources (PJ105322.6, PJ113921.7, P154432.4) are among the 200 \Planck\ sources targeted as part of the ``must-do" Director's Discretionary Time (DDT) program \citep{canameras15}.  Excluding the three sources specifically targeted by the \Herschel\ DDT program, only two out of 19 PCCS high-redshift sources we identified have a chance coverage by the existing \Herschel\ archive, and this supports our main motivation for using the \Planck\ survey data itself for finding the brightest high-redshift IR sources.

\smallskip
\noindent{\bf WISE Photometry}.  Near-IR colour is one of the key discriminators for heavily obscured starburst galaxies in our high-redshift candidate selection (see \S~\ref{sec:OIR}).  The \WISE\ photometry used in this process comes from the \WISE\ All-Sky Survey \citep{wright10}.  The relative sensitivity of the longer wavelength channels degrade rapidly with increasing wavelength -- the $5\sigma$ detection limits of the catalog are 0.068, 0.098, 0.86, \& 5.4 mJy in the 3.4 $\mu$m, 4.6 $\mu$m, 11 $\mu$m, \& 22 $\mu$m bands, according to the Explanatory Supplement to the WISE
All-Sky Data Release Products\footnote{\url{http://wise2.ipac.caltech.edu/docs/release/allsky/expsup/index.html}}.  At the same time, the mid-IR SEDs of these \Planck-selected high-redshift dusty starbursts are so steep and so bright that nearly all of them are also substantially detected in the 11   $\mu$m and the 22 $\mu$m bands as well -- see Table~\ref{tab:OIR} and also Table~5 by \citet{harrington16}.

\smallskip
\noindent{\bf 20 cm Radio Continuum}.  The radio-FIR correlation is a well established global relation for star forming galaxies, rooted in the formation and evolution of massive stars \citep[e.g.,][]{condon92,yun02}.  If these \Planck-selected high-redshift galaxies with extreme luminosity are powered by star formation, they are also expected to follow the same radio-FIR correlation and should also be bright and detectable in radio continuum even by relatively shallow surveys.  The 20 cm radio continuum photometry listed in Table~\ref{tab:OIR} comes from either the NRAO VLA Sky Survey \citep[NVSS,][]{condon98} or the VLA Faint Images of the Radio Sky at Twenty-Centimeters Survey \citep[FIRST,][]{becker95}.  The majority of the high-redshift \Planck\ sources are detected by the shallower, all-sky NVSS survey.  Those located in the FIRST survey coverage are detected  more securely with a higher resolution and SNR.  Most detections and upper limits are consistent with their following the local radio-FIR correlation, as shown in  Figure~\ref{fig:SEDs}, and suggest that the nature of the luminosity for these galaxies is consistent with being powered by a pure starburst in nearly all cases.

\begin{table*}
\caption{Best Fit Parameters of the Starburst SED Template and modified blackbody models for each source with derived dust and ISM masses}
\label{tab:chisq}
\begin{tabular}{@{}lcccccccc}
\hline
Source ID & $\mu L_{\rm IR,SB}^\dagger$ & $\mu SFR_{\rm SB}$ & $T_{\rm d} $    & $\mu L_{\rm IR,BB}^{\dagger \dagger}$ & $\mu SFR_{\rm BB}$   & $\mu M_{\rm d}$ & $\mu M_{\rm ISM} $ & $\mu_{\rm IR}^{\dagger \dagger \dagger}$ \\
		& ($10^{14}  L_\odot)$ & ($M_\odot$ yr$^{-1}$)  & (K) & ($10^{14}  L_\odot$) &	($M_\odot$ yr$^{-1}$)  &  ($10^{10} M_\odot$)  & ($10^{10} M_\odot$) & \\		
\hline
PJ011646.8 & $1.2\pm0.3$ & 9,001 & -- &  -- & -- & $2.08\pm0.41$ & $817\pm162$ & $7.0\pm0.3$ (1) \\ 	
PJ014341.2 & $0.10\pm0.02$ & 1,064 & $52.8_{-14.6}^{+39.4}$ & $ 0.10 \pm 0.43 $ & 1,064 & $0.12\pm0.02$& $49\pm9$ & $7.1\pm2.2$ (1) \\ 		
PJ022634.0 & $2.7\pm0.6$ & 23,147 & $--$ & $ -- $ & -- & $2.77 \pm 0.56 $ & $1074 \pm 214$ & -- \\ 		
PJ030510.6 & $1.3\pm0.2$ & 12,715 & $--$ & $ -- $ & -- & $1.51 \pm 0.31$ & $592 \pm 119$ & $2.2\pm0.6$ (1) \\ 		
PJ074852.6 & $1.6\pm0.3$ & 13,806 & $--$ & $ -- $ & -- & $1.81 \pm 0.36 $ & $703 \pm 140$ & -- \\ 		
PJ084650.1 & $1.1\pm0.3$ & 14,790 & $--$ & $ -- $ & -- & $2.31 \pm 0.46 $ & $908 \pm 181$ & -- \\ 		
PJ105322.6 & $2.5\pm0.5$ & 16,735 & $47.4_{-6.2}^{+5.8}$ & $ 2.24 \pm 0.08 $ & 23,816 & $5.34\pm1.07 $ & $2071\pm414$ & $7.6\pm0.5$ (2) \\ 				
PJ112713.4 & $0.25\pm0.04$ & 1,845  & $ --$ & $ -- $ & --  & $ 0.41  \pm  0.08 $ &  $ 82 \pm 16 $ & -- \\	
PJ113805.5 & $0.24\pm0.04$ & 1,872 & $--$ & $ -- $ & -- & $0.30\pm0.06 $ & $119\pm24$ & $2.8\pm0.4$ (1) \\ 	
PJ113921.7 & $1.2\pm0.2$ & 9,131 & $47.4_{-5.6}^{+3.6}$ & $ 0.79 \pm 0.13 $ & 8,450 & $1.47 \pm 0.30 $ & $576 \pm 115$ & $4.8\pm0.4$ (1) \\ 	
PJ114038.5 & $0.026\pm0.005$ & 221 & $--$ & $ -- $ & -- & $--$ & $--$ & -- \\ 	
PJ114329.5 & $0.010\pm0.003$ & 81 & $--$ & $ -- $ & -- & $-- $ & $--$ & -- \\ 	
PJ132217.5 & $1.2\pm0.2$ & 9,131 & $--$ & $ -- $ & -- & $0.42\pm0.08 $ & $163\pm32$ & -- \\ 	
PJ132630.3 & $0.75\pm0.16$ & 4,677 & $51.6_{-5.8}^{+9.2}$ & $ 0.56 \pm 0.10 $ & 5,982 & $0.83\pm0.15$ & $327\pm65$ & $4.3\pm0.6$ (1) \\ 	
PJ132934.2 & $1.5\pm0.3$ & 12,028 & $--$ & $ -- $ & -- & $1.63 \pm 0.32 $ & $640 \pm 128$ & $11\pm2$ (3) \\ 	
PJ133634.9 & $1.8\pm0.4$ & 17,944 & $--$ & $ -- $ & -- & $1.74 \pm 0.35 $ & $674 \pm 135$ & -- \\ 	
PJ141230.5 & $3.1\pm0.7$ & 26,567 & $--$ & $ -- $ & -- & $--$ & $--$ & -- \\ 	
PJ144653.2 & $0.28\pm0.07$ & 1,910 & $--$ & $ -- $ & -- & $0.53 \pm 0.11 $ & $206 \pm 41$ & $3.9\pm0.7$ (1) \\ 	
PJ144958.6 & $1.0\pm0.2$ & 8,821 & $--$ & $ -- $ & -- & $1.17 \pm 0.23 $ & $460 \pm 92$ & $10.1\pm5.2$ (1) \\ 	
PJ154432.4 & $0.63\pm0.13$ & 5,261 & $41.4_{-4.2}^{+7.0}$ & $ 0.50 \pm 0.10 $ & 5,332 & $0.97 \pm 0.19 $ & $377 \pm 75$ & $14.7\pm0.8$ (2) \\ 	
PJ231356.6 & $0.61\pm0.13$ & 6,383 & $--$ & $ -- $ & -- & $0.75 \pm 0.15 $ & $290 \pm 58$ & $6.1\pm1.2$ (1) \\ 	
\hline
\end{tabular}
\break
$^\dagger$  $L_{\rm IR}$ is the far-infrared luminosity derived by integrating between 8-1000 $\micron\ $ in wavelength \citep{sanders96}. \\ $^{\dagger\dagger}$  The modified blackbody SED model used to derive $\mu L_{\rm IR,BB}$ is described in Appendix~C by \citet{harrington16}. \\
$^{\dagger\dagger\dagger}$  Magnification factor $\mu_{\rm IR}$ for dust or gas derived from lens modelling (see \S~\ref{sec:lensing}).  The references for the estimates: (1) Kamieneski et al. (in prep); (2) \citet{canameras18}; (3) \citet{diazsanchez17}. \\
\end{table*}

\subsection{Modified Blackbody Model \label{sec:BB}}

We explore constraints on the infrared luminosity and characterise the dust emission for these systems by analysing their SEDs using a modified blackbody model, which is a widely used characterisation of dust emission in galaxies.  We adopt the functional form based on the derivation by \citet{yun02} for the modified blackbody model, and a detailed description of our method is given by \citet{harrington16}  in Appendix~C.   This approach was successful in deriving apparent IR luminosity and characteristic dust temperature for the eight \Planck\ sources identified using the \Herschel\ data studied by \citet{harrington16}, yielding consistent estimates of IR luminosity and SFR as the template SED analysis.  

Obtaining meaningful estimates of dust temperature $T_{\rm d}$ and IR luminosity is not as successful for those objects with only the \Planck\ photometry.  The primary reasons for this poor outcome are two-fold: (1) a low SNR of the \Planck\ photometry; and (2) the lack of constraints on the short wavelength side ($\lambda<200\, \mu$m) of the dust peak.  Photometry along the Rayleigh-Jeans (R-J) part of the dust spectrum alone offers little constraints on $T_{\rm d}$ or the total IR luminosity.  Even in cases where the \Planck\ 857 GHz (350 \micron) photometry probes the dust peak, a large uncertainty in the {\em absolute} photometry and the resulting low S/N ratio (e.g., $SNR\le3$, see Table~\ref{tab:smm}) lead to poor constraints on $T_{\rm d}$ and $L_{\rm IR}$.  In contrast, any available \Herschel\ 250 \micron\ photometry can provide a critical constraint on the dust peak, yielding a much stronger constraint on the dust SED, as shown by the magenta solid lines in Figure~\ref{fig:SEDs}.  

For those sources with secure \Herschel\ and AzTEC 1100 \micron\ photometry, the derived $T_{\rm d}$ ranges between 41 and 53 K, with IR luminosity and SFR in good agreement with the results from the template SED analysis described in the next section.  A more careful examination of Figure~\ref{fig:SEDs} shows that the modified BB models  fall slightly below the dusty star forming galaxy templates in the mid-IR range in nearly every case, reflecting the limitation of assuming a single temperature to characterise a dust SED, and the derived IR luminosity and SFR should be considered lower limits.

\subsection{Template SED Analysis \label{sec:SED}}

Even for an object without a strong constraint on the dust peak, IR luminosity and SFR can be estimated by adopting an SED, granted it is a plausible one for the object in consideration.  There is growing evidence that SEDs of luminous infrared galaxies vary with luminosity and redshift, but a remarkable similarity is also seen among objects of similar luminosity and SFR, even with the presence of an AGN \citep[e.g.,][]{kirkpatrick12}.  For these \Planck-selected DSFGs, template SEDs representing an ensemble of dust obscured young star clusters by \citet{efstathiou00} offer a good fit for the eight galaxies studied by \citet{harrington16}, and we adopt the same SED templates here for the analysis of their IR luminosity and SFR.

As shown in Figure~\ref{fig:SEDs}, the template SEDs by  \citet{efstathiou00} provide a good model for the observed SEDs of the 19 \Planck\ selected high-redshift sources as well from 1 \micron\ to 20 cm, spanning 5 decades in wavelength.  The photometric measurements in optical and near-IR bands are sometimes dominated by the foreground lensing galaxy or galaxies, and most of them are adequately modelled by adding a second SED component of an elliptical galaxy \citep{polletta07}.  While these model SEDs are not perfect, they are sufficient to offer a good estimate of the apparent IR luminosity and SFR.

The apparent IR luminosity and SFR\footnote{The star formation rates in  Table~\ref{tab:chisq} are calculated using the empirical calibration by \citet{kennicutt98}, corrected for the Kroupa IMF [i.e., $SFR=L_{\rm IR}/(9.4\times 10^9 L_\odot)\, M_\odot /{\rm yr}$].} derived from the template SED analysis are summarized in Table~\ref{tab:chisq}. They span the ranges $(0.1-3.1)\times 10^{14}L_\odot$ and $(1-27)\times10^3 M_\odot$ yr$^{-1}$, respectively, for the 19 high-redshift sources analysed.   For the five sources with \Herschel\ photometry, the model SEDs and their derived quantities (IR luminosity and SFR) are in good agreement between the modified blackbody modelling and the template SED analysis.  

The radio continuum portion of the template SEDs shown in Figure~\ref{fig:SEDs} is a model extension of the dusty starburst SEDs by \citet{efstathiou00} using the radio-IR correlation \citep[see review by][]{condon92,yun02}.  Over 98\% of the local star forming galaxies selected in the IR follow this relation closely \citep{yun01}, and the majority of these \Planck-selected DSFGs also show 20 cm radio continuum detection or upper limits that are consistent with the local relation.  \citet{yun01} have also shown that about 1\% of local SFGs are a factor of a few to 10 times brighter in radio continuum than the expected relation, indicating the presence of a weak radio AGN activity, and this fraction rises to 10\% for $L_{\rm IR}\gtrsim 10^{11}L_\odot$.  A few \Planck-selected sources (PJ022633.9, PJ074852.6, PJ231356.6) show a similar radio-excess with 3-10 times stronger radio continuum, indicating a possible presence of a weak radio AGN.  These sources show remarkably little sign of any AGN activity (see \S~\ref{sec:AGN}), and this radio-excess might be the only clue that some of these sources might be hosting a heavily obscured AGN \citep[also see][]{geach15}.  Our high resolution radio continuum imaging study using the VLA has shown, however, that the excess radio continuum is associated with the nuclei of the foreground lensing galaxy, rather than the submillimetre source, at least in some of the cases (Kamieneski et al., in prep.).

\begin{figure}
\includegraphics[width=1.0\columnwidth]{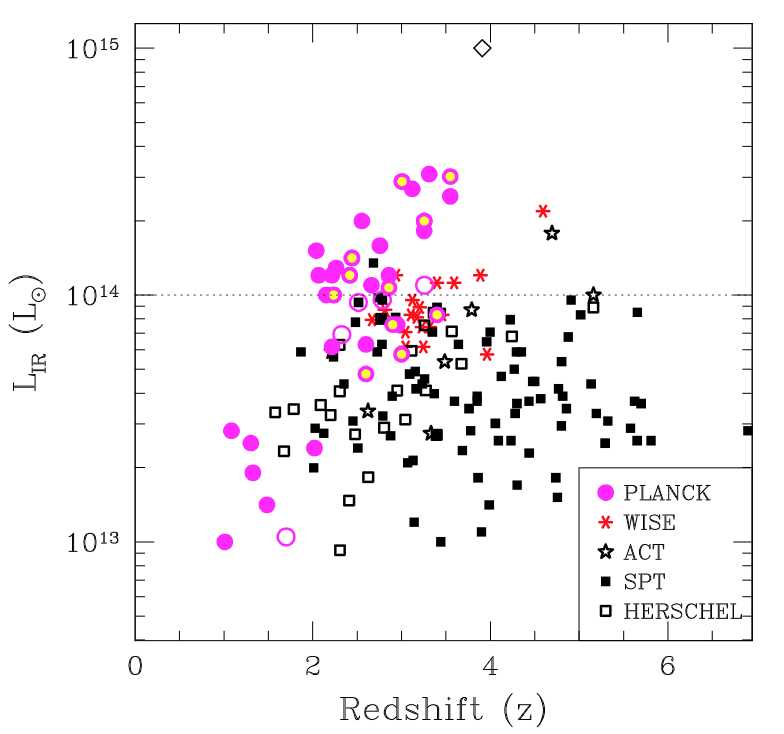}
\caption{Comparison of IR (8-1000 \micron) luminosity of the \Planck\ sources \citep[filled magenta circles, from this study as well as by][]{canameras15,harrington16} with those of strongly lensed SMGs identified by the \SPT\ \citep{vieira13,weiss13,reuter20}, \ACT\ \citep{su17,gralla20},  and the \Herschel\ \citep{bussmann13,wardlow13,bakx18}.  The four \Planck-\WISE\ sources previously identified as luminous DSFGs are shown with empty magenta circles while those identified previously by \citet{canameras15} are shown by thicker circle filled with a yellow dot.  The most luminous \WISE\ sources \citep{tsai15} are shown as stars, while the $z=3.91$ lensed infrared QSO APM 08279+5255 \citep{irwin98,rowanrobinson00}, the most luminous object known in the IR, is shown as a diamond.  Seventeen out of the 25 most luminous known IR sources in the entire sky are from our survey. }
\label{fig:LFIR}
\end{figure}

\section{Discussion}
\label{sec:discussion}

\subsection{Extreme IR Luminosity of \Planck\ DSFGs \label{sec:LIR}}

\citet{harrington16} have found their eight high-redshift \Planck\ sources identified using archival \Herschel\ data are among the most luminous IR sources known, with apparent IR luminosity $L_{8-1000\mu} \ge10^{14} L_\odot$ in the majority of the cases \citep[also see][]{canameras15}.   They match or exceed the IR luminosities of the \WISE\ selected HyLIRGs, which are the most luminous IR galaxy population previously identified \citep{tsai15}.   When the 19 new high-redshift \Planck\ sources from this study are added, as shown in Figure~\ref{fig:LFIR}, it becomes clearer that the \Planck-selected DSFGs overlap broadly with the \Herschel, \WISE, \ACT, and \SPT\ selected DSFGs in their apparent IR luminosity, but they occupy the top range in $L_{\rm IR}$ at each redshift range out to $z\approx4$. 
The majority of the \Planck-selected DSFGs cluster above $L_{\rm IR}\gtrsim 10^{14} L_\odot$, and the \Planck\ selected DSFGs are in fact the dominant population (17 out of 25, or 68\%) among all known objects with $L_{\rm IR}\gtrsim 10^{14} L_\odot$.  Adding the latest compilations of the \SPT\ \citep{reuter20}, \ACT\ \citep{su17,gralla20}, and \Herschel\ \citep{bussmann13,wardlow13,bakx18} sources in the comparison makes it even clearer that these \Planck\ selected DSFGs are on average 4 to 10 times more luminous than the lensed DSFGs identified by the other surveys.  
The lensed $z=3.91$ QSO APM~08279+5255\footnote{Detailed modelling of dust continuum and warm molecular gas emission by \citet{weiss07} suggests that the bulk of its bolometric luminosity arises from a compact (100-350 parsec) radius region surrounding the AGN, which is magnified by an effective magnification of 60-120.} \citep{irwin98,weiss07} stands out as a true exception in this comparison.

The \ACT\ and \SPT\ sources extend out to a higher redshift of $z\sim6$ while no \Planck\ source is identified at $z>4$ yet.  This is likely reflecting the redshift bias in the sample selection:  \Planck\ and \Herschel\ sources are selected primarily at 857 \& 545 GHz bands and are thus preferentially at lower redshifts compared with \ACT\ and \SPT\ sources selected at 220 \& 150 GHz (see \citet{bethermin15} and \citet{casey18} for discussions about the selection effects imposed by wavelength selection). 
The $z\ge4$ luminous DSFG population is also rarer in number, and a larger survey is needed to find them. Our redshift survey of the full sample of the \Planck\ selected sources and analysis of their physical properties should reveal whether the \Planck\ and other surveys probe the same parent population. 

There is a second grouping of \Planck-selected DSFGs in the redshift range of $z=1-2$ with $L_{\rm IR}\ge 10^{13}L_\odot$.  While they are not as luminous as many of the $z>2$ DSFGs shown in Figure~\ref{fig:LFIR}, they are still some of the most luminous galaxies known at $z<2$. The {\em absence} of DSFGs with $L_{\rm IR}\ge 10^{14}L_\odot$ in the same redshift range is noteworthy.  A flux-limited survey such as this is generally not biased against bright objects, and our \Planck\ colour selection (see \S~\ref{sec:Pcolmag}) is designed to ensure a detection if they exist.  Gravitational lensing can introduce an element of a bias against nearby objects because a massive foreground object is required for strong gravitational lensing. The magnitude of this effect is heavily dependent on assumptions going into computing a lensing optical depth, as shown by the broad range of predictions by \citet{blain98b}, \citet{hezaveh11}, \citet{bethermin15}, and \citet{strandet16}.  The fraction of  \Planck-selected lensed DSFGs in the redshift range $z=1-2$ is $\approx20$\% of the total, nearly identical to that of the \Herschel-selected DSFGs \citep{amvrosiadis18}. If the two $z=1$ sources with a single line redshift are actually $z=2$ sources (see Table~\ref{tab:RSR}), then the source fraction is reduced to $\approx15$\% of the total. 

Another possibility is that DSFGs with $L_{\rm IR}\ge 10^{14}L_\odot$ are too rare to be found in the volume of the lower redshift range.  The co-moving volume of the redshift range between $z=1$ and $z=2$ is about $30\%$ of the entire co-moving volume enclosed within $z=4$, and one can expect $\sim$7 DSFGs with $L_{\rm IR}\ge 10^{14}L_\odot$ in our sample between $z=1$ and $z=2$ if they occur with the same spatial density.  While a small sample size prevents us from drawing a firm conclusion, this analysis suggests that these extreme luminosity DSFGs are indeed absent during the past 10 Gyr of the cosmic history, in line with the known decline in the cosmic star formation rate density between $z=2$ and now \citep{madau14,scoville17,zavala21}. 

So, what exactly are these extremely luminous DSFGs identified by the \Planck\ survey? Modelling of the extragalactic millimetre and submillimetre source populations such as by \citet{blain98b} and \citet{negrello07} has suggested that a strongly lensed population of high redshift DSFGs would dominate the bright end of the source counts.  On the other hand, the most luminous high redshift objects discovered by previous all-sky IR surveys such as  APM~08279+5255 \citep{irwin98}, FSC~10214+4724 \citep{rowanrobinson91}, and H1413+117 \citep{barvainis92} are all IR QSOs, as are the more recently discovered population of extremely luminous infrared sources by the \WISE\ survey \citep{tsai15}.  Therefore, there are at least two different potential explanations for the extreme apparent luminosity of these \Planck-selected high redshift sources, and we explore these scenarios further here.

\begin{figure}
\includegraphics[width=0.9\columnwidth]{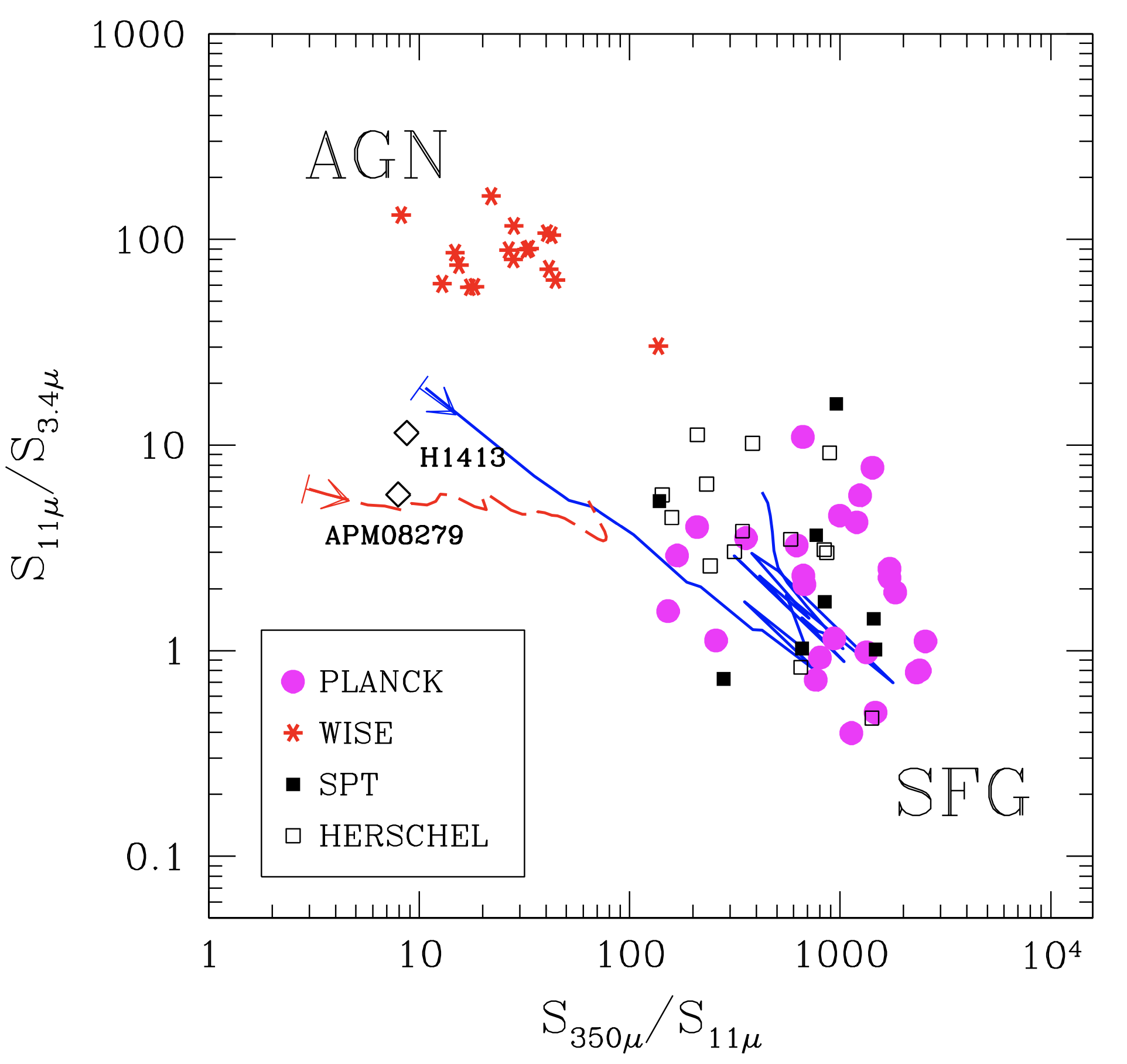}
\caption{Mid-IR to FIR colour diagnostic plot for star forming and AGN galaxies.  This is essentially the same plot as Figure~9 by \citet{harrington16} but including all \Planck-selected high-redshift sources and using the 350 \micron\ to 11 \micron\ colour for the x-axis.  All symbols are the same as Figure~\ref{fig:LFIR}.   Colour tracks for the z = 2 SF template (solid blue line) and the featureless power-law AGN template (long dashed red line) by 
 \citet{kirkpatrick13} are shown, starting from z = 0.5 and ending at z = 4, as marked by the arrows. The high-redshift IRQSOs APM 08279+5255 \citep{irwin98} and H1413+117 \citep{barvainis92} are also shown for reference.}
\label{fig:AGNcolor}
\end{figure}

\subsubsection{IR QSO Scenario \label{sec:AGN}}

There are at least two different potential sources of luminosity for these extreme luminosity objects: (1) a massive starburst powered by young stars and (2) an accreting SMBH.  Both processes might be at work simultaneously in almost all cases at some level, and either process might dominate the radiative output of a galaxy at any given time.  The luminosity of a starburst produced by OB stars averaged over their main sequence lifetime yields
\begin{equation}
  L_{\rm SB}=4.9\times 10^{43}\, [\frac{\dot{M}}{M_\odot \, {\rm yr}^-1}] \ {\rm erg~s}^{-1}
\end{equation}
where $\dot{M}$ is the gas consumption rate by star formation \citep[see Eq.~5 by][]{scoville83}.  If only the luminosity and lifetime of O stars are considered, time-averaged $L_{\rm SB}$ can be $\sim5$ times larger (N. Scoville, private communication).  
In comparison, accretion luminosity of a SMBH is  
\begin{equation}
  L_{\rm BH} = \epsilon \dot{M} c^2 = 6\times10^{45}\, [\frac{\epsilon}{0.1}][\frac{\dot{M}}{M_\odot \, {\rm yr}^-1}] \ {\rm erg~s}^{-1}
\end{equation} 
where $\epsilon$ is the radiative efficiency and $\dot{M}$ is the mass accretion rate. Therefore, an accreting BH can produce luminosity 20-100 times more efficient than a starburst at any stage.  On the other hand, BH accretion is spatially limited to the parsec-scale immediate sphere of influence, while a starburst can occur over 100s of parsecs to several kpc regions simultaneously.  As a result, detailed numerical modelling of a rapidly accreting BH within a merger-driven starburst, such as by \citet{hopkins06,hopkins08}, have shown that the peak luminosity of a starburst and that of an SMBH fuelled by the same inflowing gas streams are comparable in magnitude, and some care might be needed to distinguish their respective luminosity contributions.

The most luminous IR sources with  $L_{\rm IR}\gtrsim 10^{14} L_\odot$ known prior to the discovery of these \Planck-selected DSFGs were all IR QSOs, whereas a  galaxy undergoing star formation has a plausible physical upper limit of $L_{\rm IR}\approx (2-3)\times 10^{13} L_\odot$ \citep[$SFR \approx 2000-3000\, M_\odot$ yr$^{-1}$, see][]{harrington16}.  Since an accreting BH can be highly efficient in converting mass to luminosity, a heavily obscured luminous AGN activity can offer a natural explanation for these \Planck-selected DSFGs with $L_{\rm IR}\gtrsim10^{14} L_\odot$, as is the case for the \WISE\ HyLIRGs \citep{tsai15}.  A merger-driven starburst and subsequent, or simultaneous, fuelling of a luminous AGN has been suggested as a natural explanation for the high frequency of ULIRGs with warm IR colour and Seyfert spectra \citep{sanders88}.  Numerical simulations of this scenario \citep{hopkins06,hopkins08} have yielded a plausible cosmological framework for the connection between gas-rich galaxy mergers and quasar activity.  This mechanism also serves as the prototype of the BH-host coevolution driven by quasar-mode feedback \citep[see review by][]{kormendy13}. 

The mid-IR to far-IR colour diagnostic analysis for IR AGNs first introduced by \citet{kirkpatrick12} was later adopted for the \WISE\ bands by \citet{harrington16} to explore the frequency of IR QSOs among the \Planck-selected DSFGs.  The same analysis repeated with better statistics in Figure~\ref{fig:AGNcolor}, and we find that {\em all} of our \Planck-selected DSFGs have a cold IR SED of star forming galaxies with little evidence for any enhanced mid-IR warm dust emission characteristic of an enshrouded AGN activity.  The individual SEDs shown in Figure~\ref{fig:SEDs} also support this conclusion, although \WISE\ 11 \& 22 \micron\ bands are the only measurements sensitive to the power-law IR AGN emission.  This is a rather surprising and somewhat unexpected outcome, given their remarkable luminosity and the widely adopted merger-driven starburst-AGN co-evolution model \citep[][]{hopkins06,hopkins08}.  In fact, one of the key predictions of the AGN co-evolution model is that up to 90\% of the luminous quasar phase, fuelled by the high gas accretion that also powers the starburst, should remain obscured by dust, presumably leading to their identification as IR QSOs such as APM 08279+5255 or as power-law mid-IR sources such as the \WISE-selected HyLIRGs found on the top left corner of Figure~\ref{fig:AGNcolor}.  

A generic feature in these models is a 10-20 Myr delay between the peak starburst and the start of the quasar phase \citep[e.g., see Figure~1 by][]{hopkins08}, and one possible explanation is that these \Planck-selected DSFGs are seen exclusively during this brief phase prior to the quasar phase.  On the other hand, this time scale is an order of magnitude shorter than the gas depletion time scale and the UV/optical radiation lifetime for a simple stellar population responsible for dust heating, and objects in this phase should be  ten times rarer than the objects in the AGN phase if one were to take this model results literally.  A large fraction of these DSFGs might contain a luminous AGN if this SB-AGN co-evolution scenario occurs ubiquitously.  The \WISE-selected HyLIRGs are largely distinct from the \Planck\ DSFGs in Figure~\ref{fig:AGNcolor}, and a simplistic interpretation of the  co-evolution scenario is not supported by these data.

Drawing a parallel to the X-ray regime, one possible explanation is that the ``quasar activity" associated with these \Planck\ DSFGs is so deeply embedded in dust that little or no activity is visible even at mid- to far-IR range \citep[$N_{\rm H} \gg 10^{24-25}$ cm$^{-2}$, see][]{draine89}, with uniform and complete blockage of light.  Since AGN activity is likely limited to the central several parsec region, it might require just one well placed dense gas/dust clump to hide the luminous activity.

Neither the \Planck-selection nor the \WISE-selection is finding many objects bridging the gap between the AGN and DSFGs in Figure~\ref{fig:AGNcolor}.  This might be an important clue if the popular co-evolution model applies to these hyperluminous IR galaxy populations.  The one object falling in the bridge region, W1026$-$0529, was initially reported as a $z\sim3$ HyLIRG by \citet{tsai15} and has a substantial AGN and cold dust components in its SED.  However, it was recently shown to be a lower redshift ($z=0.82$) source with a much lower IR luminosity \citep[$L_{\rm IR}\sim 10^{13}L_\odot$,][]{penney20}, and this might be a misplaced object in this comparison.  The \WISE-selected HyLIRGs are on average $\sim$3 times fainter than the \Planck-selected DSFGs, and one possible way to connect them in the co-evolution scenario is that the AGN-dominated \WISE\ sources are seen after much of the starburst activity has shutdown.  This scenario is also consistent with the average gas column densities of $N_{\rm H} \approx 10^{23-24}$ cm$^{-2}$ inferred from the measured extinction for the \WISE-selected HyLIRGs by \citet{Assef15}.  This is more than 10 times greater absorption than the typical dividing line between Type 1 and Type 2 AGN, but it is still 10 times less than the gas column required to hide the AGN completely in the IR. The Eddington ratios for their mass accretion rates derived by \citet{wu18} are close to unity, suggesting that they represent a transitional, high accretion phase between obscured and unobscured quasars, as expected in the later stages in the co-evolution scenario.

In summary, these extremely luminous IR sources without any visible signs of quasar activity seems to be the norm, rather than exceptions, among the 31 \Planck-selected DSFGs. This result contrasts with the theoretical expectation from the popular starburst-AGN co-evolution and quasar-mode feedback scenario by \citet{hopkins06,hopkins08} and the observational trend of a higher AGN fraction among DSFGs with increasing luminosity \citep[$L_{\rm IR}\ge10^{11-12} L_\odot$; e.g.,][]{treister10}.  The correction for lensing amplification should lower the intrinsic luminosity of these \Planck\ DSFGs by an order of magnitude or more (see next section), but the absence of evidence for an accreting AGN in the IR is still unexpected.  An additional factor, such as an extreme opacity with $N_{\rm H} \gtrsim 10^{24}$ cm$^{-2}$, is needed to bridge the gap between the model and the data.  Future observations at wavelengths between 50 \& 100 \micron\ and hard X-ray observations at $\ge30-100$ keV energy or a reflected component \citep[e.g., Fe K line,][]{teng09} are crucial for setting strong constraints on any energetic contribution by a deeply embedded AGN.  

\subsubsection{Strong Gravitational Lensing \label{sec:lensing}}

\begin{figure}
\includegraphics[width=0.9\columnwidth]{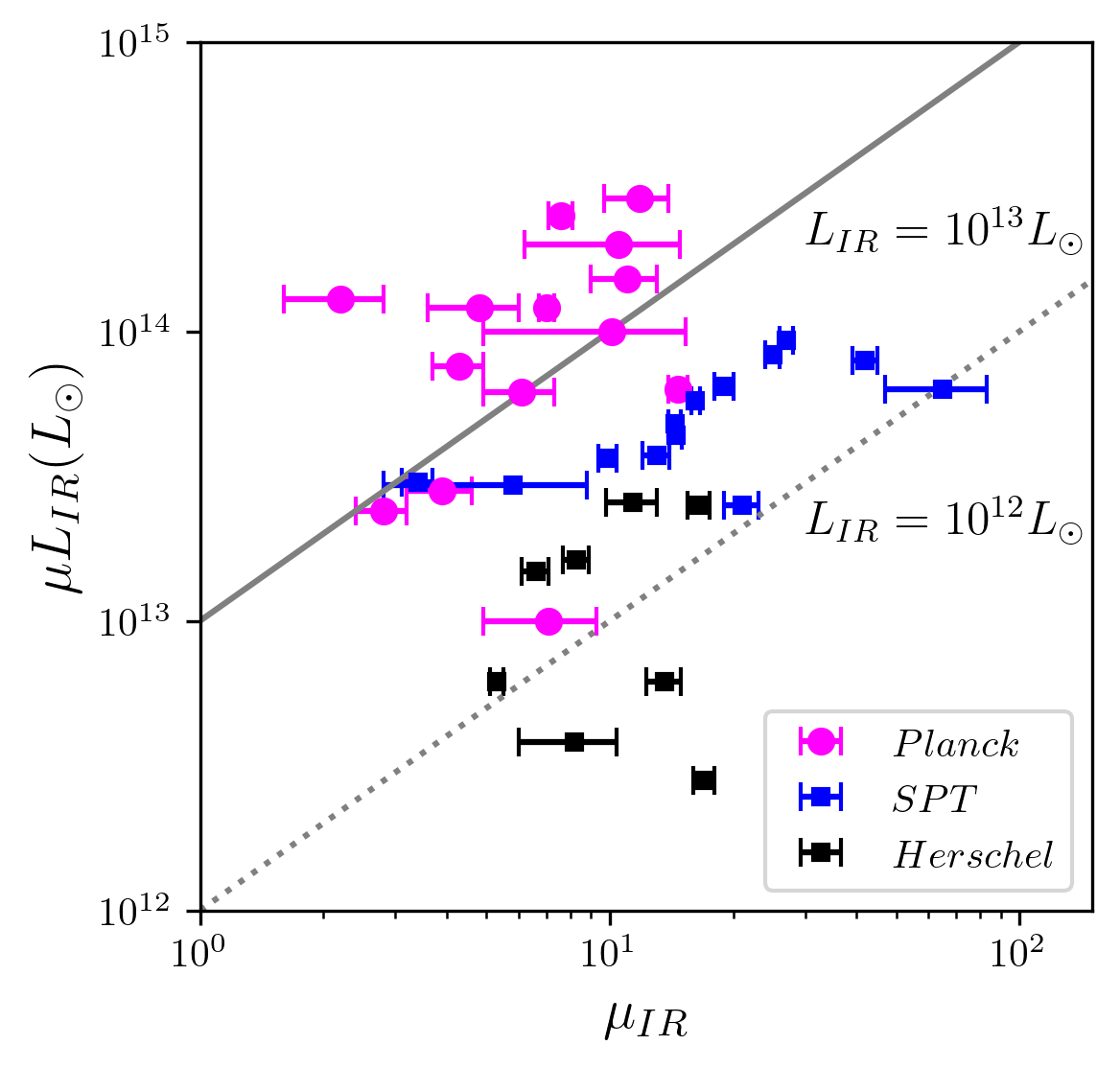}
\caption{Apparent IR luminosity of the \Planck, \Herschel, and \SPT\ selected lensed DSFGs as a function of their total lensing magnification $\mu_{\rm IR}$. The total magnification factors derived from detailed gravitational lens modelling are shown in magenta filled circles (see Table~\ref{tab:chisq}), and they are compared with equivalent quantities for \Herschel\ and \SPT\ selected lensed DSFGs (shown in filled squares) by \citet{bussmann13,bussmann15,dye15,dye18,rybak15,spilker16,massardi18}. The dotted and solid lines represent a source with intrinsic IR luminosity of $L_{\rm IR}=10^{12} L_\odot$ and $10^{12} L_\odot$, respectively. }
\label{fig:maglfir}
\end{figure}

Star formation is a self-regulated process where gas supply is modulated by the mechanical and radiative feedback from young stars.  In the most extreme starbursts, radiation pressure can limit the luminosity density and star formation density to $\sim10^{13} L_\odot$ kpc$^{-2}$ and $\sim10^3 M_\odot$ yr$^{-1}$ kpc$^{-2}$, respectively \citep{thompson05}.  Such ``Eddington-limited" starburst conditions are observed in local ULIRGs with $L_{\rm IR}\gtrsim10^{12} L_\odot$ but limited only to their central 100 parsec regions.  In more extreme high redshift SMGs with $L_{\rm IR}\gtrsim10^{13} L_\odot$, multiple such regions are found embedded in a more diffuse, kpc scale structure \citep[e.g.,][]{iono16,tadaki18}.  An even more luminous object with $L_{\rm IR}\gtrsim10^{14} L_\odot$ would require a correspondingly larger gas/SB structure spanning at least several to 10 kpc in size and an even larger cold gas reserve ($M_{\rm ISM}\gtrsim 10^{12} M_\odot$) and gas inflow rate ($\dot{M}_{\rm gas}\gtrsim 10^4 M_\odot$ yr$^{-1}$).

Gravitationally lensed objects are common at the bright end of the submillimetre source counts \citep[e.g.,][]{blain98a,blain98b,negrello10}.  Strong lensing is likely important for these \Planck-selected sources as well, and the ALMA and HST images of the 12 new \Planck\ sources shown in Figure~\ref{fig:ALMAstamps} clearly support the lensing hypothesis with their characteristic Einstein ring or arc morphology (see \S~\ref{sec:ALMA}).  The angular extent of the submillimetre continuum and stellar light ranges between 1 and 10 arcseconds, making them some of the largest Einstein rings or arcs ever found \citep[Lowenthal et al., in prep. -- also see][]{frye19}.  They are fully resolved by the HST ($\theta\approx0.1\arcsec$) and the ALMA ($\theta\approx0.4\arcsec$) in almost all cases.  
A detailed discussion of lensing models and interpretations requires a significant amount of additional data and extensive technical descriptions, and they will be presented elsewhere (Kamieneski et al., in prep.).  

The total magnification for the IR continuum $\mu_{\rm IR}$ derived from the ALMA 260 GHz continuum images ranges between 2.2 and 14.7 (a median value of $\sim$7) for the \Planck-selected DSFGs with high resolution ALMA data (see Table~\ref{tab:chisq} and Fig.~\ref{fig:maglfir}).  Dust continuum imaged by the ALMA corresponds to a rest frame wavelength of $350\, \mu$m at $z=2$, and the derived $\mu_{\rm IR}$ values are reasonable estimates of the total magnification factor for their IR luminosity and gas mass.  They are similar to the range of total magnification derived for the strongly lensed DSFGs selected in the rest-frame IR by \Herschel\ and \SPT\ surveys \citep{bussmann13,bussmann15,dye15,dye18,rybak15,spilker16,massardi18}.  The lensing magnification derived for some of the \SPT\ selected sources by \citet{spilker16} are slightly larger in comparison, and this likely reflects some systematic differences in model assumptions or input data.  For example, \citet{enia18} and others have found that extended sources with multiple emission components lead to a lower magnification while a bright, compact emitting region can be magnified by a larger factor.  

A comparison of apparent total IR luminosity as a function of their total lensing magnification factor $\mu_{\rm IR}$ shown in Figure~\ref{fig:maglfir} indicates that the \Planck-selected DSFGs have 3-10 times higher apparent IR luminosity because their {\em intrinsic} luminosity is larger. 
With the exception of 1 or 2 sources, nearly all of the \Planck-selected DSFGs lie along or above the long solid line representing a DSFG with an intrinsic luminosity $L_{\rm IR}=10^{13} L_\odot$.  A natural explanation is that the \Planck-selected DSFGs are the examples of strongly lensed DSFGs at the tip of their luminosity function while the majority of the \Herschel\ and \SPT\ lensed DSFGs represent the more numerous, lower luminosity sources with $(1-3)\times 10^{12} L_\odot$.

Whether these \Planck\ selected DSFGs are simply the most luminous but otherwise ordinary examples of the known population of DSFGs or something entirely different remains an interesting question.  By adopting a source count model for the unlensed far-IR or submillimetre source population and a distribution of lensing optical depth computed from a halo mass function, one can predict the frequency of extremely bright DSFG population, as was done by \citet{blain98a,blain98b} and \citet{negrello10}.  For example, Figure~1 by \citet{negrello10} shows that all sources brighter than 100 mJy at $500\, \mu$m wavelength band should be strongly lensed versions of otherwise normal background DSFGs.  At the integrated source density of 0.01 deg$^{-2}$  for the \Planck\ DSFGs, even the strongly lensed DSFG population is expected to decrease rapidly as it approaches the 300 mJy level according to this model.  In comparison, the majority of the $500\, \mu$m flux density of the 24 \Planck\ DSFGs listed in Table~\ref{tab:smm} are larger than 300 mJy, with a median of 426 mJy.   Noise bias and source blending can lead to an over-estimate of flux density, and two sources (Cosmic Eyelash, PJ090403.9) are associated with significantly higher PCCS flux density compared with their \Herschel/SPIRE $500\, \mu$m photometry among the sources with both \Planck\ and \Herschel\ measurements.  On the other hand, the remaining five out of seven sources show consistent photometry, and their \Herschel/SPIRE $500\, \mu$m flux density (341 to 771 mJy) exceeds the model prediction. This apparent discrepancy might not be hugely significant as the details of these predictions depend on the precise modelling of the lensing optical depth \citep{blain98a,blain98b}.  Finding a significant number of new sources with exceedingly bright submillimetre continuum in our ongoing follow-up study may require a new analysis of the lensing statistics and the nature of the DSFG population.

\subsection{Gas Mass and Consumption Time \label{sec:gasmass}}

Since the amount of total cold gas is one of the key drivers for the star formation rate in individual galaxies (``Kennicutt-Schmidt Law") and for the cosmic star formation density evolution \cite[e.g.,][]{yun09,walter14,scoville17}, methods to determine cold gas masses of distant galaxies are subjects to continued research and improvement.  Detecting gas mass tracers such as CO lines and dust continuum with sufficient S/N ratios is difficult and time-consuming for high redshift systems, and both the measurements and their gas mass calibrations are frequently subject to large uncertainties.  

The CO lines and dust continuum associated with the \Planck-selected DSFGs reported here and by \citet{canameras15} and \citet{harrington16} are significantly brighter than the most systems studied at similar redshifts previously, and the resulting high S/N measurements enable highly reliable and consistent gas mass estimates  (see \citet{harrington20} for a more complete review of this topic).  With this improved sense of reliability, we can examine their gas consumption times and compared with them with those of other populations of high redshift DSFGs.

\begin{figure}
\includegraphics[width=0.8\columnwidth]{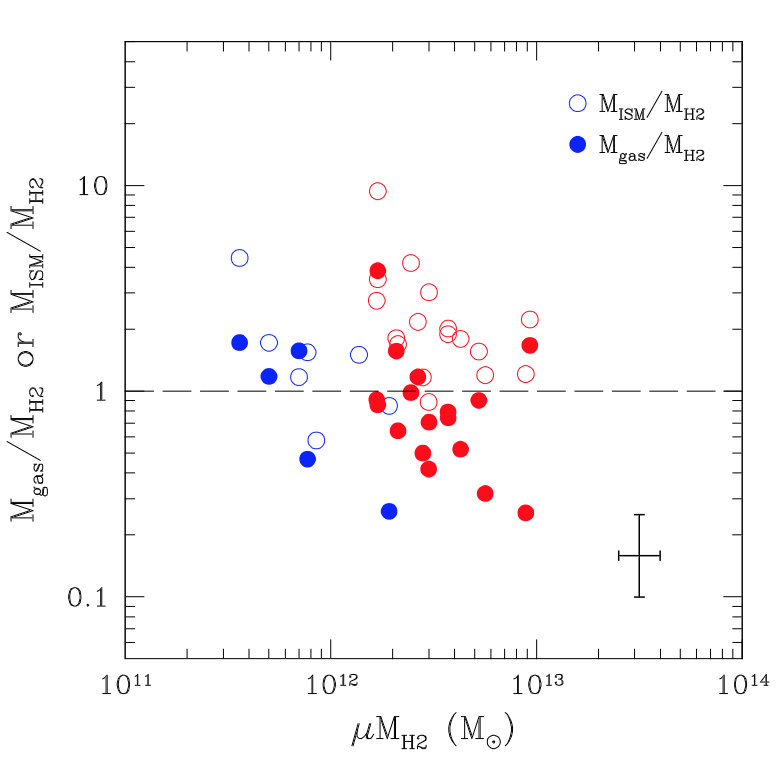}
\caption{Comparison of total molecular gas mass derived from the multi-transition data using the full radiative transfer model \citep[$M_{\rm gas}$,][]{harrington20}, $M_{\rm ISM}$ (gas mass derived from 1 mm continuum), and $M_{\rm H2}$ (molecular gas mass derived from RSR derived CO line luminosity). The molecular gas mass $\mu M_{\rm H2}$ shown on the x-axis is the apparent gas mass without correcting for the magnification factor (not known yet in all cases).  Red and blue symbols represent $M_{\rm H2}$ masses derived from CO (3--2) and CO (2--1) line, respectively.  The long dashed line marks the ``$M_{\rm ISM}/M_{\rm H2}=1$" relation, indicating the one-to-one correspondence between the ISM mass and the CO-derived molecular gas mass.  The error bar shown on the bottom right corner illustrates the typical measurement uncertainty associated with these quantities.
}
\label{fig:comparemass}
\end{figure}

\subsubsection{Gas Mass Estimates from CO Lines \label{sec:gasmassCO}}

One or more CO lines detected by the RSR for the 24 \Planck-selected sources listed in Table~\ref{tab:RSR} can yield a redshift and line luminosity $L'_{\rm CO}$ for each line detected, and this CO line luminosity can be used to estimate the total molecular gas mass $M_{\rm H2}$ by adopting a set of standard assumptions as described in \S~\ref{sec:RSR}.  
A major uncertainty in this gas mass calibration due to gas excitation can be directly addressed by measuring the full CO rotational ladder and analysing the data using a detailed radiative transfer model.  To address this issue, we have recently conducted a survey of the CO rotational transitions and [C I] lines for a sample of 24 \Planck-selected DSFGs (including the 16 sources discussed in this paper) using the Green Bank Telescope (GBT), IRAM 30-m telescope, and the APEX telescope \citep{harrington20}. A total of 160 CO lines and 37 [C~I] lines are analysed to derive gas excitation and bulk properties using two different non-LTE radiative transfer models. The radiative transfer model assuming a turbulence driven log-normal gas density distribution that simultaneously fits the CO and [C~I] lines as well as the dust continuum is new and one of the most innovative models to date, and readers are invited to examine that paper for important technical details. The total molecular gas masses derived from this analysis are shown as ``$M_{\rm gas}$" in Table~\ref{tab:RSR}. The data and the analysis presented by Harrington et al. confirm that the CO line excitation, even when measured with a high SNR ($\gtrsim10-20$), can vary considerably among different galaxies. 

A comparison of the total molecular gas masses derived from the multi-transition data using a full radiative transfer model \citep[$M_{\rm gas}$,][]{harrington20} with the molecular gas masses derived from the RSR-derived CO line luminosity ($M_{\rm H2}$, see \S~\ref{sec:RSR}) is shown in Figure~\ref{fig:comparemass}.  The agreement between $M_{\rm gas}$ and $M_{\rm H2}$ is very good on average with a median ratio of $M_{\rm gas}/M_{\rm H2} \approx 0.9$ (with a scatter of 0.35 dex). \citet{harrington20} found an average <$\alpha_{\rm CO}$> value of $3.4\pm2.1$, with a mean value of 4.2 for the galaxies with the best dust photometry and CO/[CI] line coverage, and this is very close to our adopted value of 4.3 \citep{bolatto13} for the analysis of the RSR data. This explains the good agreement between the two derived molecular gas mass estimates and much of the scatter in the measured gas mass ratios. The uncertainty in the line excitation as well as the source-to-source variation likely dominates the scatter in the derived mass ratios.  This comparison suggests that molecular gas masses derived from a single CO line adopting the standard assumptions (\S~\ref{sec:RSR}) are generally reliable, although any individual estimate may be off by a factor of 2 or more \citep[see Figure~13 by][]{harrington20}.

\subsubsection{Gas Mass Estimates from 1 mm Continuum \label{sec:gasmass1mm}}

The total dust mass $M_{\rm d}$ and the ``ISM mass" $M_{\rm ISM}$ derived from the observed 1.1 mm AzTEC photometry or ALMA 260 GHz photometry for assumed dust temperature of $T_{\rm d}=25$ K are listed in the last two columns of Table~\ref{tab:chisq}.  The ISM mass $M_{\rm ISM}$ is derived from the empirical relation between the optically thin dust continuum in the Rayleigh-Jeans regime and total cold gas mass through an empirical calibration of total gas-to-dust mass ratio \citep{scoville16,scoville17}:
\begin{equation}
	M_{\rm ISM}(M_\odot) = \frac{1.78\times 10^4  D_{\rm L}^{2}}{(1+z)^{4.8}}[\frac{S_{\rm \nu,obs}}{\rm mJy}][\frac{353\, {\rm GHz}}{\nu_{\rm obs}}]^{3.8} [\frac{\Gamma_{0}}{\Gamma_{\rm RJ}}]
\end{equation}
where $D_{\rm L}$ is the luminosity distance in megaparsec, $S_{\rm \nu,obs}$ is the measured flux density in mJy (for $\lambda_{\rm rest} \ge 250$ \micron), and the last term is the R-J correction factor.  We adopt the same dust emissivity $\beta=1.8$ to be consistent with the Scoville et al. derivation.  

This total ISM mass $M_{\rm ISM}$ is a widely used quantity because 1 mm continuum is much easier and faster to measure using the ALMA or a broadband photometer such as AzTEC, compared with CO line measurements.  However, $M_{\rm ISM}$ is also subject to systematic uncertainties depending on adopted $T_{\rm d}$ and $\beta$, as well as CO and dust abundances.  A comparison of the total ISM mass ($M_{\rm ISM}$) with the molecular gas mass derived from CO line luminosity ($M_{\rm H2}$) in Figure~\ref{fig:comparemass} shows that the ISM masses derived from the 1 mm dust continuum are systematically larger with a median ratio of  $M_{\rm ISM}/M_{\rm H2}\approx 1.5$ ($\sigma= 0.15$ in dex), nearly independent of the apparent gas mass.  

One plausible explanation for this systematic difference is that the CO to H$_2$ conversion relation used: the value Scoville et al. adopted is 50\% larger than the $\alpha_{\rm CO}$ we adopted from \citet{bolatto13}.\footnote{See \citet{scoville16} for a detailed recent review of this CO to H$_2$ conversion factor.}  However, this is only a part of the complete story.  For example, $M_{\rm ISM}\equiv M_{\rm HI} + M_{\rm H2}$ includes both {\em atomic} and molecular gas masses and is always expected to be larger than $M_{\rm H2}$ alone. The molecular gas mass fraction is expected to be near unity among these massive gas-rich galaxies \citep{scoville17}, unlike in the local late type galaxies where $M_{\rm HI}$ dominates the gas mass \citep[e.g.,][]{keres05}.   Similarly, adopting dust temperature of $T_{\rm d}=35$ K (might be appropriate for gas/dust heated by an intense starburst) instead of the mass-weighted dust temperature $T_{\rm d}=25$ K, the derived $M_{\rm ISM}$ drops by a factor $\approx2$ \citep{harrington16}.
The comparison of different gas mass estimates shown in Figure~\ref{fig:comparemass} suggests that there is a {\em systematic} uncertainty of order 50\% in total gas masses estimated using the different methods.

\begin{figure}
\includegraphics[width=0.9\columnwidth]{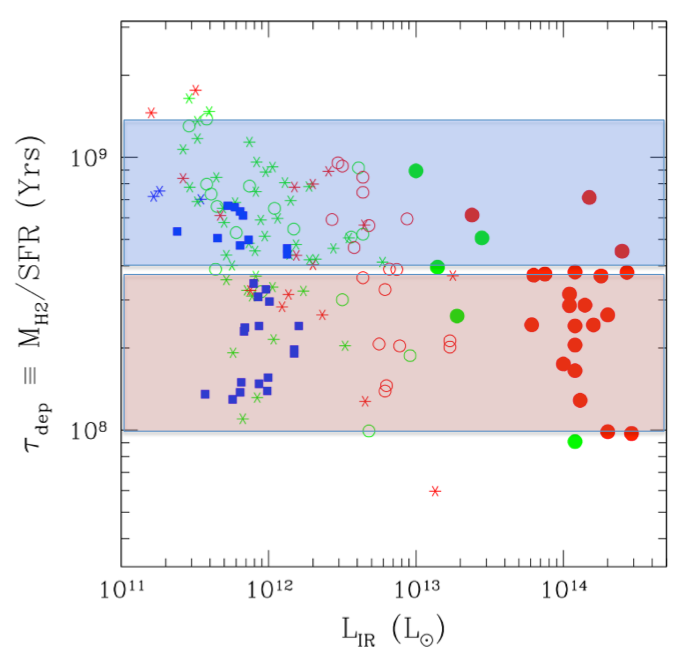}
\caption{Comparison of gas depletion time $\tau_{\rm dep} \equiv M_{\rm H2}/SFR$  as a function of apparent IR luminosity.  \Planck-selected sources are shown as filled green ($z<2$) and red ($z\ge2$) circles.  They are compared with representative samples of stellar mass selected ``main sequence" star forming galaxies \citep[stars;][]{magdis12,tacconi13}, submm/IR selected starbursts \citep[SMGs, in empty circles;][]{magnelli12}, and $z=0$ IR luminous galaxies \citep[LIRGs/ULIRGs, in blue squares;][]{chung09}.  The red and blue shaded regions mark the mean ranges of $\tau_{\rm dep}$ for galaxies on the star forming main sequence and starburst galaxies with 10 times the SFR for a characteristic stellar mass of $5\times 10^{10} M_\odot$ \citep[see Figure~10 by][]{scoville17}.
}
\label{fig:gastime}
\end{figure}

\subsubsection{Gas Consumption/Depletion Time \label{sec:gastime}}

 The gas depletion times $\tau_{\rm dep}$ ($\equiv M_{\rm H2}/SFR$)\footnote{Gas depletion time is defined as $\tau_{\rm dep} \equiv \frac{1}{2} M_{\rm H2}/SFR$ in some papers, assuming 1/2 of the gas is lost to feedback/outflow, but we ignore this order unity coefficient since the coupling efficiency between feedback and ISM is poorly understood.} for the \Planck-selected DSFGs are shown in Figure~\ref{fig:gastime}, and they cover the same range as those of the unlensed DSFGs and star forming main sequence (MS) galaxies studied by \citet{magnelli12}, \citet{magdis12}, and \citet{tacconi13} -- also, see the review by \citet{tacconi20}.  This may indicate that they share the same underlying physical processes that drive and regulate star formation in these galaxies, despite their 10 to 100 times larger apparent IR luminosity.  A more detailed comparison with the ranges of $\tau_{\rm dep}$ characterising the main sequence star forming galaxies and starburst galaxies at $z=1-4$ (shaded regions in Figure~\ref{fig:gastime}) suggests that these \Planck-selected DSFGs mostly show gas depletion times more characteristic of a starburst population \citep[$\tau_{\rm dep}=(1-3)\times 10^8$ yrs, similar to the local ULIRGs,][]{chung09} than the star forming "main sequence" population.  Accounting for the magnification factor of $\mu\sim10$, the intrinsic IR luminosities of these \Planck-selected DSFGs ($1-3\times10^{13}L_\odot$, see Figure~\ref{fig:maglfir}) are near the top end of the IR luminosity of the submm/IR selected DSFGs studied by \citet{magdis12} and others, and their gas consumption times also agree well.

The similar gas depletion times derived for the \Planck-selected DSFGs with those of the local ULIRGs and high redshift DSFGs has several interesting implications.  Firstly, this result indicates that differential magnification associated with lensing does not play a significant role here.  In theory, CO and dust continuum emission could be amplified by different amounts and could lead to a larger scatter in $\tau_{\rm dep} \equiv M_{\rm H2}/SFR$.  The observed scatter is similar to or tighter than those of the unlensed galaxies, and this might be a natural consequence of lensing magnifying areas larger than the spatial scale where the correlation between star formation and gas density is known to hold \citep[0.5-1 kiloparsec; see][and references therein]{schinnerer19}.  This comparison also supports the claim by \citet{scoville17} that the lack of dependence of SF efficiency ($1/\tau_{\rm dep}$) on galaxy mass is the result of star formation process dictated by the internal structure of the GMCs, rather than global galaxy properties.  Additionally, this comparison also supports the absence of a substantial AGN contribution to the IR luminosity for these \Planck-selected DSFGs discussed in \S~\ref{sec:AGN}, as any significant IR luminosity by a dust obscured AGN would lead to an over-estimate of SFR and correspondingly shortened $\tau_{\rm dep}$.

\begin{figure*}
\includegraphics[height=7cm]{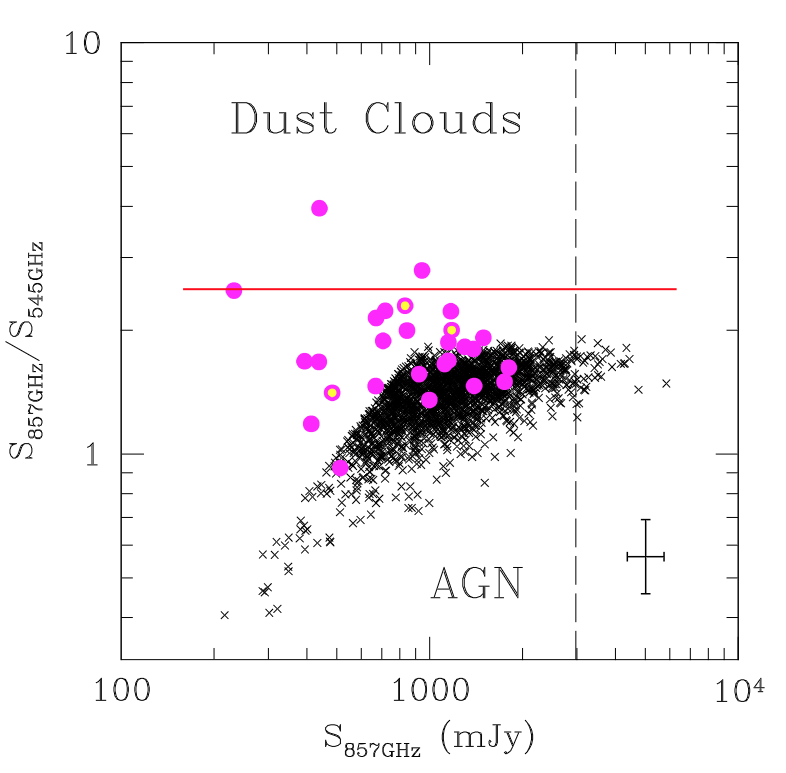}
\includegraphics[height=7cm]{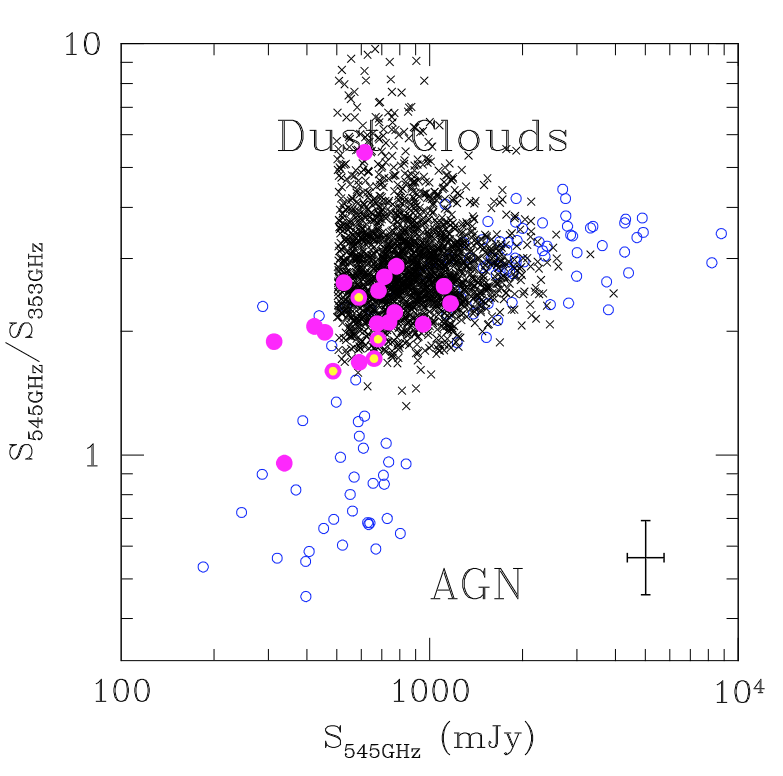}
\caption{Plots of flux density ratios as a function flux density as in Figure~\ref{fig:Pcolmag} but this time comparing the \Planck\ PHZ sources \citep[small black dots,][]{planck39}. All confirmed high-redshift PCCS sources and a subsample from the \Planck-\Herschel\ study by \citet{canameras15} are shown with large filled and empty magenta circles. As in Figure~\ref{fig:Pcolmag}, the red line on the left panel is the flux density ratio used to define our sample, and the empty blue circles on the right panel are 857 GHz dropout sources which are either very cold dust sources or flat spectrum AGNs. Typical uncertainties are shown on the bottom right corner of each figure.
}
\label{fig:Pcolmag2}
\end{figure*}

\subsection{Comparison with the \Planck\ High-Redshift Source Catalog (PHZ) \label{sec:PHZ}}

The \Planck\ science team compiled a list of 2151 high-redshift (PHZ) source candidates with $S_{\rm 545GHz}\ge500$ mJy in the cleanest 26\% of the sky using a component-separation procedure applied to the \Planck\ and IRAS data \citep{planck39}.  Rather than starting from the compact source catalogs PCCS1 or PCCS2, the PHZ sources are identified from the analysis of the \Planck\ survey data itself.  Therefore, the PHZ selection  diverges from our approach from this starting point.  The stated intent of the PHZ selection is ``to select the most luminous cold submillimetre sources with spectral energy distributions peaking between 353 and 857 GHz at 5$\arcmin$ resolution".  However, the details of the colour selections are also substantially different from ours (\S~\ref{sec:PlanckFilter}), and the readers are referred to the PHZ paper for the details of their candidate selection.  Their final list of 2151 PHZ source candidates is an order of magnitude larger in size compared with our final candidate list covering nearly the same area (see \S~\ref{sec:Planck} and Table~\ref{tab:sourcestat}).  The PHZ catalog is substantially different from the PCCS catalogs, and this is an important factor to consider for this comparison with our sample (see below).

The source density of the PHZ catalog, $\sim$0.2 per deg$^{-2}$, is much higher than the known source density of the submillimetre and infrared source populations at the sample flux density level ($\ge$500 mJy at 545 GHz), and the \Planck\ Collaboration has suggested that the majority of these PHZ sources with estimated IR luminosity of $L_{\rm IR}=(0.5-7)\times 10^{14} L_\odot$ are actually blends of multiple objects, rather than single strongly lensed objects.  By analysing the \Herschel\ follow-up observations of the 228 \Planck\ high-z source candidates \citep{planck26}, the \Planck\ Collaboration estimates that 3\% of the PHZ sources are strongly lensed DSFGs at $z=2-4$ while the remaining sources are overdensities consisting of 10 or more closely clustered DSFGs.  \citet{martinache18} have reported a significant overdensity of \Spitzer/IRAC sources associated with 82 PHZ proto-cluster candidates identified by the \Herschel\ data, and detailed studies of a few individual PHZ sources have demonstrated an overdensity in the redshift space as well \citep{florescacho16,kneissl19,koyama21}.  However, based on a statistical analysis and modelling, \citet{negrello17a} have argued that the majority of these PHZ sources are positive Poisson fluctuations of the number of dusty sources within the \Planck\ beam, rather than being individual clumps of physically bound galaxies.

When we first started our high-redshift source search in the \Planck\ database in 2013, we were unaware that the \Planck\ collaboration science team was conducting its own search for high-redshift galaxies in the full survey data.  The idea of candidate selection by analysing spectral energy distribution and colour selection is broadly similar, but the differences in the parent \Planck\ catalogs and source filtering resulted in two rather different final catalogs of high-redshift source candidates.  Examining these details should yield an interesting insight into the nature of these candidate sources.

Comparisons of the 857 GHz to 545 GHz and 545 GHz to 353 GHz flux density and flux density ratios, shown in Figure~\ref{fig:Pcolmag2}, clearly demonstrate how the two samples differ in their sample definitions.  The comparison of the $S_{\rm 857GHz}/S_{\rm 545GHz}$ flux density ratio as a function of 857 GHz flux density (left panel) shows that the overlap between the PHZ sample (small dots) and our confirmed high-redshift DSFGs (large dots) is not very good.  This is consistent with our finding only five (PJ020941.3, PJ084650.1, PJ113921.7, PJ144958.6, PJ154432.4) of our 27 PCCS-selected $z\ge1$ DSFGs \citep[from this study and][]{harrington16} with a counterpart within a 150\arcsec\ search radius in the PHZ catalogue.  
This figure demonstrates that the exclusion of the majority of our sources in the PHZ catalogue is the direct consequence of the colour and flux cuts used by the PHZ catalogue generation: (1) $S_{\rm 857GHz}/S_{\rm 545GHz} < 2$; and (2) and $S_{\rm 545GHz}>500$ mJy \citep[see][]{planck39}.  The combined effect of these two selection criteria leads to a fan-shaped distribution of PHZ sources on the left panel, and {\em both} of these selection filters play a role in excluding the majority of our PCCS-selected high-redshift DSFGs. Among the sources detected in both 545 \& 353 GHz bands, the overlap is much better as seen on the right panel.  Unlike the first case, the flux density cut of $S_{\rm 545GHz}>500$ mJy has only a minor impact here, presumably because the dust peak shifts into the \Planck\ 545 GHz band at $z\ge2$.  

As noted by the PHZ team in their paper, the overlap between the PHZ and PCCS catalogues is extremely small, with a total of only 35 sources in common.  The number of 857 GHz drop-out sources in the PCCS catalogs is also very small ($\sim$100 total, see \S~\ref{sec:PlanckFilter}), and the filtering for ``cold" SED sources adopted for the PHZ catalog must be an important factor behind this discrepancy between the PCCS and the PHZ catalogs. To be clear, the PHZ catalog sources are not exclusively 857 GHz drop-outs, as is obvious in the left panel of Figure~\ref{fig:Pcolmag2}. However, the PHZ sample selection for cold dust colour is directly responsible for excluding the majority of the PCCS-selected PASSAGES sources.

The \Planck\ Collaboration paper also suggests that the PHZ catalog is complementary to the PCCS catalogs by picking out the faintest and coldest objects at high latitude.  However, the comparisons shown in Figure~\ref{fig:Pcolmag2} does not fully support this claim in that their $S_{\rm 857GHz}/S_{\rm 545GHz} < 2$ colour selection combined with the flux density cutoff of $S_{\rm 545GHz}>500$ mJy is the primary driver behind excluding the majority of the strongly magnified DSFGs identified by our PASSAGES project. In fact, the ranges of the 857 \& 545 GHz flux density are very similar.  

In summary, our filtering of the PCCS catalogs and the sample selection criteria used to identify the ``cold" dusty sources for the PHZ catalog are similar enough that they could potentially recover the same objects, but they actually do not.  The comparison of the two catalogs in Figure~\ref{fig:Pcolmag2} shows that important detailed differences in the PHZ sample definition are leading to the exclusion of the majority of the strongly lensed DSFGs we identified from the PCCS catalogs in the PHZ catalog.  It will be interesting to see whether future searches for strong lensed DSFGs in the PHZ catalog turns up a different population of colder dust sources compared with those found by the PASSAGES project.

\section{Conclusions}

We report the identification of 22 high redshift luminous dusty star forming galaxies from a candidate list constructed using the \Planck\ Catalog of Compact Sources \citep[PCCS;][]{planck13} and a carefully designed colour and magnitude selection algorithm that utilises the all-sky \Planck\ and \WISE\ photometry data.  Their redshifts of  $z=1.1-3.3$ are confirmed through AzTEC 1.1 mm continuum imaging and CO spectroscopy obtained using the Redshift Search Receiver on the Large Millimeter Telescope.  Combined with the eight high-redshift \Planck\ sources identified using the archival \Herschel\ data in our pilot study \citep{harrington16}, this study increases the total number of high-redshift sources drawn from the PCCS to 30.  Our ALMA 1.1 mm continuum imaging and photometry show that all 12 \Planck\ sources observed at $\theta\sim0.4\arcsec$ resolution are strongly lensed with characteristic ring or arc morphology.  Two of the PCCS sources are shown to be $z\approx0.2$ ULIRGs, and the 3 pairs of sources found in the directions of massive foreground clusters appear to be multiple images of the same object or associated objects based on their common CO redshifts. 

The apparent IR luminosity of the $z\ge1$ \Planck-selected DSFGs identified in our PASSAGES project ranges between 0.1 and 3.1 times $10^{14} L_\odot$ with a median value of $1.2\times10^{14}L_\odot$, making them some of the most luminous galaxies ever known.  The \Planck-selected DSFGs from this study, along with those from our pilot study by \citet{harrington16} and those similarly identified by \citet{canameras15}, account for 17 out of 25 (68\%) of {\em all} known objects with $L_{\rm IR}\ge10^{14} L_\odot$.  In addition, the five \Planck-selected DSFGs with $L_{\rm IR}\ge10^{13}L_\odot$ at $z<2$ are also some of the most luminous objects found in this redshift range.  Their molecular gas mass $M_{\rm H2}$ derived from the measured CO line luminosity ranges between 0.7 and 9.3 times $10^{12}M_\odot$, uncorrected for their unknown lensing magnification, and they are in a good agreement with the cold ISM mass $M_{\rm ISM}$ derived from the 1 mm continuum photometry.  Correcting for their mean magnification of $\mu\approx10$, their luminosity and gas masses are similar to those of the most luminous SMGs discovered by millimetre and submillimetre surveys of the cosmology deep fields, suggesting these \Planck-selected DSFGs are the magnified versions of the most luminous DSFGs found at the same redshifts.
Their gas depletion times $\tau_{\rm dep}$ ($\equiv M_{\rm H2}/SFR$) range between 100 and 500 Myr, the same as those of  the unlensed DSFGs and star forming main sequence galaxies studied by \citet{magnelli12}, \citet{magdis12}, and \citet{tacconi13}.  This common SF efficiency independent of luminosity and galaxy mass further supports the idea that the underlying SF process is dictated by the internal structure of the GMCs, rather than global galaxy properties \citep{scoville17}.

An interesting outcome from the analysis of the spectral energy distributions of these extremely luminous \Planck-selected DSFGs is the absence of any detectable AGN activity, unlike the hyperluminous infrared galaxies discovered by other all-sky IR surveys done by \IRAS\ and \WISE.  Growth of SMBH concurrent with a triggered starburst during the period of high gas mass accretion has been proposed by \citet{sanders88} based on the high frequency of warm IR sources among the most luminous ULIRGs in the local universe, and nearly simultaneous growth of SMBH and associated luminous AGN activity is one of the key predictions for a rapid galaxy growth driven by a gas-rich merger \citep{hopkins06,hopkins08}.   In contrast to this popular scenario, both the SED analysis of individual galaxies shown in Figure~\ref{fig:SEDs} and the mid-to-far IR colour diagnostic analysis shown in Figure~\ref{fig:AGNcolor} suggest that the nearly entire luminosity of these \Planck\ sources is powered by star formation alone, in line with the mid-IR and X-ray studies that have found detectable AGN activities in less than 20 to 30\% of high redshift DSFGs \citep[see review by][]{casey14}.  A gas column density exceeding $10^{24-25}$ cm$^{-2}$ can obscure any AGN activity even in the IR, but the gas and dust has to cover both the AGN activity and any structures cleared by associated feedback process, with little room for any clumpy geometry.  Future IR ($\lambda_{\rm rest}=30-100\,\mu$m) and hard X-ray ($>30-100$ keV) observations are crucial for setting strong constraints on any energetic contribution by deeply embedded AGN activity.

Lastly, the comparison of our sample with the \Planck\ High-redshift Source Catalog \citep[PHZ;][]{planck39} shows  remarkably little overlap.  Important differences in the PHZ sample definition are leading to the exclusion of the majority of strongly lensed DSFGs we identified from the PCCS catalogs in the PHZ catalog. It will be interesting to see what any future follow-up studies may reveal about the nature of these PHZ sources.  

Confirming whether the majority of the 118 \Planck-selected high redshift DSFG candidates are real is an interesting next step in the PASSAGES project.  A much larger sample of \Planck-selected DSFGs can offer a potentially important insight into the nature of the submillimetre and far-IR source populations and the effects of gravitational lensing with significantly improved statistics.  Obtaining a larger sample can also offer more interesting constraints on the growth of SMBHs among galaxies with the highest intrinsic SFR.  Gravitational lensing offers a rare opportunity to probe 10-100 parsec spatial scales inaccessible to the current generation of astronomical facilities.  Finding the best laboratories for probing physical processes that govern the rapid growth and subsequent shutoff of star formation in the early epochs is also important next goal.

\section*{Acknowledgments}
Authors thank N. Z. Scoville for his insightful discussions on the comparison of luminosity produced by a starburst and a SMBH.  We also benefited from useful discussions with R. Snell and J. Vieira.   MSY thanks D. F. Gallup for his assistance during the initial development of the project and J. L. Yun for producing some of the plots shown here.  
This work would not have been possible without the long-term financial support from the Mexican Science and Technology Funding Agency, CONACYT (Consejo Nacional de Ciencia y Tecnolog\'{i}a) during the construction and early operational phase of the Large Millimeter Telescope Alfonso Serrano, as well as support from the US National Science Foundation via the University Radio Observatory program, the Instituto Nacional de Astrof\'{i}sica, \'{O}ptica y Electr\'{o}nica (INAOE) and the University of Massachusetts, Amherst (UMass). The UMass LMT group acknowledges support from NSF URO and ATI grants  (AST-0096854, AST-0215916, AST-0540852, and AST-0704966) for the LMT project and the construction of the RSR and AzTEC.  
A.M. thanks support from Consejo Nacional de Ciencia y Technolog\'ia (CONACYT) project A1-S-45680.  DB, KH, and RC would like to acknowledge support from a William Bannick Student Travel Grant.  We are grateful to all of the LMT observers from Mexico and UMass who took data for this project. This publication makes use of data products from the Wide-field Infrared Survey Explorer, which is a joint project of the University of California, Los Angeles, and the Jet Propulsion Laboratory/California Institute of Technology, funded by the National Aeronautics and Space Administration. This work is based in part on observations made with the Herschel Space Observatory, which is an ESA space observatory with science instruments provided by European-led Principal Investigator consortia and with important participation from NASA, and the Planck, which is a European Space Agency mission with significant NASA involvement.  
This research has made use of the NASA/ IPAC Extra-galactic Database (NED) which is operated by the Jet Propulsion Laboratory, California Institute of Technology, under contract with the National Aeronautics and Space Administration. This research has made use of the NASA/IPAC Infrared Science Archive, which is funded by the National Aeronautics and Space Administration and operated by the California Institute of Technology.
This paper makes use of the following ALMA data: ADS/JAO.ALMA\#2017.1.01214.S. ALMA is a partnership of ESO (representing its member states), NSF (USA) and NINS (Japan), together with NRC (Canada), MOST and ASIAA (Taiwan), and KASI (Republic of Korea), in cooperation with the Republic of Chile. The Joint ALMA Observatory is operated by ESO, AUI/NRAO and NAOJ.  The National Radio Astronomy Observatory is a facility of the National Science Foundation operated under cooperative agreement by Associated Universities, Inc.

\section*{Data Availability}
The data underlying this article are available in the article and in its online supplementary material. Other datasets were derived from sources in the public domain, such as the \ALMA\ and \HST\ data archives as well as NASA/IPAC IRSA,






\bibliographystyle{mnras}
\bibliography{references} 

%



\appendix

\section{Photometry of Candidate Sources Undetected by AzTEC}

AzTEC follow-up observations of the 30 sources identified by the \Planck-\WISE\ selection are described in \S~\ref{sec:AzTEC}.  The 13 sources clearly detected as bright 1.1 mm continuum sources are summarized in Table~\ref{tab:smm}.  The remaining 17 sources are undetected by AzTEC, and the details of those observations and their sensitivity are summarized in Table~\ref{tab:NoAzTEC}.  These initial targets for AzTEC observations were drawn from a much larger parent list than our ``high priority" candidates.  There are no obvious indications in the \Planck\ data that would set these non-detections apart from the detected sources.  At least some of these non-detections appear to be compact Galactic cirrus clouds that passed through our initial colour-magnitude filters, as our strategy was all along to keep the initial filtering simple and to identify high redshift sources using LMT follow-up observations (see discussions in \S~\ref{sec:Planck}).

\begin{table*}
 \caption{Summary of Sources Undetected by AzTEC on LMT}
 \label{tab:NoAzTEC}
 \begin{tabular}{@{}lccccccc}
  \hline
  ID & RA & DEC & \multicolumn{3}{c}{AzTEC} \\
  & (J2000) & (J2000) & Dates & Int. Time (mins) & $S_{1100\mu}^a$ \\
  \hline
  PJ012508.1 & 01h25m08.1s  & $+$29d22m55s & 2015-02-05 & 10 & $\le7.8$ mJy &  \\  
  PJ042847.0 & 04h28m47.0s  & $-$27d30m58s & 2016-01-24 & 10 & $\le6.7$ mJy &  \\  
  PJ082721.1 & 08h27m21.1s  & $+$60d16m08s & 2015-04-06 & 10 & $\le4.2$ mJy &  \\
  PJ083622.6 & 08h36m22.6s  & $+$23d40m18s & 2016-01-22 & 10 & $\le7.1$ mJy &  \\  
  PJ084759.7 & 08h47m59.7s  & $+$17d42m12s & 2014-11-10 & 10 & $\le3.2$ mJy &  \\  
  PJ091626.2 & 09h16m26.2s  & $+$18d55m35s & 2015-04-05 & 10 & $\le3.9$ mJy &  \\  
  PJ093055.1 & 09h30m55.1s  & $+$46d48m57s & 2015-04-05, 2015-04-06 & 20 & $\le3.3$ mJy &  \\  
  PJ094609.1 & 09h46m09.1s  & $+$00d43m12s & 2016-01-23 & 10 & $\le8.6$ mJy &  \\  
  PJ094921.3 & 09h49m21.3s  & $+$14d50m28s & 2016-01-22 & 10 & $\le7.9$ mJy &  \\  
  PJ095326.1 & 09h53m26.1s  & $+$70d53m40s & 2016-01-24 & 10 & $\le7.1$ mJy &  \\  
  PJ100023.1 & 10h00m23.1s  & $+$12d15m21s & 2015-04-06 & 10 & $\le4.1$ mJy &  \\  
  PJ101804.5 & 10h18m04.5s  & $+$28d12m05s & 2016-02-05 & 10 & $\le9.3$ mJy &  \\  
  PJ103146.9 & 10h31m46.9s  & $-$22d12m44s & 2015-02-24 & 10 & $\le4.0$ mJy &  \\  
  PJ105248.6 & 10h52m48.9s  & $+$42d15m43s & 2015-02-24 & 10 & $\le4.2$ mJy &  \\  
  PJ112025.4 & 11h20m25.4s  & $+$05d53m33s & 2016-02-05 & 10 & $\le5.8$ mJy &  \\  
  PJ112716.7 & 11h27m16.7s  & $+$42d28m39s & 2016-01-23 & 10 & $\le7.9$ mJy &  \\  
  PJ114329.5 & 11h43m29.5s  & $+$68d01m07s & 2016-01-23 & 10 & $\le6.8$ mJy &  \\  
    \hline
\end{tabular}
\break
      $^a$ $S_{1100\mu}$ listed is a $3\sigma$ upper limit based on the photometry over a 1.5 arcmin diameter region centred on the source position.\\
 \end{table*}

\section{Notes on Individual Sources \label{sec:sources}}

In this section we briefly summarise each of the 13 high-redshift \Planck\ sources which have not been studied previously (Table~\ref{tab:smm}). Complementing the eight sources identified in \citet{harrington16}, there are 24 additional high-redshift \Planck\ sources identified with the selection criteria presented in this study. 

\medskip
\noindent{\bf P011646.8:}
The derived source redshift, $z = 2.125$, is based on the CO (3--2) line detected with the RSR and subsequent spectroscopic confirmation by \citet{harrington20}. This source is detected in two \Planck\ bands and is also imaged in 260 GHz continuum using ALMA.    The HST/WFC3 1.6 \micron\ image reveals a $\sim6\arcsec$ diameter, nearly complete double-Einstein ring.  No redshift information is available for the lensing galaxy at the moment. The ALMA 260 GHz image spatially resolves the two dominant lensing arc features in dust continuum.  

\medskip
\noindent{\bf PJ014341.2:}
A photometric redshift support from the combined SED analysis of the \Planck, \Herschel\ and ALMA measurements suggests that the single line detected by the RSR measurement corresponds to a CO (2--1) line emission at a source redshift, $z = 1.095$. The ALMA 260 GHz observation reveals a compact galaxy-galaxy lensing system with a partially complete Einstein ring with a diameter of $\sim1.5\arcsec$. The foreground lensing SDSS galaxy has a spectroscopic redshift of $z = 0.594$. 

\medskip
\noindent{\bf PJ022634.0:}
With a \Planck\ detection of $S_{350} \sim 1800$ mJy, this system is the brightest in our sample. Its apparent SED peaking in the \Planck\ 350 \micron\ band suggests this to be a low redshift ($z\le2$) source, but our subsequent spectroscopic follow-up observations by \citet{harrington20} have shown that the single line detected by the RSR is a CO (3--2) line at $z = 3.120$. The HST/WFC3 1.6 \micron\ image (Lowenthal et al., in prep) shows an Einstein ring with a diameter of $\sim3\arcsec$ centred on a large elliptical galaxy with a photometric redshift of $z=0.41$ \citep{wen15}. 

\medskip
\noindent{\bf PJ030510.6:}
The derived source redshift, $z = 2.263$, is based on the CO (3--2) line detection with the RSR, and this is confirmed by our follow-up spectroscopy \citep{harrington20}. It is detected in two \Planck\ bands, and ALMA 260 GHz measurements recover the total flux density measured by AzTEC 1.1 mm observations. The ALMA resolves the source into two separate and compact peaks with a less than $1\arcsec$ separation.  
The HST/WFC3 1.6 \micron\ image shows one stellar component well centred on the brighter 260 GHz peak while the second stellar component is clearly offset from the second 260 GHz peak.  No redshift is available for either stellar galaxies. 

\medskip
\noindent{\bf PJ074851.7 \& PJ074852.6:}
This cluster lensed system at $z=2.755$ has two \WISE-selected candidates, PJ074851.7 and PJ074852.6.  The two colour-selected \WISE\ sources are targeted separately using the RSR, and a CO (3--2) line is detected in both objects with a redshift of $z = 2.755$.  These two sources are seen as a single \Planck\ source, with $S_{350}\sim 1000$ mJy.  The HST/WFC3 1.6 \micron\ image shows several lensed arc features in a cluster field, and one of the lensed arclets extends $15-20\arcsec$ in length. The source redshift of PJ074852.6 has been verified with our follow-up spectroscopy campaign \citep{harrington20}.  The foreground cluster PSZ2 G157.43+30.34 has a published redshift of 0.42 by \citet{khatri16} and 0.402 by \citet{amodeo18}.

\medskip
\noindent{\bf PJ084648.6 \& PJ084650.1:}
This cluster lensed system at $z=2.66$ has two \WISE-selected candidates, PJ084648.6 and PJ084650.1, and they are both detected in CO (3--2) line by the RSR at $z=2.664$ and $z=2.661$, respectively.  The HST/WFC3 1.6 \micron\ image as well as ALMA 260 GHz continuum image show them to be two distinct lensed arcs.  Because they are both associated with the same \Planck\ point source ($S_{350} \sim 1750$ mJy), they are treated as a single \Planck\ source in this paper. The source redshift of PJ084650.1 has been verified with our subsequent spectroscopic follow-up observations by \citet{harrington20}. 
The redshift of the foreground cluster is somewhat uncertain since few spectroscopic redshifts are available in this field.  The bright elliptical galaxy located between the two lensed arcs is an SDSS source with a photometric redshift of $z\sim 0.4$.

\medskip
\noindent{\bf PJ105322.6:}
The two CO lines detected in the RSR spectrum leads to an unambiguous redshift of $z=3.549$ for this 857 GHz drop-out \Planck\ source.  This is the brightest 1.1 mm sources detected with AzTEC ($202\pm 30$ mJy) and has a fully mapped rest frame far-IR SED because this is one of the \Planck\ sources studied using \Herschel\ (PLCK\_G145.2+50.9).   The SMA 850 \micron\ continuum imaging by \citet{canameras15} shows a nearly complete Einstein ring with $\sim5\arcsec$ radius.  Using their \HST\ F110W and F160W band imaging and ground-based spectroscopy, \citet{frye19} have shown that the lensing mass is a small compact group of red galaxies at $z=0.837$.

\medskip
\noindent{\bf PJ112713.4:}
This system has only a single \Planck\ detection, with $S_{350} \sim 200$ mJy, and was not observed by AzTEC or by ALMA. The single line detected by the RSR observations is tentatively interpreted as a CO (2--1) line at $z = 1.303$, and this is later confirmed by our spectroscopic follow-up observations \citep{harrington20}.  The HST/WFC3 1.6 \micron\ image shows an Einstein ring with $\sim1\arcsec$ diameter centred on a compact galaxy with an SDSS photometric redshift of $z=0.42$, suggesting a galaxy-galaxy lensing (Lowenthal et al., in prep).  

\medskip
\noindent{\bf PJ113805.5:}
The single line detected by the RSR is interpreted to be a CO (2--1) line at a redshift of $z = 2.019$ based on photometric support using the SED analysis, and this redshift is confirmed by our spectroscopic follow-up observations \citep{harrington20}. This source is detected in all 3 \Planck\ bands and by ALMA, with $S_{350} \sim 200$ mJy. It is likely lensed by a single foreground elliptical galaxy with an SDSS photometric redshift of $z=0.52$, as the ALMA 260 GHz continuum image shows a compact ($\le1.5\arcsec$) elongated structure centred on the optical galaxy seen in the HST/WFC3 1.6 \micron\ image.  

\medskip
\noindent{\bf PJ113921.7:}
PJ113921.7 is another high-redshift \Planck\ source studied with a \Herschel\ follow-up study (PLCK\_G231.3+72.2) that we independently recovered from the PCCS using our \WISE\ colour selection and AzTEC photometry.  The CO (3--2) line redshift of $z=2.858$ is derived from our RSR observations and the photometric support, and this redshift agrees well with those  measured by other multi-line studies \citep{canameras18,nesvadba19,harrington20}.  Our ALMA 260 GHz continuum image shows a compact ring of emission with $\sim1.5\arcsec$ diameter, centred around a compact galaxy seen in our HST/WFC3 1.6 \micron\ image (Lowenthal et al., in prep).  No visible counterpart to the ALMA 260 GHz continuum emission is seen in the HST image, and the redshift of the lensing galaxy is unknown.

\medskip
\noindent{\bf PJ114038.5:}
The colour-selected \WISE\ source J114038.5+53215 associated with this PCCS object shows a compact red core surrounded by a blue irregular structure, characteristic of a recent merger system in the SDSS images.  It was previously identified as one of the infrared bright galaxies found by the Infrared Astronomical Satellite (``IRAS 2Jy Sample"), with a redshift of $z=0.092$ \citep{strauss92}.  The SDSS database offers no spectroscopic data, but it reports a photometric redshift of $0.127\pm0.029$.  Our RSR spectrum shows a single emission line at 98.932 GHz, and this corresponds to a spectroscopic redshift of $z=0.1652\pm0.0001$ if the line is interpreted as that of CO (1--0) transition.  While these redshifts do not match exactly, they are in a reasonable agreement.  Considering its identification as an \IRAS\ 2Jy Sample source along with its relatively bright radio continuum ($16.8\pm0.3$ mJy at 1.4 GHz in the FIRST database),  we conclude that PJ114038.5 is a low-z galaxy at $z=0.1652$.  

\medskip
\noindent{\bf PJ114329.5:}
The colour-selected \WISE\ source J114329.5+680107 associated with this PCCS object is associated with a well resolved SDSS galaxy with red colour and a smooth morphology.  The SDSS spectrum shows strong emission lines and bright continuum, typical of a star forming galaxy at $z=0.212$.  Our RSR spectrum shows a single emission line at 95.124 GHz, and this corresponds to a spectroscopic redshift of $z=0.2118\pm0.0001$ assuming the line is that of CO (1--0) transition.  Therefore, we conclude that PJ114329.5 is another example of a foreground low-z galaxy.

\medskip
\noindent{\bf PJ132217.5:}
The single CO (2--1) line detected with the RSR yields a redshift of $z = 2.068$ with the photometric redshift support, and this is confirmed by our spectroscopic follow-up observations \citep{harrington20}. The unresolved \Planck\ detection peaks at $S_{350}\sim 700$ mJy, and the 260 GHz continuum is resolved into multiple clumps with the $0.4\arcsec$ resolution ALMA image.  The lensed source morphology detected by ALMA (also a stellar Einstein ring detected in the HST/WFC3 1.6 \micron\ image) is nearly $20\arcsec$ in diameter, and some of the 260 GHz flux might have been resolved out by ALMA, given its large angular extent.  Multiple foreground galaxies are seen in the HST/WFC3 image, and a group potential might be responsible for the lensing (Lowenthal et al., in prep).  No spectroscopic redshifts are available for the foreground galaxies, and the brightest SDSS galaxy inside the Einstein ring has a photometric redshift of $z\sim0.53$.

\medskip
\noindent{\bf PJ132630.3:}
PJ132630.3 is a \Planck\ high-redshift candidate source that is found in the \Herschel-ATLAS survey area, and it is one of the faintest \Planck\ sources detected in just 2 bands, with $S_{350}\sim 300$ mJy.  The single CO (3--2) line detected with the RSR yields a redshift of $z = 2.951$ with the photometric redshift support, and this redshift is confirmed by subsequent multi-line observations \citep{yang17,harrington20}.  This source is also identified as one of the brightest \Herschel\ source found in the H-ATLAS survey with the same redshift, and \citet{bussmann13} derived a lensing model using their SMA 340 GHz continuum image and the lensing galaxy redshift of 0.786, with a derived magnification factor of $\mu=4.1\pm0.3$ for the 340 GHz continuum.  Our higher resolution ALMA 260 GHz image shows a similar morphology as their SMA 340 GHz image, which is two images straddling the central lensing galaxy.

\medskip
\noindent{\bf PJ132934.2 \& PJ132935.3:}
PJ132934.2 and PJ132935.3 are two \WISE-selected high-redshift \Planck\ candidates, associated with a single \Planck\ source detected in all 3 \Planck\ bands.  The two colour-selected \WISE\ sources are targeted separately using the RSR, and a CO (2--1) line is detected in both objects with a redshift of $z = 2.040$.  Our ALMA 260 GHz continuum image shows that PJ132934.2 is significantly brighter than PJ132935.3 in submillimetre continuum, and the HST/WFC3 1.6 \micron\ image shows they are both gravitationally lensed arcs associated with a massive foreground cluster.  These sources were discovered independently by \citet{diazsanchez17} who also used the \Planck\ and \WISE\ data (``Cosmic Eyebrow"). Their multi-wavelength observations and modelling found that these two sources are the same object lensed by the foreground cluster at $z=0.44$ \citep{dannerbauer19}.

\medskip
\noindent{\bf PJ133634.9:}
The two bright CO lines detected by the RSR are CO (3--2) and CO (4--3) lines at $z=3.254$, and this interpretation is further supported by our spectroscopic follow-up observations \citep{harrington20}. It is detected in all 3 \Planck\ bands and by AzTEC, with $S_{350} \sim 650$ mJy. 
The HST/WFC3 1.6 \micron\ image reveals a pair of compact galaxies with an SDSS photometric redshift of $z=0.26$, surrounded by a ring of nebulosity with $\sim1\arcsec$ diameter (Lowenthal et al., in prep). 

\medskip
\noindent{\bf PJ141230.5:}
The two bright lines detected by the RSR are interpreted as CO (3--2) and CO (4--3) lines at $z=3.310$, and this conclusion is further supported by the SED analysis for the measured photometry in the 3 \Planck\ bands as well as the \WISE\ and VLA/FIRST 1.4 GHz photometry (see Fig.~\ref{fig:SEDs}).  The foreground (likely lensing) SDSS galaxy has only a photometric redshift of $0.54\pm0.07$.  However, a clear over-density of similarly red galaxies is seen in the SDSS image centred on the \WISE\ source position, including two galaxies with a SDSS spectroscopic redshift of $z=0.588$ within a 30$\arcsec$ radius, and there is a high likelihood that the foreground lensing source is also at $z=0.588$.  

\medskip
\noindent{\bf PJ144653.2:}
This system has \Planck, AzTEC and ALMA measurements of the dust SED, with a peak \Planck\ detection of $S_{350} \sim 1150$ mJy.  Given its brightness in the rest frame far-IR, the single line detected by the RSR was initially thought to be a CO line at $z\approx2$.  However, our follow-up spectroscopy has shown that the RSR detection is of the CO (2--1) line at $z = 1.084$ \citep{harrington20}. The HST/WFC3 image shows a late type galaxy with an SDSS spectroscopic redshift of $z=0.493$.  The clearly resolved ALMA continuum peaks form a $2\arcsec$ diameter ring centred on the galaxy, suggesting a galaxy-galaxy lensing.

\medskip
\noindent{\bf PJ144958.6:}
The RSR detected CO (3--2) line corresponds to a source redshift of $z = 2.153$, and this is confirmed by our spectroscopic follow-up observations \citep{harrington20}.  The \Planck\ detection peaks at $S_{350} \sim 700$ mJy, and the ALMA 260 GHz continuum observations have recovered the complete flux density of the AzTEC 1.1 mm measurements. The stellar light in the HST/WFC3 image and the ALMA 260 GHz continuum both trace the same 
$\sim10\arcsec$ long lensed arc, with some notable differences -- only parts of the stellar arc are associated with 260 GHz continuum. Therefore, this source is clearly spatially resolved by lensing and is likely a multiple system.  A galaxy group (of unknown redshift) contributes to the foreground gravitational potential, resulting in one of the largest lens features in our sample. 

\medskip
\noindent{\bf PJ154432.4:}
PJ154432.4 is another high-redshift \Planck\ source studied with a \Herschel\ follow-up study (PLCK\_G080.2+49.8) that we independently recovered from the PCCS using our \WISE\ colour selection and AzTEC photometry.  The CO (3--2) line redshift of $z=2.5989\pm0.0003$ derived from our RSR observations and the photometric support agrees well with the redshift reported by \citet[][``$z_{\rm source}=2.6$"]{canameras15}. The extension and morphology of the 1.1 mm continuum imaged by AzTEC instrument (see Figure~\ref{fig:AzTEC}) matches the higher resolution SMA 850 $\mu$m image by Ca{\~n}ameras et al. and suggests lensing by a foreground potential with a mass exceeding $10^{13} M_\odot$. The SDSS spectrum of the foreground lensing galaxy shows emission lines at $z=0.672$ while Ca{\~n}ameras et al. reported a redshift of ``$z_{\rm fg,sdss}=0.5$" for the foreground galaxy.
The wealth of photometry found in the \Planck\ and \Herschel\ database as well as our own AzTEC photometry define the SED nicely with a modified blackbody temperature of $41.4^{+7.0}_{-4.2}$ K (see Fig.~\ref{fig:SEDs} and Table~\ref{tab:chisq}), and the VLA/FIRST 1.4 GHz photometry is consistent with this object following the radio-FIR correlation \citep{condon92,yun01}.

\medskip
\noindent{\bf PJ231356.6:}
The bright line detected by the RSR is interpreted as the CO (3--2) line at $z = 2.215$, and this is confirmed by our spectroscopic follow-up \citep{harrington20}.  It was detected in two \Planck\ bands and was imaged by our ALMA 260 GHz continuum observations.  PJ231356.6 is located in the Atacama Cosmology Telescope (ACT) Equatorial Survey field and is identified as one of the 30 brightest DSFGs with computed deboosted flux density of $S_{148GHz}=3.2^{+1.7}_{-1.4}$ mJy, $S_{218GHz}=16.8^{+3.8}_{-4.4}$ mJy, $S_{277GHz}=26.8^{+5.3}_{-5.3}$ mJy \citep{gralla20}.   The HST/WFC3 image shows possibly multiple lensed objects, only some associated with the ALMA 260 GHz continuum,  centred around a large elliptical galaxy with an SDSS spectroscopic redshift of $z=0.560$.


\bsp	
\label{lastpage}
\end{document}